\documentclass[journal, 10pt]{IEEEtran}
\PassOptionsToPackage{dvipsnames,table}{xcolor}
\usepackage[utf8]{inputenc}% optional T1 font encoding
\usepackage[english]{babel}
\usepackage[ruled, vlined]{algorithm2e}
\usepackage{amsthm}
\usepackage{algpseudocode}
\usepackage{amssymb}
\usepackage{colortbl}
\usepackage{arydshln}
\usepackage{blkarray, bigstrut}
\usepackage{bm}
\usepackage{booktabs}
\usepackage{braket}
\usepackage{cite}
\usepackage{cuted}
\usepackage{enumitem}
\usepackage{graphicx}
\usepackage{glossaries}
\usepackage{indentfirst}
\usepackage[mathscr]{euscript}
\usepackage[switch]{lineno}
\usepackage{mathrsfs}
\usepackage{mathtools}
\usepackage{multirow}
\usepackage{multicol}
\usepackage{makecell}
\usepackage{lipsum}
\usepackage{xfrac}
\usepackage{pifont}
\usepackage{pgfplots}
\usepackage{pgfplotstable}
\pgfplotsset{compat=1.16}
\usepackage{filecontents}
\usepackage{adjustbox}
\usepackage{rotating}
\usepackage{float}
\usepackage{stfloats}
\usepackage[skip=2pt,font=footnotesize]{caption}
\usepackage{subcaption}
\usepackage{url}
\usepackage{tikz}
\usetikzlibrary{positioning}
\usepackage{verbatimbox}
\usetikzlibrary{patterns}
\usetikzlibrary{spy}
\usetikzlibrary{shapes, backgrounds,calc, snakes}
\usetikzlibrary{decorations.pathreplacing}
\usetikzlibrary{matrix}
\usepackage[normalem]{ulem}
\usepackage[compact]{titlesec}
\usepackage{thmtools}
\usepackage{balance}
\usepackage{totpages}
\usepackage[nodisplayskipstretch]{setspace}
\newsavebox{\twosubbox}

\pgfplotsset{
    colormap={parula}{
        rgb255=(53,42,135)
        rgb255=(15,92,221)
        rgb255=(18,125,216)
        rgb255=(7,156,207)
        rgb255=(21,177,180)
        rgb255=(89,189,140)
        rgb255=(165,190,107)
        rgb255=(225,185,82)
        rgb255=(252,206,46)
        rgb255=(249,251,14)
    },
}

\declaretheoremstyle[
spaceabove=5pt, spacebelow=4pt,
headfont=\normalfont\bfseries,
notefont=\mdseries, notebraces={(}{)},
bodyfont=\normalfont,
postheadspace=0.5em,
qed=\qedsymbol,
shaded={rulecolor=black,
rulewidth=1pt, bgcolor={rgb}{1,1,1}}
]{mystyle}

\declaretheorem[style=mystyle]{proposition}
\usetikzlibrary{external}
\tikzexternalenable

\newcommand*\circled[1]{\tikz[baseline=(char.base)]{
            \node[shape=circle,draw,inner sep=2pt] (char) {#1};}}

\makeatletter
\def\newmaketag{%
	\def\maketag@@@##1{\hbox{\m@th\normalfont\normalsize##1}}%
}
\makeatother

\definecolor{blue1}{HTML}{b3cde0}
\definecolor{blue2}{HTML}{6497b1}
\definecolor{blue3}{HTML}{005b96}
\definecolor{blue4}{HTML}{03396c}
\definecolor{blue5}{HTML}{011f4b}

\definecolor{dcolor1}{HTML}{253494}
\definecolor{dcolor2}{HTML}{636363}
\definecolor{dcolor3}{HTML}{fcbba1}
\definecolor{dcolor4}{HTML}{fb6a4a}
\definecolor{dcolor5}{HTML}{cb181d}
\definecolor{dcolor6}{HTML}{67000d}
\definecolor{dcolor7}{HTML}{9ecae1}
\definecolor{dcolor8}{HTML}{4292c6}
\definecolor{dcolor9}{HTML}{08519c}

\definecolor{layer1}{HTML}{003f5c}
\definecolor{layer2}{HTML}{444e86}
\definecolor{layer3}{HTML}{955196}
\definecolor{layer4}{HTML}{dd5182}
\definecolor{layer5}{HTML}{ff6e54}
\definecolor{layer6}{HTML}{ffa600}

\definecolorset{rgb}{}{}{%
AliceBlue,.94,.972,1;%
AntiqueWhite,.98,.92,.844;%
Aqua,0,1,1;%
Aquamarine,.498,1,.83;%
Azure,.94,1,1;%
Beige,.96,.96,.864;%
Bisque,1,.894,.77;%
Black,0,0,0;%
BlanchedAlmond,1,.92,.804;%
Blue,0,0,1;%
BlueViolet,.54,.17,.888;%
Brown,.648,.165,.165;%
BurlyWood,.87,.72,.53;%
CadetBlue,.372,.62,.628;%
Chartreuse,.498,1,0;%
Chocolate,.824,.41,.116;%
Coral,1,.498,.312;%
CornflowerBlue,.392,.585,.93;%
Cornsilk,1,.972,.864;%
Crimson,.864,.08,.235;%
Cyan,0,1,1;%
DarkBlue,0,0,.545;%
DarkCyan,0,.545,.545;%
DarkGoldenrod,.72,.525,.044;%
DarkGray,.664,.664,.664;%
DarkGreen,0,.392,0;%
DarkGrey,.664,.664,.664;%
DarkKhaki,.74,.716,.42;%
DarkMagenta,.545,0,.545;%
DarkOliveGreen,.332,.42,.185;%
DarkOrange,1,.55,0;%
DarkOrchid,.6,.196,.8;%
DarkRed,.545,0,0;%
DarkSalmon,.912,.59,.48;%
DarkSeaGreen,.56,.736,.56;%
DarkSlateBlue,.284,.24,.545;%
DarkSlateGray,.185,.31,.31;%
DarkSlateGrey,.185,.31,.31;%
DarkTurquoise,0,.808,.82;%
DarkViolet,.58,0,.828;%
DeepPink,1,.08,.576;%
DeepSkyBlue,0,.75,1;%
DimGray,.41,.41,.41;%
DimGrey,.41,.41,.41;%
DodgerBlue,.116,.565,1;%
FireBrick,.698,.132,.132;%
FloralWhite,1,.98,.94;%
ForestGreen,.132,.545,.132;%
Fuchsia,1,0,1;%
Gainsboro,.864,.864,.864;%
GhostWhite,.972,.972,1;%
Gold,1,.844,0;%
Goldenrod,.855,.648,.125;%
Gray,.5,.5,.5;%
Grey,.5,.5,.5;%
Green,0,.5,0;%
GreenYellow,.68,1,.185;%
Honeydew,.94,1,.94;%
HotPink,1,.41,.705;%
IndianRed,.804,.36,.36;%
Indigo,.294,0,.51;%
Ivory,1,1,.94;%
Khaki,.94,.9,.55;%
Lavender,.9,.9,.98;%
LavenderBlush,1,.94,.96;%
LawnGreen,.488,.99,0;%
LemonChiffon,1,.98,.804;%
LightBlue,.68,.848,.9;%
LightCoral,.94,.5,.5;%
LightCyan,.88,1,1;%
LightGoldenrodYellow,.98,.98,.824;%
LightGray,.828,.828,.828;%
LightGreen,.565,.932,.565;%
LightGrey,.828,.828,.828;%
LightPink,1,.712,.756;%
LightSalmon,1,.628,.48;%
LightSeaGreen,.125,.698,.668;%
LightSkyBlue,.53,.808,.98;%
LightSlateGray,.468,.532,.6;%
LightSlateGrey,.468,.532,.6;%
LightSteelBlue,.69,.77,.87;%
LightYellow,1,1,.88;%
Lime,0,1,0;%
LimeGreen,.196,.804,.196;%
Linen,.98,.94,.9;%
Magenta,1,0,1;%
Maroon,.5,0,0;%
MediumAquamarine,.4,.804,.668;%
MediumBlue,0,0,.804;%
MediumOrchid,.73,.332,.828;%
MediumPurple,.576,.44,.86;%
MediumSeaGreen,.235,.7,.444;%
MediumSlateBlue,.484,.408,.932;%
MediumSpringGreen,0,.98,.604;%
MediumTurquoise,.284,.82,.8;%
MediumVioletRed,.78,.084,.52;%
MidnightBlue,.098,.098,.44;%
MintCream,.96,1,.98;%
MistyRose,1,.894,.884;%
Moccasin,1,.894,.71;%
NavajoWhite,1,.87,.68;%
Navy,0,0,.5;%
OldLace,.992,.96,.9;%
Olive,.5,.5,0;%
OliveDrab,.42,.556,.136;%
Orange,1,.648,0;%
OrangeRed,1,.27,0;%
Orchid,.855,.44,.84;%
PaleGoldenrod,.932,.91,.668;%
PaleGreen,.596,.985,.596;%
PaleTurquoise,.688,.932,.932;%
PaleVioletRed,.86,.44,.576;%
PapayaWhip,1,.936,.835;%
PeachPuff,1,.855,.725;%
Peru,.804,.52,.248;%
Pink,1,.752,.796;%
Plum,.868,.628,.868;%
PowderBlue,.69,.88,.9;%
Purple,.5,0,.5;%
Red,1,0,0;%
RosyBrown,.736,.56,.56;%
RoyalBlue,.255,.41,.884;%
SaddleBrown,.545,.27,.075;%
Salmon,.98,.5,.448;%
SandyBrown,.956,.644,.376;%
SeaGreen,.18,.545,.34;%
Seashell,1,.96,.932;%
Sienna,.628,.32,.176;%
Silver,.752,.752,.752;%
SkyBlue,.53,.808,.92;%
SlateBlue,.415,.352,.804;%
SlateGray,.44,.5,.565;%
SlateGrey,.44,.5,.565;%
Snow,1,.98,.98;%
SpringGreen,0,1,.498;%
SteelBlue,.275,.51,.705;%
Tan,.824,.705,.55;%
Teal,0,.5,.5;%
Thistle,.848,.75,.848;%
Tomato,1,.39,.28;%
Turquoise,.25,.88,.815;%
Violet,.932,.51,.932;%
Wheat,.96,.87,.7;%
White,1,1,1;%
WhiteSmoke,.96,.96,.96;%
Yellow,1,1,0;%
YellowGreen,.604,.804,.196}

\definecolorset{rgb}{}{}{%
AntiqueWhite1,1,.936,.86;%
AntiqueWhite2,.932,.875,.8;%
AntiqueWhite3,.804,.752,.69;%
AntiqueWhite4,.545,.512,.47;%
Aquamarine1,.498,1,.83;%
Aquamarine2,.464,.932,.776;%
Aquamarine3,.4,.804,.668;%
Aquamarine4,.27,.545,.455;%
Azure1,.94,1,1;%
Azure2,.88,.932,.932;%
Azure3,.756,.804,.804;%
Azure4,.512,.545,.545;%
Bisque1,1,.894,.77;%
Bisque2,.932,.835,.716;%
Bisque3,.804,.716,.62;%
Bisque4,.545,.49,.42;%
Blue1,0,0,1;%
Blue2,0,0,.932;%
Blue3,0,0,.804;%
Blue4,0,0,.545;%
Brown1,1,.25,.25;%
Brown2,.932,.23,.23;%
Brown3,.804,.2,.2;%
Brown4,.545,.136,.136;%
Burlywood1,1,.828,.608;%
Burlywood2,.932,.772,.57;%
Burlywood3,.804,.668,.49;%
Burlywood4,.545,.45,.332;%
CadetBlue1,.596,.96,1;%
CadetBlue2,.556,.898,.932;%
CadetBlue3,.48,.772,.804;%
CadetBlue4,.325,.525,.545;%
Chartreuse1,.498,1,0;%
Chartreuse2,.464,.932,0;%
Chartreuse3,.4,.804,0;%
Chartreuse4,.27,.545,0;%
Chocolate1,1,.498,.14;%
Chocolate2,.932,.464,.13;%
Chocolate3,.804,.4,.112;%
Chocolate4,.545,.27,.075;%
Coral1,1,.448,.336;%
Coral2,.932,.415,.312;%
Coral3,.804,.356,.27;%
Coral4,.545,.244,.185;%
Cornsilk1,1,.972,.864;%
Cornsilk2,.932,.91,.804;%
Cornsilk3,.804,.785,.694;%
Cornsilk4,.545,.532,.47;%
Cyan1,0,1,1;%
Cyan2,0,.932,.932;%
Cyan3,0,.804,.804;%
Cyan4,0,.545,.545;%
DarkGoldenrod1,1,.725,.06;%
DarkGoldenrod2,.932,.68,.055;%
DarkGoldenrod3,.804,.585,.048;%
DarkGoldenrod4,.545,.396,.03;%
DarkOliveGreen1,.792,1,.44;%
DarkOliveGreen2,.736,.932,.408;%
DarkOliveGreen3,.635,.804,.352;%
DarkOliveGreen4,.43,.545,.24;%
DarkOrange1,1,.498,0;%
DarkOrange2,.932,.464,0;%
DarkOrange3,.804,.4,0;%
DarkOrange4,.545,.27,0;%
DarkOrchid1,.75,.244,1;%
DarkOrchid2,.698,.228,.932;%
DarkOrchid3,.604,.196,.804;%
DarkOrchid4,.408,.132,.545;%
DarkSeaGreen1,.756,1,.756;%
DarkSeaGreen2,.705,.932,.705;%
DarkSeaGreen3,.608,.804,.608;%
DarkSeaGreen4,.41,.545,.41;%
DarkSlateGray1,.592,1,1;%
DarkSlateGray2,.552,.932,.932;%
DarkSlateGray3,.475,.804,.804;%
DarkSlateGray4,.32,.545,.545;%
DeepPink1,1,.08,.576;%
DeepPink2,.932,.07,.536;%
DeepPink3,.804,.064,.464;%
DeepPink4,.545,.04,.312;%
DeepSkyBlue1,0,.75,1;%
DeepSkyBlue2,0,.698,.932;%
DeepSkyBlue3,0,.604,.804;%
DeepSkyBlue4,0,.408,.545;%
DodgerBlue1,.116,.565,1;%
DodgerBlue2,.11,.525,.932;%
DodgerBlue3,.094,.455,.804;%
DodgerBlue4,.064,.305,.545;%
Firebrick1,1,.19,.19;%
Firebrick2,.932,.172,.172;%
Firebrick3,.804,.15,.15;%
Firebrick4,.545,.1,.1;%
Gold1,1,.844,0;%
Gold2,.932,.79,0;%
Gold3,.804,.68,0;%
Gold4,.545,.46,0;%
Goldenrod1,1,.756,.145;%
Goldenrod2,.932,.705,.132;%
Goldenrod3,.804,.608,.112;%
Goldenrod4,.545,.41,.08;%
Green1,0,1,0;%
Green2,0,.932,0;%
Green3,0,.804,0;%
Green4,0,.545,0;%
Honeydew1,.94,1,.94;%
Honeydew2,.88,.932,.88;%
Honeydew3,.756,.804,.756;%
Honeydew4,.512,.545,.512;%
HotPink1,1,.43,.705;%
HotPink2,.932,.415,.655;%
HotPink3,.804,.376,.565;%
HotPink4,.545,.228,.385;%
IndianRed1,1,.415,.415;%
IndianRed2,.932,.39,.39;%
IndianRed3,.804,.332,.332;%
IndianRed4,.545,.228,.228;%
Ivory1,1,1,.94;%
Ivory2,.932,.932,.88;%
Ivory3,.804,.804,.756;%
Ivory4,.545,.545,.512;%
Khaki1,1,.965,.56;%
Khaki2,.932,.9,.52;%
Khaki3,.804,.776,.45;%
Khaki4,.545,.525,.305;%
LavenderBlush1,1,.94,.96;%
LavenderBlush2,.932,.88,.898;%
LavenderBlush3,.804,.756,.772;%
LavenderBlush4,.545,.512,.525;%
LemonChiffon1,1,.98,.804;%
LemonChiffon2,.932,.912,.75;%
LemonChiffon3,.804,.79,.648;%
LemonChiffon4,.545,.536,.44;%
LightBlue1,.75,.936,1;%
LightBlue2,.698,.875,.932;%
LightBlue3,.604,.752,.804;%
LightBlue4,.408,.512,.545;%
LightCyan1,.88,1,1;%
LightCyan2,.82,.932,.932;%
LightCyan3,.705,.804,.804;%
LightCyan4,.48,.545,.545;%
LightGoldenrod1,1,.925,.545;%
LightGoldenrod2,.932,.864,.51;%
LightGoldenrod3,.804,.745,.44;%
LightGoldenrod4,.545,.505,.298;%
LightPink1,1,.684,.725;%
LightPink2,.932,.635,.68;%
LightPink3,.804,.55,.585;%
LightPink4,.545,.372,.396;%
LightSalmon1,1,.628,.48;%
LightSalmon2,.932,.585,.448;%
LightSalmon3,.804,.505,.385;%
LightSalmon4,.545,.34,.26;%
LightSkyBlue1,.69,.888,1;%
LightSkyBlue2,.644,.828,.932;%
LightSkyBlue3,.552,.712,.804;%
LightSkyBlue4,.376,.484,.545;%
LightSteelBlue1,.792,.884,1;%
LightSteelBlue2,.736,.824,.932;%
LightSteelBlue3,.635,.71,.804;%
LightSteelBlue4,.43,.484,.545;%
LightYellow1,1,1,.88;%
LightYellow2,.932,.932,.82;%
LightYellow3,.804,.804,.705;%
LightYellow4,.545,.545,.48;%
Magenta1,1,0,1;%
Magenta2,.932,0,.932;%
Magenta3,.804,0,.804;%
Magenta4,.545,0,.545;%
Maroon1,1,.204,.7;%
Maroon2,.932,.19,.655;%
Maroon3,.804,.16,.565;%
Maroon4,.545,.11,.385;%
MediumOrchid1,.88,.4,1;%
MediumOrchid2,.82,.372,.932;%
MediumOrchid3,.705,.32,.804;%
MediumOrchid4,.48,.215,.545;%
MediumPurple1,.67,.51,1;%
MediumPurple2,.624,.475,.932;%
MediumPurple3,.536,.408,.804;%
MediumPurple4,.365,.28,.545;%
MistyRose1,1,.894,.884;%
MistyRose2,.932,.835,.824;%
MistyRose3,.804,.716,.71;%
MistyRose4,.545,.49,.484;%
NavajoWhite1,1,.87,.68;%
NavajoWhite2,.932,.81,.63;%
NavajoWhite3,.804,.7,.545;%
NavajoWhite4,.545,.475,.37;%
OliveDrab1,.752,1,.244;%
OliveDrab2,.7,.932,.228;%
OliveDrab3,.604,.804,.196;%
OliveDrab4,.41,.545,.132;%
Orange1,1,.648,0;%
Orange2,.932,.604,0;%
Orange3,.804,.52,0;%
Orange4,.545,.352,0;%
OrangeRed1,1,.27,0;%
OrangeRed2,.932,.25,0;%
OrangeRed3,.804,.215,0;%
OrangeRed4,.545,.145,0;%
Orchid1,1,.512,.98;%
Orchid2,.932,.48,.912;%
Orchid3,.804,.41,.79;%
Orchid4,.545,.28,.536;%
PaleGreen1,.604,1,.604;%
PaleGreen2,.565,.932,.565;%
PaleGreen3,.488,.804,.488;%
PaleGreen4,.33,.545,.33;%
PaleTurquoise1,.732,1,1;%
PaleTurquoise2,.684,.932,.932;%
PaleTurquoise3,.59,.804,.804;%
PaleTurquoise4,.4,.545,.545;%
PaleVioletRed1,1,.51,.67;%
PaleVioletRed2,.932,.475,.624;%
PaleVioletRed3,.804,.408,.536;%
PaleVioletRed4,.545,.28,.365;%
PeachPuff1,1,.855,.725;%
PeachPuff2,.932,.796,.68;%
PeachPuff3,.804,.688,.585;%
PeachPuff4,.545,.468,.396;%
Pink1,1,.71,.772;%
Pink2,.932,.664,.72;%
Pink3,.804,.57,.62;%
Pink4,.545,.39,.424;%
Plum1,1,.732,1;%
Plum2,.932,.684,.932;%
Plum3,.804,.59,.804;%
Plum4,.545,.4,.545;%
Purple1,.608,.19,1;%
Purple2,.57,.172,.932;%
Purple3,.49,.15,.804;%
Purple4,.332,.1,.545;%
Red1,1,0,0;%
Red2,.932,0,0;%
Red3,.804,0,0;%
Red4,.545,0,0;%
RosyBrown1,1,.756,.756;%
RosyBrown2,.932,.705,.705;%
RosyBrown3,.804,.608,.608;%
RosyBrown4,.545,.41,.41;%
RoyalBlue1,.284,.464,1;%
RoyalBlue2,.264,.43,.932;%
RoyalBlue3,.228,.372,.804;%
RoyalBlue4,.152,.25,.545;%
Salmon1,1,.55,.41;%
Salmon2,.932,.51,.385;%
Salmon3,.804,.44,.33;%
Salmon4,.545,.298,.224;%
SeaGreen1,.33,1,.624;%
SeaGreen2,.305,.932,.58;%
SeaGreen3,.264,.804,.5;%
SeaGreen4,.18,.545,.34;%
Seashell1,1,.96,.932;%
Seashell2,.932,.898,.87;%
Seashell3,.804,.772,.75;%
Seashell4,.545,.525,.51;%
Sienna1,1,.51,.28;%
Sienna2,.932,.475,.26;%
Sienna3,.804,.408,.224;%
Sienna4,.545,.28,.15;%
SkyBlue1,.53,.808,1;%
SkyBlue2,.494,.752,.932;%
SkyBlue3,.424,.65,.804;%
SkyBlue4,.29,.44,.545;%
SlateBlue1,.512,.435,1;%
SlateBlue2,.48,.404,.932;%
SlateBlue3,.41,.35,.804;%
SlateBlue4,.28,.235,.545;%
SlateGray1,.776,.888,1;%
SlateGray2,.725,.828,.932;%
SlateGray3,.624,.712,.804;%
SlateGray4,.424,.484,.545;%
Snow1,1,.98,.98;%
Snow2,.932,.912,.912;%
Snow3,.804,.79,.79;%
Snow4,.545,.536,.536;%
SpringGreen1,0,1,.498;%
SpringGreen2,0,.932,.464;%
SpringGreen3,0,.804,.4;%
SpringGreen4,0,.545,.27;%
SteelBlue1,.39,.72,1;%
SteelBlue2,.36,.675,.932;%
SteelBlue3,.31,.58,.804;%
SteelBlue4,.21,.392,.545;%
Tan1,1,.648,.31;%
Tan2,.932,.604,.288;%
Tan3,.804,.52,.248;%
Tan4,.545,.352,.17;%
Thistle1,1,.884,1;%
Thistle2,.932,.824,.932;%
Thistle3,.804,.71,.804;%
Thistle4,.545,.484,.545;%
Tomato1,1,.39,.28;%
Tomato2,.932,.36,.26;%
Tomato3,.804,.31,.224;%
Tomato4,.545,.21,.15;%
Turquoise1,0,.96,1;%
Turquoise2,0,.898,.932;%
Turquoise3,0,.772,.804;%
Turquoise4,0,.525,.545;%
VioletRed1,1,.244,.59;%
VioletRed2,.932,.228,.55;%
VioletRed3,.804,.196,.47;%
VioletRed4,.545,.132,.32;%
Wheat1,1,.905,.73;%
Wheat2,.932,.848,.684;%
Wheat3,.804,.73,.59;%
Wheat4,.545,.494,.4;%
Yellow1,1,1,0;%
Yellow2,.932,.932,0;%
Yellow3,.804,.804,0;%
Yellow4,.545,.545,0;%
Gray0,.745,.745,.745;%
Green0,0,1,0;%
Grey0,.745,.745,.745;%
Maroon0,.69,.19,.376;%
Purple0,.628,.125,.94}

\pgfplotsset{
    colormap={my colormap}{
        rgb255=(251, 106, 74),
        rgb255=(203, 24, 29),
    },
}

\def\pgfplotsinvokeiflessthan#1#2#3#4{%
    \pgfkeysvalueof{/pgfplots/iflessthan/.@cmd}{#1}{#2}{#3}{#4}\pgfeov
}%
\def\pgfplotsmulticmpthree#1#2#3#4#5#6\do#7#8{%
    \pgfplotsset{float <}%
    \pgfplotsinvokeiflessthan{#1}{#4}{%
        #7%
    }{%
        \pgfplotsinvokeiflessthan{#4}{#1}{%
            #8%
        }{%
            \pgfplotsset{float <}%
            \pgfplotsinvokeiflessthan{#2}{#5}{%
                #7%
            }{%
                \pgfplotsinvokeiflessthan{#5}{#2}{%
                    #8%
                }{%
                    \pgfplotsset{float <}%
                    \pgfplotsinvokeiflessthan{#3}{#6}{%
                        #7%
                    }{%
                        #8%
                    }%
                }%
            }%
        }%
    }%
}%

\titlespacing{\section}{0pt}{0.5ex}{0.5ex}
\titlespacing{\subsection}{0pt}{0.25ex}{0ex}
\titlespacing{\subsubsection}{0pt}{0.25ex}{0ex}

\renewcommand\arraystretch{1}

\DeclareMathSizes{11.5pt}{11pt}{6pt}{5pt} 	% Reduce math font size

\makeatletter
\patchcmd{\env@cases}{1.2}{0.9}{}{}
\makeatother

\makeatletter
\newlength\minalignvsep

\def\align@preamble{%
   &\hfil
    \setboxz@h{\@lign$\m@th\displaystyle{##}$}%
    \ifnum\row@>\@ne
    \ifdim\ht\z@>\ht\strutbox@
    \dimen@\ht\z@
    \advance\dimen@\minalignvsep
    \ht\strutbox\dimen@
    \fi\fi
    \strut@
    \ifmeasuring@\savefieldlength@\fi
    \set@field
    \tabskip\z@skip
   &\setboxz@h{\@lign$\m@th\displaystyle{{}##}$}%
    \ifnum\row@>\@ne
    \ifdim\ht\z@>\ht\strutbox@
    \dimen@\ht\z@
    \advance\dimen@\minalignvsep
    \ht\strutbox@\dimen@
    \fi\fi
    \strut@
    \ifmeasuring@\savefieldlength@\fi
    \set@field
    \hfil
    \tabskip\alignsep@
}
\makeatother

\everypar{\looseness=-1}

\makeatletter
\def\ps@IEEEtitlepagestyle{%
  \def\@oddfoot{\mycopyrightnotice}%
  \def\@oddhead{\hbox{}\@IEEEheaderstyle\leftmark\hfil\thepage}\relax
  \def\@evenhead{\@IEEEheaderstyle\thepage\hfil\leftmark\hbox{}}\relax
  \def\@evenfoot{}%
}
\def\mycopyrightnotice{%
  \begin{minipage}{\textwidth}
  \centering \scriptsize \textcolor{red}{
  Copyright~\copyright~20xx IEEE. Personal use of this material is permitted. Permission from IEEE must be obtained for all other uses, in any current or future media, including\\reprinting/republishing this material for advertising or promotional purposes, creating new collective works, for resale or redistribution to servers or lists, or reuse of any copyrighted component of this work in other works by sending a request to pubs-permissions@ieee.org.}
  \end{minipage}
}
\makeatother

\begin{document}

\title{\huge{RadiOrchestra: Proactive Management of Millimeter-wave Self-backhauled Small Cells via Joint Optimization of Beamforming, User Association, Rate Selection, and Admission Control}}

\author{\normalsize{
\IEEEauthorblockN{Luis F. Abanto-Leon\IEEEauthorrefmark{2}, Arash Asadi\IEEEauthorrefmark{2}, Andres Garcia-Saavedra\IEEEauthorrefmark{3}, Gek Hong (Allyson) Sim\IEEEauthorrefmark{2}, Matthias Hollick\IEEEauthorrefmark{2}} \\ 
\IEEEauthorblockA{\IEEEauthorrefmark{2}Technische Universit\"{a}t Darmstadt, Germany, \IEEEauthorrefmark{3}NEC Laboratories Europe GmbH}  \\ \IEEEauthorrefmark{2}\{labanto, aasadi, asim, mhollick\}@seemoo.tu-darmstadt.de, \IEEEauthorrefmark{3}andres.garcia.saavedra@neclab.eu
}
}

% make the title area
\maketitle

\begin{abstract}
	Millimeter-wave self-backhauled small cells are a key component of next-generation wireless networks. Their dense deployment will increase data rates, reduce latency, and enable efficient data transport between the access and backhaul networks, providing greater flexibility not previously possible with optical fiber. Despite their high potential, operating dense self-backhauled networks optimally is an open challenge, particularly for radio resource management (RRM). This paper presents, \textsf{RadiOrchestra}, a holistic RRM framework that models and optimizes beamforming, rate selection as well as user association and admission control for self-backhauled networks. The framework is designed to account for practical challenges such as hardware limitations of base stations (e.g., computational capacity, discrete rates), the need for adaptability of backhaul links, and the presence of interference. Our framework is formulated as a nonconvex mixed-integer nonlinear program, which is challenging to solve. To approach this problem, we propose three algorithms that provide a trade-off between complexity and optimality. Furthermore, we derive upper and lower bounds to characterize the performance limits of the system. We evaluate the developed strategies in various scenarios, showing the feasibility of deploying practical self-backhauling in future networks. 
\end{abstract}


\IEEEpeerreviewmaketitle

\section{Introduction} \label{section_introduction} 

Network \emph{densification}, through the deployment of small cells, is indispensable to meet the increasing user demands for emerging wireless services \cite{ge2016:5g-ultra-dense-cellular-networks}. Small cells are realized by low-cost
radio access nodes, known as small base stations (SBSs), that provide wireless connectivity to undersized geographical areas \cite{an2017:achieving-sustainable-ultra-dense-heterogeneous-networks-5g}. SBSs are strategically installed in close proximity to the end users, bolstering the quality of experience and improving the radio access network (RAN) performance. \emph{In this way, dense small cell deployments are expected to increase data rates, maintain low latency, extend coverage and support a large number of users, thereby enabling the rollout of a wide range of new services.}

As small cell deployments become denser, more efficient forms of backhauling data traffic between SBSs and the core network will be needed \cite{wang2015:backhauling-5g-small-cells-radio-resource-management-perspective}. Optical fiber has been the predominant means for this task, but its installation and maintenance are costly. Self-backhauling, standardized under the name of \emph{integrated access and backhaul} (IAB) \cite{3gpp.38.175}, is an innovative technology that promises to reduce costs by sharing the wireless spectrum in time/frequency/space between RAN and backhaul links \cite{5gamericas2020:innovations-5g-backhaul-technologies}. Small cells with self-backhauling capabilities benefit from a tight integration of access and backhaul functions, leading to high reconfigurability and facilitating self-adaptation to a wide range of cases. 

Self-backhauled small cells require wide bandwidth to cope with the growing access-backhaul traffic. The millimeter-wave spectrum offers the necessary bandwidth to meet this requirement but it poses challenges, e.g., limited transmission range. Fortunately, recent advances in beamforming \cite{larsson2014:massive-mimo-next-generation-wireless-systems} have overcome the physical drawbacks of millimeter-waves by taking advantage of the small antennas size that have enabled large antenna arrays. \emph{Thus, millimeter-wave self-backhauled small cell networks, realized by multi-antenna SBSs, will play a key role in next-generation wireless networks. Their dense deployment will reduce costs and enable efficient transport of massive data traffic between access and backhaul networks. In addition, the flexibility of millimeter-wave self-backhauled small cells will provide higher adaptability in various topologies and network conditions, previously not possible with fiber.}

Despite consensus on the potential of millimeter-wave self-backhauling, designing an optimal system remains an open research challenge \cite{polese2020:integrated-access-backhaul-5g-mmwave-networks-potential-challenges}, which requires efficient radio resource management (RRM) across the access and backhaul networks. To date, the body of work in this area often overlooks practical challenges inherent to realistic wireless communications systems, such as discrete modulations and coding schemes (MCSs), or low computing capabilities of SBSs. \emph{Our work is motivated by the absence of holistic RRM frameworks providing a realistic model and a practical solution for millimeter-wave self-backhauled small cells deployments.} In the following, we introduce these challenges and put them in perspective with the literature.

% Challenge M1
\noindent {\textbf{{Challenge 1}: Scalable self-backhauling design.}} The majority of prior works relies on point-to-point links, e.g., \cite{hur2013:mmwave-beamforming-wireless-backhaul-access-small-cell-networks, pan2017:joint-precoding-rrh-selection-user-centric-green-mimo-cran}, between macro base station (MBS) and SBSs, which is unscalable in dense SBS deployments. The scalability issue is addressed in a handful of works, e.g., \cite{chen2019:user-centric-joint-access-backhaul-full-duplex-self-backhauled-wireless-networks, lei2018:noma-aided-interference-management-full-duplex-self-backhauling-hetnets} assume that SBSs are capable of multi-layer successive interference cancellation (SIC). While this assumption simplifies traffic transport, it involves heavy computational tasks (i.e., SIC) not suited for SBSs. \emph{Thus, to keep SBS economical for the operators, it is necessary to reduce the computational burden of SBSs by developing practical backhauling mechanisms.}

% Challenge M2
\noindent {\textbf{{Challenge 2}: Adaptive backhaul capacity.}} Although self-backhauling relies on wireless media, whose capacity is inherently highly variable due to noise and interference, the assumption of unlimited or fixed capacity prevails in many prior works, e.g., \cite{pan2017:joint-precoding-rrh-selection-user-centric-green-mimo-cran, huang2016:joint-scheduling-beamforming-coordination-cran-qos-guarantees, dinh2021:energy-efficient-resource-allocation-optimization-fran-outdated-csi, nguyen2020:nonsmooth-optimization-algorithms-multicast-beamforming-content-centric-fran}. \emph{However, it is necessary to consider the capacity limitation of backhaul links as well as their variability in real systems.}

% Challenge M3
\noindent {\textbf{{Challenge 3}: User association.}} It is conventionally assumed that users are served by a single SBS \cite{chen2016:green-full-duplex-self-backhaul-energy-harvesting-small-cell-networks-massive-mimo, dinh2021:energy-efficient-resource-allocation-optimization-fran-outdated-csi} or by all SBSs within a given range \cite{pan2017:joint-precoding-rrh-selection-user-centric-green-mimo-cran, huang2016:joint-scheduling-beamforming-coordination-cran-qos-guarantees}. While these assumptions simplify the problem formulation and solution, they are neither realistic nor optimal. \emph{Thus, a general scheme is needed where users are associated to multiple SBSs in a flexible manner without considering extremes cases.}

% Challenge M4
\noindent {\textbf{{Challenge 4}: Admission control.}} Many works assume that all users can be served simultaneously \cite{nguyen2020:nonsmooth-optimization-algorithms-multicast-beamforming-content-centric-fran, chen2019:robust-multigroup-multicast-beamforming-design-backhaul-limited-cran, kwon2019:joint-user-association-beamforming-mmwave-udn-wireless-backhaul}, which is unrealistic due to limitations in power, number of antennas or RF chains. \emph{Admission control (or user scheduling) is crucial to guarantee the quality of service requirements for at least a subset of admitted users, thereby circumventing unfeasibility issues.}

% Challenge M5
\noindent {\textbf{{Challenge 5}: Discrete data rates.}} It is usually assumed that data rates are continuous-valued, e.g., \cite{chen2019:user-centric-joint-access-backhaul-full-duplex-self-backhauled-wireless-networks, vu2017:joint-load-balancing-interference-mitigation-5g-heterogeneous-networks, kwon2019:joint-user-association-beamforming-mmwave-udn-wireless-backhaul, hur2013:mmwave-beamforming-wireless-backhaul-access-small-cell-networks, hu2017:joint-fronthaul-multicast-beamforming-user-centric-clustering-downlink-crans}. However, in practice they are limited to a number of possible choices, i.e., finite set of MCSs. \emph{It is critical to consider the discreteness of rates since results obtained from solving problems for continuous values cannot be easily applied to real systems and are not expected to work properly.}

% Table I
% Table: Literature review
\begin{table*}[!t]
	\fontsize{6.5}{6}\selectfont
	\setlength\tabcolsep{2.9pt}
	\centering
	\caption{Categorization of related work}
	\label{table_related_literature}
	\begin{tabular}{c c c c c c c c c c c c c c}
		\toprule
		\multirow{3}{*}{\centering{\makecell{\\ Approach }}} &
		\multirow{3}{*}{\centering{\makecell{ \\ Solution}}} &
		 \multirow{3}{*}{\centering{\makecell{ \\ Spectrum}}} & \multirow{3}{*}{\centering{\makecell{\\ Network }}} & \multicolumn{5}{c}{Access network} & \multicolumn{5}{c}{Backhaul network} \\ 
		\cmidrule(lr){5-9}
		\cmidrule(lr){10-14}
		& & & & \makecell{Topology} & Beamforming & \makecell{User \\ association} & \makecell{Rate \\ selection} & \makecell{Admission \\ control} & Topology & Link & Medium & Beamforming & \makecell{Rate \\ selection} \\ 
		\midrule
		\midrule
		
		% Work 3
		\centering{\cite{cheng2012:dynamic-rate-adaptation-multiuser-beamforming-mixed-integer-conic-programming, cheng2015:joint-discrete-rate-adaptation-beamforming-mixed-integer-conic-programming}} & 
		% Solution
		{\centering {Joint}} &
		% Spectrum
		{\centering {Sub-6GHz}} &
		% Network
		{\centering {Single-SBS}} &
		% Access | Topology
		{\centering {Unicast}} &
		% Access | Beamforming
		{\centering {\ding{51}}} & 
		% Access | User association
		{\centering {\textcolor{gray}{\ding{55}}}} & 
		% Access | Rate selection
		{\centering {\ding{51}}} & 
		% Access | Admission control
		{\centering {\ding{51}}} & 
		% Backhaul | Topology
		{\centering {N/A}} & 
		% Backhaul | Link
		{\centering {N/A}} & 
		% Backhaul | Medium
		{\centering {N/A}} & 
		% Backhaul | Beamforming
		{\centering {N/A}} & 
		% Backhaul | Rate selection
		{\centering {N/A}} \\
		%\midrule

		% Work 4
		\centering{\cite{cheng2013:joint-network-optimization-beamforming-comp-transmissions-mixed-integer-conic-programming, nguyen2017:optimal-dynamic-point-selection-power-minimization-multiuser-comp, sanjabi2014:optimal-joint-base-station-assignment-beamforming-heterogeneous-networks, ghauch2018:user-assignment-cran-systems-algorithms-bounds}} & 
		% Solution
		{\centering {Joint}} &
		% Spectrum
		{\centering {Sub-6GHz}} &
		% Network
		{\centering {Multi-SBS}} &
		% Access | Topology
		{\centering {Unicast}} & 
		% Access | Beamforming
		{\centering {\ding{51}}} & 
		% Access | User association
		{\centering {Many}} & 
		% Access | Rate selection
		{\centering {\textcolor{gray}{\ding{55}}}} & 
		% Access | Admission control
		{\centering {\textcolor{gray}{\ding{55}}}} & 
		% Backhaul | Topology
		{\centering {N/A}} & 
		% Backhaul | Link
		{\centering {N/A}} & 
		% Backhaul | Medium
		{\centering {N/A}} & 
		% Backhaul | Beamforming
		{\centering {N/A}} & 
		% Backhaul | Rate selection
		{\centering {N/A}} \\
		%\midrule

		% Work 5
		\centering{\cite{ni2018:mixed-integer-semidefinite-relaxation-joint-admission-control-beamforming-soc-outer-approximation-provable-guarantees, bandi2020:joint-user-grouping-scheduling-precoding-multicast-energy-efficiency-multigroup-multicast-systems}} & 
		% Solution
		{\centering {Decoupled}} &  
		% Spectrum
		{\centering {Sub-6GHz}} &
		% Network
		{\centering {Single-SBS}} &  
		% Access | Topology
		{\centering {Multicast}} &  
		% Access | Beamforming
		{\centering {\ding{51}}} &  
		% Access | User association
		{\centering {\textcolor{gray}{\ding{55}}}} & 
		% Access | Rate selection
		{\centering {\textcolor{gray}{\ding{55}}}} &  
		% Access | Admission control
		{\centering {\ding{51}}} &  
		% Backhaul | Topology
		{\centering {N/A}} & 
		% Backhaul | Link
		{\centering {N/A}} & 
		% Backhaul | Medium
		{\centering {N/A}} &  
		% Backhaul | Beamforming
		{\centering {N/A}} & 
		% Backhaul | Rate selection
		{\centering {N/A}} \\
		%\midrule

		% Work 5
		\centering{\cite{alizadeh2019:load-balancing-user-association-mmwave-mimo-networks}} & 
		% Solution
		{\centering {Joint}} &  
		% Spectrum
		{\centering {Millimeter-wave}} &
		% Network
		{\centering {Multi-SBS}} &  
		% Access | Topology
		{\centering {Unicast}} &  
		% Access | Beamforming
		{\centering {3D}} &  
		% Access | User association
		{\centering {Many}} & 
		% Access | Rate selection
		{\centering {\textcolor{gray}{\ding{55}}}} &  
		% Access | Admission control
		{\centering {\textcolor{gray}{\ding{55}}}} &  
		% Backhaul | Topology
		{\centering {N/A}} & 
		% Backhaul | Link
		{\centering {N/A}} & 
		% Backhaul | Medium
		{\centering {N/A}} &  
		% Backhaul | Beamforming
		{\centering {N/A}} & 
		% Backhaul | Rate selection
		{\centering {N/A}} \\
		%\midrule

		% Work 6
		\centering{\cite{pizzo2017:optimal-design-energy-efficient-mmwave-transceivers-wireless-backhaul, hur2013:mmwave-beamforming-wireless-backhaul-access-small-cell-networks}} &
		% Solution
		{\centering {Joint}} &
		% Spectrum
		{\centering {Millimeter-wave}} & 
		% Network
		{\centering {Multi-SBS}} &
		% Access | Topology
		{\centering {N/A}} & 
		% Access | Beamforming
		{\centering {N/A}} & 
		% Access | User association
		{\centering {N/A}} & 
		% Access | Rate selection
		{\centering {N/A}} & 
		% Access | Admission control
		{\centering {N/A}} & 
		% Backhaul | Topology
		{\centering {Unicast}} & 
		% Backhaul | Link
		{\centering {Adaptive}} & 
		% Backhaul | Medium
		{\centering {Wireless}} & 
		% Backhaul | Beamforming
		{\centering {2D}} & 
		% Backhaul | Rate selection
		{\centering {\textcolor{gray}{\ding{55}}}} \\ 
		%\midrule

		% Work 7
		\centering{\cite{tao2016:content-centric-sparse-multicast-beamforming-cache-enabled-cran}} & 
		% Solution
		{\centering {Joint}} &  
		% Spectrum
		{\centering {Sub-6GHz}} &
		% Network
		{\centering {Multi-SBS}} & 
		% Access | Topology
		{\centering {Multicast}} &  
		% Access | Beamforming
		{\centering {2D}} & 
		% Access | User association
		{\centering {Many}} &  
		% Access | Rate selection
		{\centering {\textcolor{gray}{\ding{55}}}} &  
		% Access | Admission control
		{\centering {\textcolor{gray}{\ding{55}}}} & 
		% Backhaul | Topology 
		{\centering {Unicast}} & 
		% Backhaul | Link
		{\centering {Fixed}} & 
		% Backhaul | Medium
		{\centering {Wired}} &  
		% Backhaul | Beamforming
		{\centering {\textcolor{gray}{\ding{55}}}} &  
		% Backhaul | Rate selection
		{\centering {\textcolor{gray}{\ding{55}}}} \\
		%\midrule

		% Work 8
		\centering \cite{chen2018:joint-base-station-clustering-beamforming-non-orthogonal-multicast-unicast-transmission-backhaul-constraints, tam2017:joint-load-balancing-interference-management-heterogeneous-networks-backhaul-capacity} & 
		% Solution
		{\centering {Joint}} & 
		% Spectrum
		{\centering {Sub-6GHz}} &
		% Network
		{\centering {Multi-SBS}} & 
		% Access | Topology
		{\centering {Both}} & 
		% Access | Beamforming
		{\centering {2D}} &  
		% Access | User association
		{\centering {Many}} &  
		% Access | Rate selection
		{\centering {\textcolor{gray}{\ding{55}}}} &  
		% Access | Admission control
		{\centering {\textcolor{gray}{\ding{55}}}} &  
		% Backhaul | Topology
		{\centering {Unicast}} & 
		% Backhaul | Link
		{\centering {Fixed}} & 
		% Backhaul | Medium
		{\centering {Wired}} & 
		% Backhaul | Beamforming
		{\centering {\textcolor{gray}{\ding{55}}}} &  
		% Backhaul | Rate selection
		{\centering {\textcolor{gray}{\ding{55}}}} \\
		%\midrule

		% Work 9
		\centering{\cite{pan2017:joint-precoding-rrh-selection-user-centric-green-mimo-cran, huang2016:joint-scheduling-beamforming-coordination-cran-qos-guarantees}} &
		% Solution
		{\centering {Decoupled}} &
		% Spectrum
		{\centering {Sub-6GHz}} &
		% Network
		{\centering {Multi-SBS}} & 
		% Access | Topology
		{\centering {Unicast}} &
		% Access | Beamforming
		{\centering {2D}} &  
		% Access | User association
		{\centering {\textcolor{gray}{\ding{55}}}} & 
		% Access | Rate selection
		{\centering {\textcolor{gray}{\ding{55}}}} & 
		% Access | Admission control
		{\centering {\ding{51}}} &  
		% Backhaul | Topology
		{\centering {Unicast}} & 
		% Backhaul | Link
		{\centering {Unbounded}} & 
		% Backhaul | Medium
		{\centering {Wired}} & 
		% Backhaul | Beamforming
		{\centering {\textcolor{gray}{\ding{55}}}} & 
		% Backhaul | Rate selection
		{\centering {\textcolor{gray}{\ding{55}}}} \\
		%\midrule

		% Work 10
		\centering{\cite{dinh2021:energy-efficient-resource-allocation-optimization-fran-outdated-csi}} &
		% Solution
		{\centering {Joint}} &
		% Spectrum
		{\centering {Sub-6GHz}} &
		% Network
		{\centering {Multi-SBS}} & 
		% Access | Topology
		{\centering {Unicast}} & 
		% Access | Beamforming
		{\centering {2D}} & 
		% Access | User association
		{\centering {One}} & 
		% Access | Rate selection
		{\centering {\textcolor{gray}{\ding{55}}}} & 
		% Access | Admission control
		{\centering {\ding{51}}} & 
		% Backhaul | Topology
		{\centering {Unicast}} & 
		% Backhaul | Link
		{\centering {Fixed}} & 
		% Backhaul | Medium
		{\centering {Wireless}} & 
		% Backhaul | Beamforming
		{\centering {\textcolor{gray}{\ding{55}}}} & 
		% Backhaul | Rate selection
		{\centering {\textcolor{gray}{\ding{55}}}} \\
		%\midrule

		% Work 11
		\centering{\cite{kwon2019:joint-user-association-beamforming-mmwave-udn-wireless-backhaul}} &
		% Solution
		{\centering {Decoupled}} &
		% Spectrum
		{\centering {Millimeter-wave}} & 
		% Network
		{\centering {Multi-SBS}} & 
		% Access | Topology
		{\centering {Unicast}} & 
		% Access | Beamforming
		{\centering {3D}} & 
		% Access | User association
		{\centering {Many}} & 
		% Access | Rate selection
		{\centering {\textcolor{gray}{\ding{55}}}} & 
		% Access | Admission control
		{\centering {\textcolor{gray}{\ding{55}}}} & 
		% Backhaul | Topology
		{\centering {Unicast}} & 
		% Backhaul | Link
		{\centering {Adaptive}} & 
		% Backhaul | Medium
		{\centering {Wireless/TDM}} & 
		% Backhaul | Beamforming
		{\centering {3D}} & 
		% Backhaul | Rate selection
		{\centering {\textcolor{gray}{\ding{55}}}} \\
		%\midrule

		% Work 12
		\centering{\cite{vu2017:joint-load-balancing-interference-mitigation-5g-heterogeneous-networks}} &
		% Solution
		{\centering {Decoupled}} &
		% Spectrum
		{\centering {Sub-6GHz}} &
		% Network
		{\centering {Multi-SBS}} &
		% Access | Topology
		{\centering {Unicast}} & 
		% Access | Beamforming
		{\centering {2D}} & 
		% Access | User association
		{\centering {One}} & 
		% Access | Rate selection
		{\centering {\textcolor{gray}{\ding{55}}}} & 
		% Access | Admission control
		{\centering {\ding{51}}} & 
		% Backhaul | Topology
		{\centering {Unicast}} & 
		% Backhaul | Link
		{\centering {Adaptive}} & 
		% Backhaul | Medium
		{\centering {Wireless/SDM}} & 
		% Backhaul | Beamforming
		{\centering {2D}} & 
		% Backhaul | Rate selection
		{\centering {\textcolor{gray}{\ding{55}}}} \\
		%\midrule

		% Work 13
		\centering{ \cite{chen2019:user-centric-joint-access-backhaul-full-duplex-self-backhauled-wireless-networks}} &
		% Solution
		{\centering {Decoupled}} &
		% Spectrum
		{\centering {Sub-6GHz}} &
		% Network
		{\centering {Multi-SBS}}&
		% Access | Topology
		{\centering {Unicast}} & 
		% Access | Beamforming
		{\centering {2D}} & 
		% Access | User association
		{\centering {Many}} & 
		% Access | Rate selection
		{\centering {\textcolor{gray}{\ding{55}}}} & 
		% Access | Admission control
		{\centering {\textcolor{gray}{\ding{55}}}} & 
		% Backhaul | Topology
		{\centering {Multicast}} & 
		% Backhaul | Link
		{\centering {Adaptive/SIC}} & 
		% Backhaul | Medium
		{\centering {Wireless/SDM}} & 
		% Backhaul | Beamforming
		{\centering {2D}} & 
		% Backhaul | Rate selection
		{\centering {\textcolor{gray}{\ding{55}}}} \\	
		%\midrule		

		% Work 14
		\centering{ \cite{chen2016:green-full-duplex-self-backhaul-energy-harvesting-small-cell-networks-massive-mimo}} &
		% Solution
		{\centering {Joint}} & 
		% Spectrum
		{\centering {Sub-6GHz}} &
		% Network
		{\centering {Multi-SBS}} &
		% Access | Topology
		{\centering {Unicast}}  & 
		% Access | Beamforming
		{\centering {2D}} & 
		% Access | User association
		{\centering One} & 
		% Access | Rate selection
		{\centering {\textcolor{gray}{\ding{55}}}} & 
		% Access | Admission control
		{\centering {\ding{51}}} & 
		% Backhaul | Topology
		{\centering {Unicast}} & 
		% Backhaul | Link
		{\centering {Adaptive}} & 
		% Backhaul | Medium
		{\centering {Wireless/SDM}} & 
		% Backhaul | Beamforming
		{\centering {2D}} & 
		% Backhaul | Rate selection
		{\centering {\textcolor{gray}{\ding{55}}}} \\
		%\midrule	

		% Work 15
		\centering{ \cite{nguyen2020:nonsmooth-optimization-algorithms-multicast-beamforming-content-centric-fran, chen2019:robust-multigroup-multicast-beamforming-design-backhaul-limited-cran}} &
		% Solution
		{\centering {Joint}} & 
		% Spectrum
		{\centering {Sub-6GHz}} &
		% Network
		{\centering {Multi-SBS}} & 
		% Access | Topology
		{\centering {Multicast}} & 
		% Access | Beamforming
		{\centering {2D}} & 
		% Access | User association
		{\centering {\textcolor{gray}{\ding{55}}}} & 
		% Access | Rate selection
		{\centering {\textcolor{gray}{\ding{55}}}} & 
		% Access | Admission control
		{\centering {\textcolor{gray}{\ding{55}}}} & 
		% Backhaul | Topology
		{\centering {Unicast}} & 
		% Backhaul | Link
		{\centering {Fixed}} & 
		% Backhaul | Medium
		{\centering {Wireless}} & 
		% Backhaul | Beamforming
		{\centering {\textcolor{gray}{\ding{55}}}} & 
		% Backhaul | Rate selection
		{\centering {\textcolor{gray}{\ding{55}}}} \\
		%\midrule

		% Work 16
		\centering{\cite{hu2018:joint-beamformer-design-wireless-fronthaul-access-links-crans}} &
		% Solution
		{\centering {Joint}} & 
		% Spectrum
		{\centering {Sub-6GHz}} &
		% Network
		{\centering {Multi-SBS}} & 
		% Access | Topology
		{\centering {Unicast}} & 
		% Access | Beamforming
		{\centering {2D}} & 
		% Access | User association
		{\centering {One}} & 
		% Access | Rate selection
		{\centering {\textcolor{gray}{\ding{55}}}} & 
		% Access | Admission control
		{\centering {\ding{51}}} & 
		% Backhaul | Topology
		{\centering {Unicast}} & 
		% Backhaul | Link
		{\centering {Adaptive}} & 
		% Backhaul | Medium
		{\centering {Wireless/SDM}} & 
		% Backhaul | Beamforming
		{\centering {2D}} & 
		% Backhaul | Rate selection
		{\centering {\textcolor{gray}{\ding{55}}}} \\
		%\midrule

		% Work 17
		\centering{\cite{hu2017:joint-fronthaul-multicast-beamforming-user-centric-clustering-downlink-crans}} &
		% Solution
		{\centering {Joint}} & 
		% Spectrum
		{\centering {Sub-6GHz}} &
		% Network
		{\centering {Multi-SBS}} & 
		% Access | Topology
		{\centering {Unicast}} & 
		% Access | Beamforming
		{\centering {2D}} & 
		% Access | User association
		{\centering {Many}} & 
		% Access | Rate selection
		{\centering {\textcolor{gray}{\ding{55}}}} & 
		% Access | Admission control
		{\centering {\textcolor{gray}{\ding{55}}}} & 
		% Backhaul | Topology
		{\centering {Multicast}} & 
		% Backhaul | Link
		{\centering {Adaptive}} & 
		% Backhaul | Medium
		{\centering {Wireless/TDM}} & 
		% Backhaul | Beamforming
		{\centering {2D}} & 
		% Backhaul | Rate selection
		{\centering {\textcolor{gray}{\ding{55}}}} \\
		%\midrule

		% Work 18
		\centering{\cite{lei2018:noma-aided-interference-management-full-duplex-self-backhauling-hetnets}} &
		% Solution
		{\centering {Joint}} & 
		% Spectrum
		{\centering {Sub-6GHz}} &
		% Network
		{\centering {Multi-SBS}} & 
		% Access | Topology
		{\centering {Unicast}} & 
		% Access | Beamforming
		{\centering {2D}} & 
		% Access | User association
		{\centering {\textcolor{gray}{\ding{55}}}} & 
		% Access | Rate selection
		{\centering {\textcolor{gray}{\ding{55}}}} & 
		% Access | Admission control
		{\centering {\textcolor{gray}{\ding{55}}}} & 
		% Backhaul | Topology
		{\centering {Unicast}} & 
		% Backhaul | Link
		{\centering {Adaptive/SIC}} & 
		% Backhaul | Medium
		{\centering {Wireless/SDM}} & 
		% Backhaul | Beamforming
		{\centering {2D}} & 
		% Backhaul | Rate selection
		{\centering {\textcolor{gray}{\ding{55}}}} \\
		%\midrule

		% Proposed
		\centering{Proposed} &
		% Solution
		{\centering {Joint}} & 
		% Spectrum
		{\centering {Millimeter-wave}} & 
		% Network
		{\centering {Multi-SBS}} & 
		% Access | Topology
		{\centering {Unicast}} & 
		% Access | Beamforming
		{\centering {3D}} & 
		% Access | User association
		{\centering {Many}} & 
		% Access | Rate selection
		{\centering {\ding{51}}} & 
		% Access | Admission control
		{\centering {\ding{51}}} & 
		% Backhaul | Topology
		{\centering {Multicast}} & 
		% Backhaul | Link
		{\centering {Adaptive}} & 
		% Backhaul | Medium
		{\centering {Wireless/SDM}} & 
		% Backhaul | Beamforming
		{\centering {3D}} & 
		% Backhaul | Rate selection
		{\centering {\ding{51}}} \\

		\bottomrule
	\end{tabular}
	\vspace{1mm}\\
	{\raggedright The connection between the MBS and SBSs is called backhaul link, which is a convention in small cells literature. However, in a cloud-RAN context, MBSs are called central processors or BBUs, SBSs are called RRHs, and the connection between MBS and SBSs are called fronthaul links. In Table \ref{table_related_literature}, we have considered both kinds of nomenclatures since the problems originated from these two contexts are essentially the same. \par}
\end{table*}

In contrast to prior art, we propose a comprehensive RRM framework that includes the challenges mentioned above, allowing us to more realistically validate millimeter-wave self-backhauled small cell deployments. Our approach makes the following novel contributions.

% Contribution C1
\noindent {\textbf{{Contribution 1:}} In Section \ref{section_system_model_problem_formulation}, we address \emph{Challenge 1} by proposing a simple yet effective \emph{clustering mechanism for SBSs and users} that results in multiple non-overlapping \emph{virtual cells} or \emph{clusters}. This allows us to exploit multigroup multicast beamforming for backhaul traffic transmissions. Our clustering approach simplifies the backhaul design and reduces hardware/computational requirements at the sending and receiving nodes.

% Contribution C2
\noindent {\textbf{{Contribution 2:}} In Section \ref{subsection_backhaul_network} and Section \ref{subsection_access_network} we model \emph{Challenge 2}, \emph{Challenge 3}, \emph{Challenge 4}, \emph{Challenge 5} considering the access-backhaul interdependencies between MBS, SBSs and users. In Section \ref{subsection_problem_formulation}, we include these challenges in our formulation to jointly optimize \emph{beamforming, user association, rate selection, admission control in the access network} and \emph{beamforming, rate selection in the backhaul network} for maximizing the access network downlink weighted sum-rate. We cast the problem as a nonconvex mixed-integer nonlinear program (MINLP), which to the best of our knowledge, has not been investigated before.

% Contribution C3
\noindent {\textbf{{Contribution 3:}} To tackle the nonconvex MINLP, we propose three formulations and their corresponding algorithms. In Section \ref{section_BNBC_MISOCP}, we recast the nonconvex MINLP as a mixed-integer second-order cone program (MISOCP), which can be solved optimally. Due to the large number of integral variables, the cost of solving the MISOCP via branch-and-cut (BnC) techniques is prohibitive. To cope with this issue, in Section \ref{section_RNP_SOCP_1} we propose a formulation solved via an iterative algorithm that tackles a SOCP at every instance. In Section \ref{section_RNP_SOCP_2}, a much simpler SOCP formulation further decreases the complexity by reducing the number of  variables, and optimizing only the beamformers gains. In particular, the complexity of the latter algorithm with respect to the former decreases roughly by a factor equal to the third power of the number of antennas at the SBS. 

% Contribution C4
\noindent {\textbf{{Contribution 4:}}  In Section \ref{section_proposed_bounds}, we derive an upper bound to provide insights on the performance gaps and trade-offs of \textsf{RadiOrchestra}. We also provide a simple lower bound marking the oerformance. We note that the upper bound is a novel problem itself that has not been investigated before. 

% Contribution C5
\noindent {\textbf{{Contribution 5:}} In Section \ref{section_simulation_results}, we examine \textsf{RadiOrchestra} exhaustively under several scenarios including transmit power, number of clusters, and channel estimation errors.

%\emph{Our framework helps to realize the true potential of self-backhauled mobile networks. However, the inherent couplings between the different optimization variables of the system result in a complex problem. In Fig. \ref{figure_overview}, we provide an overview of the steps taken to solve this problem. }

% Section 1.2: Related literature
%\subsection{Related Literature} \label{subsection_related_literature}

There is a plethora of literature on self-backhauling for sub-6GHz spectrum, e.g., \cite{hu2018:joint-beamformer-design-wireless-fronthaul-access-links-crans, hu2017:joint-fronthaul-multicast-beamforming-user-centric-clustering-downlink-crans, tam2017:joint-load-balancing-interference-management-heterogeneous-networks-backhaul-capacity}, which assume signals properties that do not work for millimeter-wave. Many works have focused on the design of either the backhaul, e.g., \cite{pizzo2017:optimal-design-energy-efficient-mmwave-transceivers-wireless-backhaul, hur2013:mmwave-beamforming-wireless-backhaul-access-small-cell-networks, yuan2020:optimal-approximation-algorithms-joint-routing-scheduling-mmwave-cellular-networks, ortiz2019:scaros-scalable-robust-self-backhauling-solution-highly-dynamic-mmwave-networks} or the access network, e.g., \cite{alizadeh2019:load-balancing-user-association-mmwave-mimo-networks,   bandi2020:joint-user-grouping-scheduling-precoding-multicast-energy-efficiency-multigroup-multicast-systems} alone. However, the growth that mobile networks are experiencing calls for heterogeneous networks with wireless backhauling, which require joint optimization. Considering linear antenna arrays, many works have optimized beamforming, e.g., \cite{vu2017:joint-load-balancing-interference-mitigation-5g-heterogeneous-networks, hu2018:joint-beamformer-design-wireless-fronthaul-access-links-crans}. However, planar arrays are capable of 3D beamforming and hence are more suitable for dense deployments. The joint optimization of beamforming and user association \emph{(Challenge 3)}, admission \emph{(Challenge 4)}, or rate selection \emph{(Challenge 5)}, generally requires solving complex nonconvex MINLPs. Thus, many works facing these challenges split the problem into stages and solve them separately. For instance, the integer variables are eliminated first by assuming a given set of scheduled users, e.g., \cite{vu2017:joint-load-balancing-interference-mitigation-5g-heterogeneous-networks, kwon2019:joint-user-association-beamforming-mmwave-udn-wireless-backhaul}. Then, the nonconvex functions are linearized and the problem is solved in the continuous domain. Although simpler to solve, variable decoupling affects optimality due to interdependencies removal. To meet the continuously growing demands, resources have to be exploited more optimally. Therefore, RRM problems need to be solved as a whole, without relying on variable partitioning which translates to inefficient radio resource usage.

After a scrupulous study of the state of the art, we found that the works most related to ours are \cite{chen2019:user-centric-joint-access-backhaul-full-duplex-self-backhauled-wireless-networks, hu2017:joint-fronthaul-multicast-beamforming-user-centric-clustering-downlink-crans}. Like us, the authors of \cite{chen2019:user-centric-joint-access-backhaul-full-duplex-self-backhauled-wireless-networks} assumed a multicast topology in the backhaul network, with a MBS transmitting multiple signals to various SBSs using multigroup multicasting beamforming (each signal carrying the data of a user). Since a single SBS may serve several users, SBSs are therefore required to decode many signal layers via SIC, which entails heavy computational burden for low-cost SBSs. Further, the decoding order of signals is known to  affect the performance, leading to potential high decoding errors and making SIC impractical, which was not evaluated in \cite{chen2019:user-centric-joint-access-backhaul-full-duplex-self-backhauled-wireless-networks}. The authors of \cite{hu2017:joint-fronthaul-multicast-beamforming-user-centric-clustering-downlink-crans} considered multiple SBS groups served in a multicast manner using time division multiplexing (TDM), i.e., each group at a time. However, as the number of clusters grows, the multiplexing time generates longer latency that is unavoidable as SBSs need to transmit coordinately to users, making it less practical. In addition, these works do not consider discrete rates, admission control, millimeter-wave spectrum and 3D beamforming. For completeness, we summarize in Table \ref{table_related_literature} the related literature on RRM for small cells.

% Figure: Overview
\begin{figure*}[!t]
	\centering
	\includegraphics[]{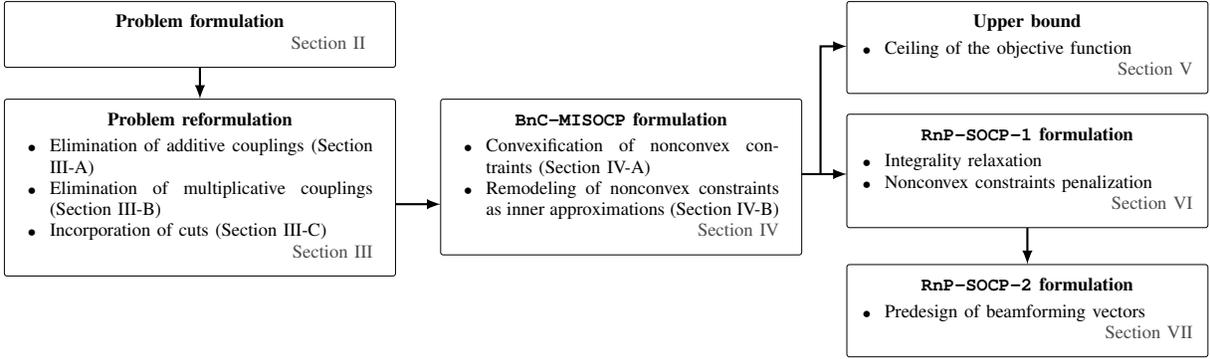}
	\vspace{1mm}
	\caption{Overview of the steps to formulate and solve the problem in \textsf{RadiOrchestra}.}
	\label{figure_overview}
\end{figure*}

% Overview
\noindent {\textbf{{Overview:}} Not surprisingly, the inherent couplings among all the different parameters of the system result in a complex problem which is difficult to address. However, our framework helps to realize the true potential of self-backhauled mobile networks, in particular in the presence of real-world constraints. To the best of our knowledge, this is the first work that has modeled an integrated access-backhaul system with such practical constraints and proposed solutions to assess its performance. The investigated problem is unique and hence existing solutions are not applicable to it. In the following, we provide an overview of the steps taken to solve our problem, from a systems design perspective as well as the mathematical treatment. }

\noindent{\emph {Systems aspect.}} 3GPP specifications for 5G leave several design choices to the operators such as spectrum allocation of backhaul and access. We leveraged these degrees of freedom to reduce the complexity of the problem while maintaining a realistic setup. The wireless nature of the access and backhaul links, coupled with the dense deployment of SBSs and users, creates a very complex interference environment. In \textsf{RadiOrchestra}, we choose an out-of-band system where backhaul and access links use different frequency bands, thus disentangling the interference between the two networks. Conventionally, the MBS sends individual backhaul signals to each SBS thus producing interference, which is handled via (point-to-point) unicast beamforming. In dense deployments this solution does not scale well due to the need to multiplex various data streams. Thus, we propose a clustering strategy where the MBS divides the SBSs into \emph{clusters}, which are served simultaneously via (point-to-multipoint) multigroup multicast beamforming. This has three advantages: $ ( i ) $ Enhancing the scalability of self-backhauling by avoiding point-to-point transmissions which cause higher interference; $ ( ii ) $ Eliminating the need for heavy signal processing (e.g., SIC operation) at SBSs \cite{chen2019:user-centric-joint-access-backhaul-full-duplex-self-backhauled-wireless-networks, lei2018:noma-aided-interference-management-full-duplex-self-backhauling-hetnets}; $ ( iii ) $ Reducing hardware requirements and costs since MBS becomes more cost-efficient only requiring as many RF chains as SBS clusters, which is far less than the point-to-point topology (i.e., dedicated RF-chain per link). 

\noindent{\emph{Problem formulation and solution.}} Considering our design choices above, we model the system and propose solutions in a series of steps that are demonstrated in Fig. \ref{figure_overview}. We formulate a RRM problem for integrated access-backhaul networks considering real-world constraints, which results in a nonconvex MINLP with entangled variables (see \textbf{Section \ref{section_system_model_problem_formulation}}). We adopt a series of procedures to simplify the structure of the nonconvex MINLP without altering its optimality. Thus, we \emph{(i)} improve its tractability by eliminating additive binary couplings and multiplicative mixed-integer couplings, and \emph{(ii)} reduce the search space by adding cuts. Although the problem structure is greatly simplified after these procedures, it still remains a nonconvex MINLP. However, its more amenable layout allows us to tailor algorithms for its solution (see \textbf{Section \ref{section_proposed_reformulation}}). We transform some of the nonconvex constraints into equivalent (convex) SOC constraints and remodel others as convex inner SOC approximations. As a result, we recast the nonconvex MINLP into a MISOCP, which can be solved optimally (see \textbf{Section \ref{section_BNBC_MISOCP}}). Although solving the proposed MISOCP guarantees an optimal solution, it requires a considerable amount of time due to the numerous integral variables. To deal with that, which translates to more branches evaluations by the BnC method, we propose a reformulation based on relaxation and penalization of the integral variables that only requires to solve iteratively a SOCP, and is guaranteed to attain a local optimum (see \textbf{Section \ref{section_RNP_SOCP_1}}). To further simplify the computational burden and expedite the solving time, we offer a much simpler reformulation that reduces the number of continuous variables, where we predesign the access and backhaul beamforming vectors and only optimize their gains. As a result, we only solve a low-complexity SOCP problem iteratively (see \textbf{Section \ref{section_RNP_SOCP_2}}). Finally, we derive an upper bound for the problem, which we use to characterize the performance of the developed algorithms (see \textbf{Section \ref{section_proposed_bounds}}).

% Section 2: Related literature
\section{System Model and Problem Formulation} \label{section_system_model_problem_formulation}

We consider that data is transported from the core network to the user equipments (UEs) via a MBS and a deployment of SBSs as shown in Fig. \ref{figure_system_model}. The SBSs are connected to the MBS through wireless backhaul links. We assume an out-of-band full-duplex access-backhaul system, i.e., the backhaul network (connecting SBSs to the MBS) and the access network (connecting UEs to SBSs) operate simultaneously employing orthogonal bands. In the following, we detail the modeling assumptions.

% Figure: System model
\begin{figure*}[!t]
	\centering
	\begin{minipage}[t]{.56\textwidth}
	  	\centering
	  	\includegraphics[scale=0.26]{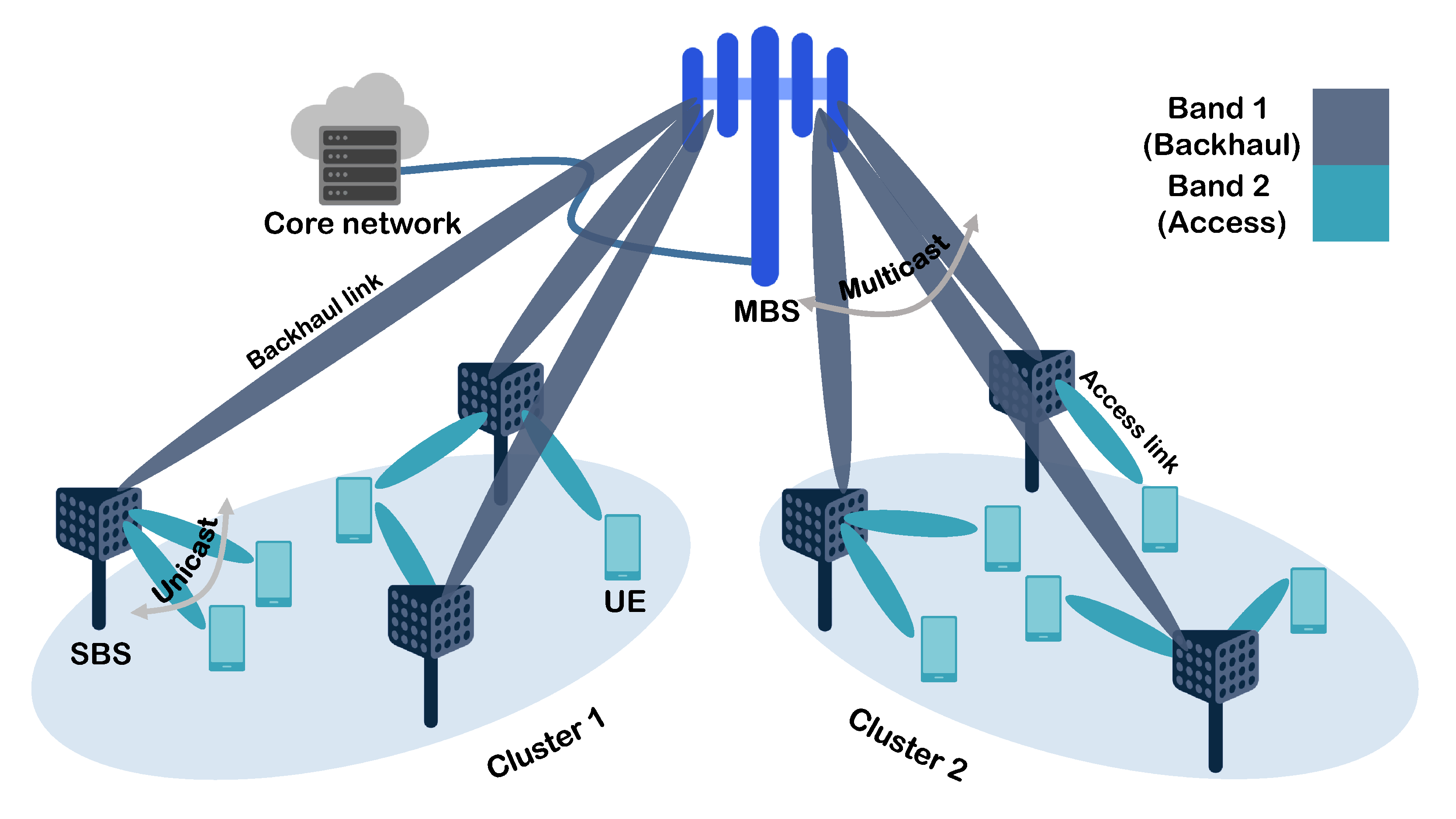}
	  	\vspace{-1mm}
	  	\caption[Caption for figure]%
	    	{Self-backhauled SBSs grouped into clusters. The backhaul exploits multigroup multicast beamforming for data sharing whereas the access network is based on distributed unicast beamforming.}
	  	\label{figure_system_model}
	\end{minipage}%
	\hfill
	% Figure: Backhaul model
	\begin{minipage}[t]{.42\textwidth}
		\centering
		\includegraphics[scale=0.26]{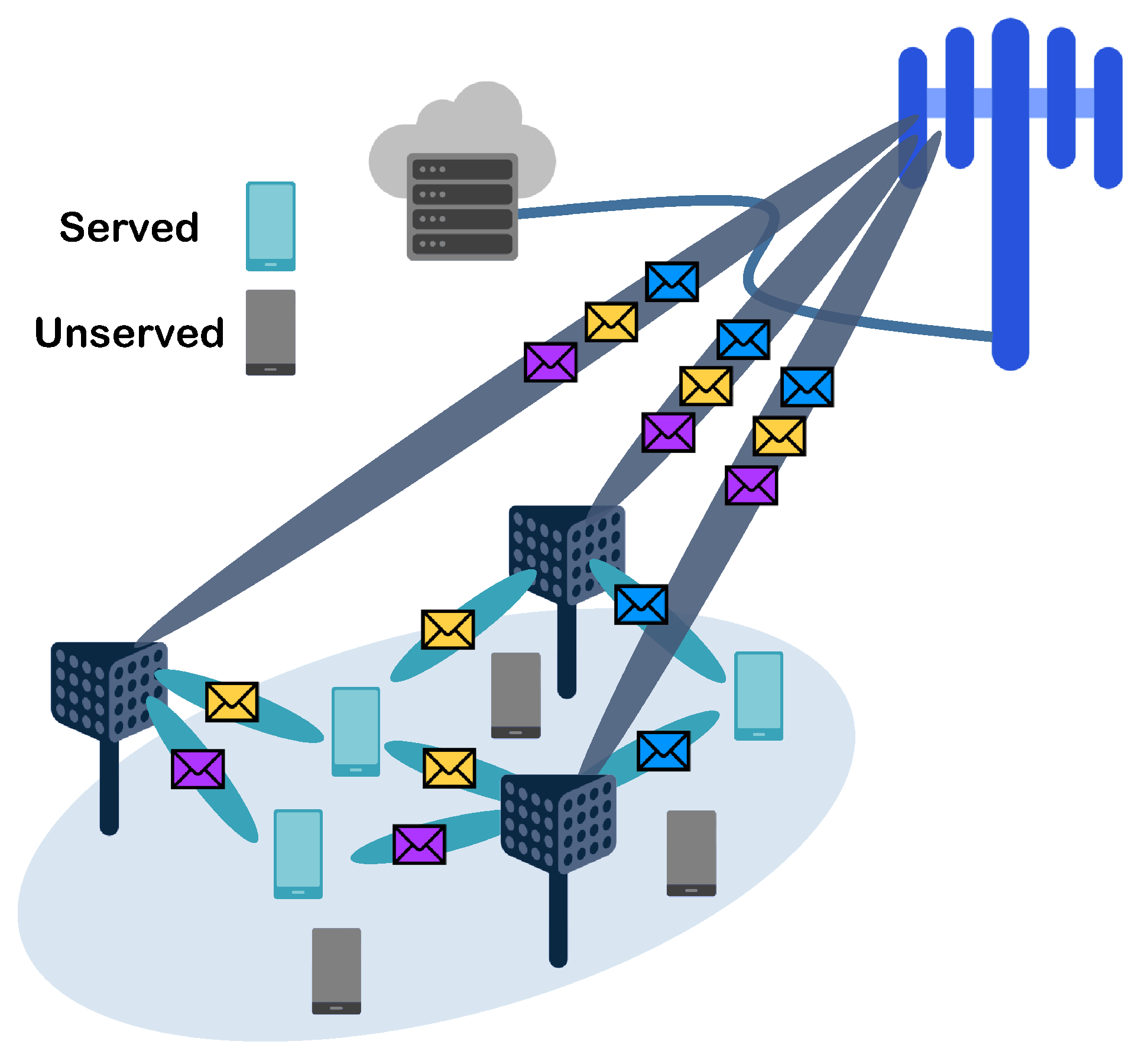}
		\vspace{-1mm}
		\caption[Caption for figure]%
			{SBS clustering allows to merge data of all the served users into one stream, minimizing interference and simplifying data decoding at the SBSs.}
		\label{figure_backhaul_model}
	\end{minipage}
\end{figure*}

% Backhaul model
\noindent {\textbf{{Backhaul model}:}} We rely on an advantageous clustering approach, where we divide the SBSs into $ L $ non-overlapping \emph{virtual cells or clusters}, each formed by $ B $ SBSs (as in distributed antenna systems). 
%This partitioning has many advantages. First, the MBS does not need to send the SBSs a data stream per each associated UE, as in \cite{chen2019:user-centric-joint-access-backhaul-full-duplex-self-backhauled-wireless-networks}, thus relieving SBSs from using SIC decoding. Second, since the MBS transmits data to each SBS cluster, this approach generates less interference compared to transmitting as many signals as UEs. Third, the MBS becomes cheaper as the MBS only requires as many RF chains as SBS clusters, which is far less than the number of UEs. 
In this way, data streams sent from the MBS to a SBS cluster contain the aggregate content for all the served UEs in that cluster, as shown in Fig. \ref{figure_backhaul_model}. The SBSs are deployed in a planned fashion and grouped based on their proximity. The antenna arrays are oriented towards the cluster center, as shown in Fig. \ref{figure_sbs_distribution}.

\begin{figure}[!h]
  	\centering
  	\includegraphics[scale=0.32]{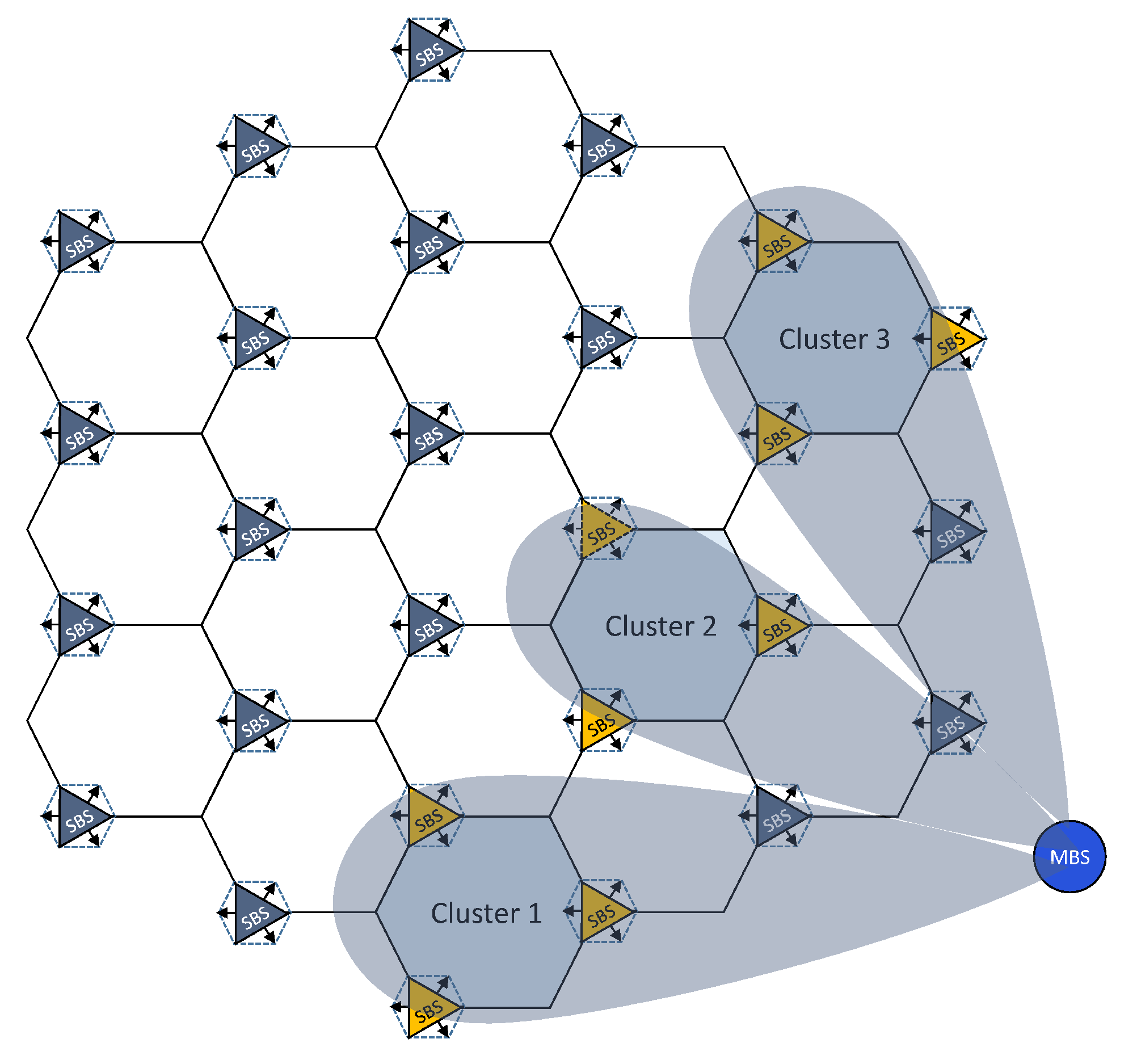}
  	\vspace{-1mm}
  	\caption[Caption for figure]%
    	{SBS distribution and clustering with a MBS transmitting multicast streams to three different clusters.}
  	\label{figure_sbs_distribution}
\end{figure}

\noindent {\textbf{{Access model}:}} Each UE is pre-associated to a SBS cluster, based on the geographical distance or a given operator policy. Without loss of generality, we assume that each cluster has $ U $ UEs. Thus, SBSs in a cluster transmit collaboratively to UEs only within that cluster. However, not all SBSs are necessarily involved in serving a particular UE, and not all UEs may be served. The information for all the served UEs is co-processed by all SBSs, thus allowing to handle interference more efficiently. 

% Channel model
\noindent {\textbf{{Channel model}:}} The backhaul links operate over a bandwidth $ W_\mathrm{BW}^\mathrm{backhaul} $ and we assume line-of-sight (LOS) connectivity since the MBS and SBSs are usually strategically installed in the planning phase. Besides, the access network operates over a bandwidth $ W_\mathrm{BW}^\mathrm{access} $ and its channels (i.e., between SBSs and UEs) exhibit multipath scattering containing both line-of-sight (LOS) and Non-line-of-sight (NLOS) components. Both access and backhaul channels are modeled according to \cite{3gpp2017:38.901}.

% Optimization model
\noindent {\textbf{{Optimization model}:}} In line with the related literature, we assume that the MBS has knowledge of the access channels between the SBSs and UEs. In particular, 3GPP specifies channel training procedures in the access network that we can rely upon. In addition, the MBS also knows the backhaul channels, i.e., between itself and the SBSs. This knowledge is even simpler to acquire than the access channels since backhaul links are rather static with small variability. In summary, the MBS collects knowledge of all the wireless channels and, accordingly, optimizes all the radio resources of the system.

\emph{For the sake of clarity, variables and parameters used in the following sections are summarized in Table \ref{table_parameters_variables}.}

% Table II
\begin{table}[!h]
 \centering
	\scriptsize
	\caption{Parameters and variables of the system}
	\begin{tabular}{|m{5.4cm} |c|}
		% Header
		\hline
		\centering {\bf Parameters and Variables} & \bf Notation	\\ 
		\hline
		% Rows
		Number of transmit antennas at the MBS and SBSs  	& $ N^{\mathrm{MBS}}_\mathrm{tx} $, $ N^{\mathrm{SBS}}_\mathrm{tx} $ \\ 
		Maximum transmit power at the MBS and SBSs			& $ P^{\mathrm{MBS}}_\mathrm{tx} $, $ P^{\mathrm{SBS}}_\mathrm{tx} $ \\ 
		Number of clusters in the system 					& $ L $ \\
		Number of UEs per cluster							& $ U $ \\
		Number of SBSs per cluster							& $ B $ \\
		Number of predefined rate/SINR values & $ J^\mathrm{UE} $, $ J^\mathrm{SBS} $ \\
		Bandwidth of the access and backhaul networks		& $ W_\mathrm{BW}^\mathrm{access} $, $ W_\mathrm{BW}^\mathrm{backhaul} $ \\
		Set of clusters										& $ \mathcal{L} = \left\lbrace 1, \cdots, L \right\rbrace $ \\
		Set of SBSs			  	& $ \mathcal{B} = \bigcup_{l \in \mathcal{L}} \mathcal{B}_l $\\
		Set of UEs				& $ \mathcal{U} = \bigcup_{l \in \mathcal{L}} \mathcal{U}_l $\\
		Set of predefined rate/SINR values at SBSs			& $ \mathcal{J}^\mathrm{SBS} $ \\
		Set of predefined rate/SINR values at UEs			& $ \mathcal{J}^\mathrm{UE} $ \\
		Set of UEs in the $l$-th cluster					& $ \mathcal{U}_l $\\
		Set of SBSs in the $l$-th cluster					& $ \mathcal{B}_l $\\
		Channel between the MBS and SBS $ b $				& $ \mathbf{g}_b $ \\
		Channel between SBS $ b $ and UE $ u $				& $ \mathbf{h}_{b,u} $ \\
		%Noise power 										& $ \sigma^2_\mathrm{UE}, \sigma^2_\mathrm{SBS} $ \\ 
		Multicast precoder from the MBS to SBS cluster $ \mathcal{B}_l $	& $ \mathbf{m}_l $ \\
		Unicast precoder from SBS $ b $ to UE $ u $					& $ \mathbf{w}_{b,u} $ \\
		Binary variable for UE rate/SINR selection & $ \alpha_{u,j} $ \\
		Binary variable for SBS rate/SINR selection & $ \beta_{l,j} $ \\
		Binary variable for UE association & $ \kappa_{b,u} $ \\
		\hline
	\end{tabular}
	\label{table_parameters_variables}
\end{table}

% Section 2.1: Backhaul Network
\subsection{Backhaul Network: Multicast Transmissions from MBS to SBSs} \label{subsection_backhaul_network}

In the backhaul network, two important aspects are dealt with. First, \emph{rate selection}, i.e. choosing appropriate data rates at which the MBS transmits information to the SBSs. Second, \emph{beamforming}, i.e. adjusting the amplitude and phases of the signals at the MBS to guarantee the selected rates.

% Section 2.1.1: Beamforming
\noindent \textbf{\textit{Beamforming:}} The MBS is equipped with a planar array of $ N^\mathrm{MBS}_\mathrm{tx} $ transmit antennas operating on Band 1 used for communication with the SBSs, which have $ N^{\mathrm{SBS}}_\mathrm{rx} = 1 $ receive antenna. The MBS transmits as many streams as clusters. Every stream contains the aggregate data for the served UEs in their respective clusters (see Fig. \ref{figure_backhaul_model}). The instantaneous multicast symbol for the SBSs in cluster $ \mathcal{B}_l $ is denoted by $ z_l $, with $ \mathbb{E} \left[ z_l\right] = 0 $ and $\mathbb{E} \left[ \left| z_l \right|^2_2 \right] = 1 $. The beamforming vector conveying $ z_l $ is denoted by $ \mathbf{m}_l $. The composite signal transmitted from the MBS to all SBS clusters is given by $ \mathbf{x}^{\mathrm{MBS}} = \sum_{l \in \mathcal{L}} \mathbf{m}_l z_l $. The received signal at SBS $ b \in \mathcal{B}_l $ is expressed as
% Equation
\begin{align} \label{equation_received_signal_SBS}
		y_b^{\mathrm{SBS}} & = \mathbf{g}_b^H \mathbf{x}^{\mathrm{MBS}} + n_b \nonumber 
		\\
		& = \underbrace{\mathbf{g}_b^H \mathbf{m}_l z_l}_\text{signal for SBS $ b $} + \underbrace{ \sum_{\substack{l' \in \mathcal{L}, l' \ne l}} \mathbf{g}_b^H \mathbf{m}_{l'} z_{l'} }_\text{interference} + \underbrace{n_b}_\text{noise},
\end{align}
where $ \mathbf{g}_b $ is the channel between SBS $ b \in \mathcal{B}_l $ and the MBS whereas $ n_b \sim \mathcal{CN} \left(0, \sigma_\mathrm{SBS}^2 \right) $ symbolizes circularly symmetric Gaussian noise. The signal-to-interference-plus-noise ratio (SINR) at SBS $ b $ is
% Equation
\begin{align} \label{equation_SINR_SBS}
	\mathrm{SINR}_b^{\mathrm{SBS}} & = \frac{ \left| \mathbf{g}_b^H \mathbf{m}_l \right|^2 } { \sum_{\substack{l' \in \mathcal{L}, l' \ne l}} \left| \mathbf{g}_b^H \mathbf{m}_{l'} \right|^2 + \sigma_\mathrm{SBS}^2 }.	
\end{align}

Since all SBSs within a cluster receive the same common information (i.e. aggregate UE content), the effective rate/SINR per cluster is determined by the SBS with the worst conditions. As a result, a more sensible means of quantifying the maximal SINR per cluster is the following $ \widetilde{\mathrm{SINR}}_l^{\mathrm{SBS}} = \min_{b \in \mathcal{B}_l} \left\lbrace \mathrm{SINR}_b^{\mathrm{SBS}} \right\rbrace, \forall l \in \mathcal{L} $.

% Remark
\textit{\textsc{Remark:} This system is known as multigroup multicast beamforming \cite{bornhorst2012:distributed-beamforming-multigroup-multicasting-relay-networks} and has been studied for transmissions from a MBS/SBS to multiple clusters of UEs. We exploit that same idea to transmit data streams from the MBS to the SBSs. We assume that the number of streams that the MBS can handle is sufficient to serve all SBS clusters, i.e. $ N^\mathrm{MBS}_\mathrm{streams} \geq L $.}

% Section 2.1.2: Rate Selection
\noindent \textbf{\textit{Rate Selection:}} In practical wireless communications systems, the set of eligible data rates is finite \cite[pp.~64]{3gpp2020:38.214}. These predefined rates are uniquely identified by their associated CQI index, and each corresponds to a specific MCS. In addition, for each rate, a minimum received SINR is required in order to ensure a target block error rate (BLER) \cite{leung2002:integrated-link-adaptation-power-control-improve-error-throughput-broadband-wireless-networks}. While the rates and MCSs are standardized, the corresponding target SINRs are usually vendor- and equipment-specific. We consider the target SINRs in \cite{kovalchukov2019:accurate-approximation-resource-request-distributions-mmwave-3gpp}, which are shown in Table \ref{table_SBS_rates_sinr} (in linear scale) and approximately exhibit increments of twice the previous rate starting from $ R^\mathrm{SBS}_1 = 0.2344 $ bps/Hz.

% Table IV
% Table
\begin{table}[h!]
	\scriptsize
	\caption{Rates and target SINR values}
	\centering
	\begin{tabular}{|c|c|c|c|}
		% Header
		\hline
		 \thead{\bf \scriptsize Coding \bf \scriptsize rate}
		 & \thead{\bf \scriptsize Rate $ R^\mathrm{SBS}_j $ {[bps/Hz]}} & \thead{\bf \scriptsize SINR} $ \Gamma^\mathrm{SBS}_j $ \\ 
		\hline
		\hline
		% Rows                                      
		120/1024 (QPSK) & 0.2344 & 0.2159\\ 
		\hline
		308/1024 (QPSK) & 0.6016 & 0.6610 \\ 
		\hline
		602/1024 (QPSK) & 1.1758 & 1.7474 \\ 
		\hline
		466/1024 (QAM) & 2.7305 & 10.6316 \\ 
		\hline
		948/1024 (QAM) & 5.5547 & 95.6974 \\ 
		\hline
	\end{tabular}
	\label{table_SBS_rates_sinr}
\end{table}

In order to assign $ R^\mathrm{SBS}_j $ to the $ l $-th SBS cluster, it is required that $ \widetilde{\mathrm{SINR}}_l^{\mathrm{SBS}} \geq \Gamma^\mathrm{SBS}_j $, $ j \in \mathcal{J}^\mathrm{SBS} $, where $ \mathcal{J}^\mathrm{SBS} $ represent the set of possible rates. To represent the rate assignment, we introduce the binary variables $ \beta_{l,j} \in \left\lbrace 0, 1 \right\rbrace $ with $ \beta_{l,j} = 1 $ denoting that the SBSs in $ \mathcal{B}_l $ are allocated $ R^\mathrm{SBS}_j $. We assume that all SBS clusters are served, which is ensured through $ \sum_{j \in \mathcal{J}^\mathrm{SBS} } \beta_{l,j} = 1, \forall l \in \mathcal{L} $ and $ N^\mathrm{MBS}_\mathrm{streams} \geq L $. Thus, to guarantee the predefined target BLER for cluster $ \mathcal{B}_l $, it must hold that $ \widetilde{\mathrm{SINR}}_l^{\mathrm{SBS}} \geq \sum_{j \in \mathcal{J}^\mathrm{SBS} } \beta_{l,j} \Gamma_j^\mathrm{SBS} $.

% Section 2.2: Access Network
\subsection{Access Network: Distributed Unicast Transmissions from SBSs to UEs} \label{subsection_access_network}

In the access network, four pivotal aspects are addressed. First, \emph{admission control}, i.e. deciding which UEs are served. Second, \emph{rate selection}, i.e. choosing data rates for the served UEs. Third, \emph{user association}, i.e. determining which subset of SBSs transmit to a served UE. Fourth, \emph{beamforming}.

% Section 2.2.1: Beamforming and User Association
\noindent \textbf{\textit{Beamforming and User Association:}} Each SBS is equipped with a planar array of $ N^\mathrm{SBS}_\mathrm{tx} $ transmit antennas operating on Band 2 and used for communication with the UEs, which have $ N^{\mathrm{UE}}_\mathrm{rx} = 1 $ receive antenna. A SBS $ b \in \mathcal{B}_l $ serving a subset of UEs in $ \mathcal{U}_l $ transmits multiple unicast signals simultaneously, each signal targeting a specific UE. The instantaneous unicast symbol for UE $ u \in \mathcal{U}_l $ is denoted by $ s_{l,u} $, with $ \mathbb{E} \left[ s_{l,u} \right] = 0 $ and $\mathbb{E} \left[ \left| s_{l,u} \right|^2_2 \right] = 1 $. In addition, the beamforming vector from SBS $ b \in \mathcal{B}_l $ transmitting $ s_{l,u} $ to UE $ u \in \mathcal{U}_l $ is denoted by $ \mathbf{w}_{b,u} $. Therefore, the composite signal that SBS $ b $ in $ \mathcal{B}_l $ sends to the UEs in $ \mathcal{U}_l $ is represented by $ \mathbf{x}_b^{\mathrm{SBS}} = \sum_{u \in \mathcal{U}_l} \mathbf{w}_{b,u} s_{l,u} \kappa_{b,u} $, where $ \kappa_{b,u} $ is a binary variable that is $ 1 $ when SBS $ b \in \mathcal{B}_l $ serves UE $ u \in \mathcal{U}_l $ and $ 0 $ otherwise. A served UE $ u \in \mathcal{U}_l $ receives its information from at least $ B_\mathrm{min} = 1 $ and at most $ B_\mathrm{max} = B $ SBSs in $ \mathcal{B}_l $. The signal received by UE $ u $ in $ \mathcal{U}_l $ is given by (\ref{equation_received_signal_UE}), where $ n_u \sim \mathcal{CN} \left(0, \sigma_\mathrm{UE}^2 \right) $ and $ \mathbf{h}_{b,u} $ represents the channel between SBS $ b $ and UE $ u $. Every UE perceives interference from within its own cluster and from neighboring clusters. The SINR at UE $ u $ in $ \mathcal{U}_l $ is defined by (\ref{equation_sinr_UE}). When $ \kappa_{b,u} = 0 $, no information is sent to the UE. The effective beamforming vector is $ \kappa_{b,u} \cdot \mathbf{w}_{b,u} $, which becomes a zero-vector for unserved UEs, thus accomplishing the association between UEs and SBSs.

\begin{figure*}[!t]
% ensure that we have normalsize text
\normalsize
% Equation
\begin{equation} \label{equation_received_signal_UE}
		y_u^{\mathrm{UE}}  = \underbrace{{\sum_{b \in \mathcal{B}_l} \mathbf{h}_{b,u}^H \mathbf{w}_{b,u} s_{l,u} \kappa_{b,u}}}_{\text{signal for UE $ u $ in cluster $ \mathcal{U}_l $}} + 
		                      \underbrace{{\sum_{b \in \mathcal{B}_l} \sum_{\substack{u' \in \mathcal{U}_l \\ u' \ne u}} \mathbf{h}_{b,u}^H \mathbf{w}_{b,u'} s_{l,u'} \kappa_{b,u'}}}_{\text{interference originated in cluster $ \mathcal{U}_l $}} + 
		                      \underbrace{{\sum_{\substack{l' \in \mathcal{L} \\ l' \ne l}} \sum_{b' \in \mathcal{B}_{l'}} \sum_{u' \in \mathcal{U}_{l'}} \mathbf{h}_{b',u}^H \mathbf{w}_{b',u'} s_{l',u'} \kappa_{b',u'}}}_{\text{aggregate interference originated in clusters $ \mathcal{U}_{l' \neq l} $}}
		                      + \underbrace{n_u}_{\text{noise}}  \\
\end{equation}
\hrulefill
% Equation
\begin{equation} \label{equation_sinr_UE}
	\mathrm{SINR}^\mathrm{UE}_u = \frac{ {\left| \sum_{b \in \mathcal{B}_l}  \mathbf{h}_{b,u}^H \mathbf{w}_{b,u} \kappa_{b,u} \right|^2 } }
					   		{
			                  { \sum_{\substack{u' \in \mathcal{U}_l \\ u' \ne u}} \Big| \sum_{b \in \mathcal{B}_l} \mathbf{h}_{b,u}^H \mathbf{w}_{b,u'} \kappa_{b,u'} \Big|^2 } + 
			                  { \sum_{\substack{l' \in \mathcal{L} \\ l' \ne l}} \sum_{u' \in \mathcal{U}_{l'}} \Big| \sum_{b' \in \mathcal{B}_{l'}} \mathbf{h}_{b',u}^H \mathbf{w}_{b',u'} \kappa_{b',u'} \Big|^2 } +
			                  \sigma_\mathrm{UE}^2
		           	   		}.
\end{equation}
\hrulefill
% Problem P'
\noindent
\begin{subequations} \label{problem_P_prima}
	\begin{align}
		% Objective
		\mathcal{P}': & \max_{
				\substack{
							\mathbf{m}_l, \mathbf{w}_{b,u},
							\alpha_{u,j}, \beta_{l,j}, \kappa_{b,u}
						 }
			   } 
		& & R_\mathrm{w-sum}^\mathrm{access} \left( \boldsymbol{\alpha} \right) \equiv \sum_{l \in \mathcal{L}} \sum_{u \in \mathcal{U}_l} \omega_u \sum_{j \in \mathcal{J}^{\mathrm{UE}}} \alpha_{u,j} R_j^\mathrm{UE} & \nonumber
		\\
		% Constraint C1
		& ~~~~~~~~~~~~ \mathrm{s.t.} & \mathrm{C_1}: ~ & \alpha_{u,j} = \left\lbrace 0,1 \right\rbrace, \forall l \in \mathcal{L}, u \in \mathcal{U}_l, j \in \mathcal{J}^\mathrm{UE}, \nonumber
		\\
		% Constraint C2
		& & \mathrm{C_2}: ~ & \sum_{j \in \mathcal{J}^{\mathrm{UE}}} \alpha_{u,j} \leq 1, \forall l \in \mathcal{L}, u \in \mathcal{U}_l, \nonumber
		\\
		% Constraint C3
		& & \mathrm{C_3}: ~ & \sum_{l \in \mathcal{L}} \left\| \mathbf{m}_l \right\|_2^2 \leq P_\mathrm{tx}^\mathrm{MBS}, \nonumber
		\\
		% Constraint C4
		& & \mathrm{\bar{C}_4}: ~ & \sum_{u \in \mathcal{U}_l} \left\| \mathbf{w}_{b,u} \kappa_{b,u} \right\|_2^2 \leq P_\mathrm{tx}^\mathrm{SBS}, \forall l \in \mathcal{L}, b \in \mathcal{B}_l, \nonumber
		\\
		% Constraint C5
		& & \mathrm{\bar{C}_5}: ~ & \mathrm{SINR}^\mathrm{UE}_u \geq \sum_{j \in \mathcal{J}^{\mathrm{UE}}} \alpha_{u,j} \Gamma_j^\mathrm{UE}, \forall l \in \mathcal{L}, u \in \mathcal{U}_l, \nonumber
		\\
		% Constraint C6
		& & \mathrm{C_6}: ~ & \kappa_{b,u} = \left\lbrace 0,1 \right\rbrace, \forall l \in \mathcal{L}, b \in \mathcal{B}_l, u \in \mathcal{U}_l, \nonumber
		\\
		% Constraint C7
		& & \mathrm{C_7}: ~ & \sum_{u \in \mathcal{U}_l} \kappa_{b,u} \leq N^\mathrm{SBS}_\mathrm{streams}, \forall l \in \mathcal{L}, b \in \mathcal{B}_l,	\nonumber
		\\	
		% Constraint C8
		& & \mathrm{C_8}: ~ & \sum_{u \in \mathcal{U}_l} \kappa_{b,u} \geq 1, \forall l \in \mathcal{L}, b \in \mathcal{B}_l,	\nonumber
		\\	
		% Constraint C9
		& & \mathrm{C_9}: ~ & \sum_{b \in \mathcal{B}_l} \kappa_{b,u} \leq B_\mathrm{max} \sum_{j \in \mathcal{J}^{\mathrm{UE}}} \alpha_{u,j}, \forall l \in \mathcal{L}, u \in \mathcal{U}_l, \nonumber
		\\
		% Constraint C10
		& & \mathrm{C_{10}}: ~ & \sum_{b \in \mathcal{B}_l} \kappa_{b,u} \geq B_\mathrm{min} \sum_{j \in \mathcal{J}^{\mathrm{UE}}} \alpha_{u,j}, \forall l \in \mathcal{L}, u \in \mathcal{U}_l, \nonumber
		\\
		% Constraint C11
		& & \mathrm{C_{11}}: ~ & \beta_{l,j} = \left\lbrace 0, 1 \right\rbrace, \forall l \in \mathcal{L}, j \in \mathcal{J}^\mathrm{SBS}, \nonumber
		\\
		% Constraint C12
		& & \mathrm{C_{12}}: ~ & \sum_{j \in \mathcal{J}^\mathrm{SBS} } \beta_{l,j} = 1, \forall l \in \mathcal{L}, \nonumber
		\\
		% Constraint C13
		& & \mathrm{C_{13}}: ~ & W_\mathrm{BW}^\mathrm{access} \sum_{u \in \mathcal{U}_l} \sum_{j \in \mathcal{J}^{\mathrm{UE}}} \alpha_{u,j} R_j^\mathrm{UE} \leq W_\mathrm{BW}^\mathrm{backhaul} \sum_{j \in \mathcal{J}^\mathrm{SBS} } \beta_{l,j} R_j^\mathrm{SBS}, \forall l \in \mathcal{L}, \nonumber
		\\
		% Constraint C14
		& & \mathrm{C_{14}}: ~ & \sum_{u \in \mathcal{U}_l} \sum_{j \in \mathcal{J}^{\mathrm{UE}}} \alpha_{u,j} = U_\mathrm{served}, \forall l \in \mathcal{L}, \nonumber
		\\
		% Constraint C15
		& & \mathrm{\bar{C}_{15}}: ~ & \widetilde{\mathrm{SINR}}_l^{\mathrm{SBS}} \geq \sum_{j \in \mathcal{J}^\mathrm{SBS} } \beta_{l,j} \Gamma_j^\mathrm{SBS}, \forall l \in \mathcal{L}, \nonumber
	\end{align}
\end{subequations}
\hrulefill
\vspace*{4pt}
\end{figure*}

% Section 2.2.2: Rate Selection and Admission Control
\noindent \textbf{\textit{Rate Selection and Admission Control:}} Similarly to Section \ref{subsection_backhaul_network}, the rate assigned to a served UE can only be one within a set of predefined values. To depict the rate selection for the UEs, we introduce the binary variables $ \alpha_{u,j} \in \left\lbrace 0, 1 \right\rbrace $. These variables perform the dual task of admission control and rate selection, which is ensured by $ \sum_{j \in \mathcal{J}^{\mathrm{UE}}} \alpha_{u,j} \leq 1, \forall l \in \mathcal{L}, u \in \mathcal{U}_l $, where $ \mathcal{J}^\mathrm{UE} $ represents the set of possible rate values. A UE $ u $ is served when $ \sum_{j \in \mathcal{J}^{\mathrm{UE}}} \alpha_{u,j} = 1 $, meaning that one rate has been assigned. Otherwise, when $ \sum_{j \in \mathcal{J}^{\mathrm{UE}}} \alpha_{u,j} = 0 $, the UE is not served. We denote the rates and target SINRs for UEs with $ R_j^{\mathrm{UE}} $ and $ \Gamma_j^{\mathrm{UE}} $, respectively. To assign $ R_j^{\mathrm{UE}} $ to UE $ u $, it is required that $ \mathrm{SINR}^\mathrm{UE}_u \geq \Gamma_j^{\mathrm{UE}} $, $ j \in \mathcal{J}^\mathrm{UE} $, for which we assume the same values shown in Table \ref{table_SBS_rates_sinr} in Section \ref{table_SBS_rates_sinr}. Further, not all UEs shall be admitted since each SBS can support up to $ N^\mathrm{SBS}_\mathrm{streams} $ streams simultaneously.

\begin{figure*}[!t]
% ensure that we have normalsize text
\normalsize
% Proposition 1
\begin{proposition}
	Due to existence of $ \mathrm{C}_1 - \mathrm{C}_2 $, constraint $ \mathrm{\bar{C}_5} $ can be equivalently rewritten as
	% Constraints
	\begin{align}
		\mathrm{C_5}: \mathrm{SINR}^\mathrm{UE}_u\geq \alpha_{u,j} \Gamma_j^\mathrm{UE}, \forall l \in \mathcal{L}, u \in \mathcal{U}_l, j \in \mathcal{J}^{\mathrm{UE}}. \nonumber 
	\end{align}
	% Proof 1
	\noindent \textit{Proof: Because of $ \mathrm{C}_2 $, there is at most one variable at a time that is $ 1 $. As a result, the SINR constraints can be decomposed into multiple constraints, each being related to only one binary variable.} 
\end{proposition} 
% Proposition 2
\begin{proposition}
	Due to existence of $ \mathrm{C_6} $, constraints $ \mathrm{\bar{C}_4} - \mathrm{C_5} $ can be equivalently rewritten as $ \mathrm{C_{17}} $, $ \mathrm{C_{18}} $, $ \mathrm{C_{19}} $, $ \mathrm{C_{20}} $, $ \mathrm{\bar{C}_{21}} $, where
	
	% Constraints
	\noindent{
	\resizebox{1.01\columnwidth}{!}{
	\begin{minipage}{1.01\columnwidth}
	\begin{align}
	  \mathrm{\bar{C}_4} - \mathrm{C_5} =
	  \begin{cases}
	  	\mathrm{C_{17}}: p_{b,u} \geq 0, \forall l \in \mathcal{L}, b \in \mathcal{B}_l, u \in \mathcal{U}_l,  \\
	   	\mathrm{C_{18}}: \sum_{u \in \mathcal{U}_l} p_{b,u} \leq P_\mathrm{tx}^\mathrm{SBS}, \forall l \in \mathcal{L}, b \in \mathcal{B}_l, \\
	   	\mathrm{C_{19}}: p_{b,u} \leq \kappa_{b,u} P_\mathrm{tx}^\mathrm{SBS}, \forall l \in \mathcal{L}, b \in \mathcal{B}_l, u \in \mathcal{U}_l, \\
	   	\mathrm{C_{20}}: \left\| \left[ 2 \mathbf{w}_{b,u}^H,  \kappa_{b,u} - p_{b,u} \right] \right\|_2 \leq \kappa_{b,u} + p_{b,u}, \forall l \in \mathcal{L}, b \in \mathcal{B}_l, u \in \mathcal{U}_l, \\
	   	\mathrm{\bar{C}_{21}}: \frac{ \big| \sum_{b \in \mathcal{B}_l} \mathbf{h}_{b,u}^H \mathbf{w}_{b,u} \big|^2 }
	     				 { \sum_{\substack{u' \in \mathcal{U}_l \\ u' \ne u}} \big| \sum_{b \in \mathcal{B}_l} \mathbf{h}_{b,u}^H \mathbf{w}_{b,u'} \big|^2 + \sum_{\substack{l' \in \mathcal{L} \\ l' \ne l}} \sum_{u' \in \mathcal{U}_{l'}} \big| \sum_{b' \in \mathcal{B}_{l'}} \mathbf{h}_{b',u}^H \mathbf{w}_{b',u'} \big|^2 + \sigma_\mathrm{UE}^2 } \geq \alpha_{u,j} \Gamma_j^\mathrm{UE}, \forall l \in \mathcal{L}, u \in \mathcal{U}_l, j \in \mathcal{J}^{\mathrm{UE}}, \nonumber \\		 	 				 
	\end{cases}  
	\end{align}
	\end{minipage}
	}}

	% Proof 2
	\noindent \textit{Proof: See Appendix \ref{appendix_proposition_2}.}
\end{proposition} 
% Proposition 3
\begin{proposition}
	Due to existence of $ \mathrm{C_1} $, constraint $ \mathrm{\bar{C}_{21}} $ can be rewritten as $ \mathrm{C_{21}} $, where
	
	% Constraints
	\begin{align}
		\mathrm{C_{21}}: \sum_{\substack{l' \in \mathcal{L}}} \sum_{u' \in \mathcal{U}_{l'}} \Big| \sum_{b' \in \mathcal{B}_{l'}} \mathbf{h}_{b',u}^H \mathbf{w}_{b',u'} \Big|^2 + \sigma_\mathrm{UE}^2 \leq \left( 1 + {\Gamma_j^\mathrm{UE} }^{-1} \right) \Big| \sum_{b \in \mathcal{B}_l} \mathbf{h}_{b,u}^H \mathbf{w}_{b,u} \Big|^2 + \left( 1 - \alpha_{u,j} \right)^2 Q_u^2, \forall l \in \mathcal{L}, u \in \mathcal{U}_l, j \in \mathcal{J}^{\mathrm{UE}}, \nonumber 
	\end{align}
	and $ Q_u^2 = P_\mathrm{tx}^\mathrm{SBS} \sum_{\substack{l' \in \mathcal{L}}} \sum_{b' \in \mathcal{B}_{l'}} \left\| \mathbf{h}_{b',u} \right\|_2^2 + \sigma^2_\mathrm{UE} $ is an upper bound for the left-hand side (LHS) term of $ \mathrm{C_{21}} $.
	
	% Proof 3
	\noindent \textit{Proof: See Appendix \ref{appendix_proposition_3}.}
\end{proposition}

% The spacer can be tweaked to stop underfull vboxes.
\vspace*{4pt}
\end{figure*}

% Problem Formulation
\subsection{Problem Formulation} \label{subsection_problem_formulation}
We investigate the problem of joint optimization of \emph{beamforming}, \emph{user association}, \emph{rate selection}, \emph{admission control} in the access network and \emph{beamforming}, \emph{rate selection} in the backhaul network aiming to maximize the weighted sum-rate at the access network (i,e., for the UEs), which is formulated as $ \mathcal{P}' $ in the previous page.

In $ \mathcal{P}' $, $ R_\mathrm{w-sum}^\mathrm{access} \left( \boldsymbol{\alpha} \right) $ denotes the weighted sum-rate achieved by all UEs in the access network. Besides, $ \omega_u $ represents the weight associated to UE $ u $, which can be adjusted by the network operator to assign different priorities, for instance, to balance fairness among UEs. Formally, the objective function is expressed as $ R_\mathrm{w-sum}^\mathrm{access} \left( \boldsymbol{\alpha} \right) \equiv W_\mathrm{BW}^\mathrm{access} \sum_{l \in \mathcal{L}} \sum_{u \in \mathcal{U}_l} \omega_u \sum_{j \in \mathcal{J}^{\mathrm{UE}}} \alpha_{u,j} R_j^\mathrm{UE} $. However, since $ W_\mathrm{BW}^\mathrm{access} $ is constant, we have redefined it as $ R_\mathrm{w-sum}^\mathrm{access} \left( \boldsymbol{\alpha} \right) \equiv \sum_{l \in \mathcal{L}} \sum_{u \in \mathcal{U}_l} \omega_u \sum_{j \in \mathcal{J}^{\mathrm{UE}}} \alpha_{u,j} R_j^\mathrm{UE} $ without altering the nature of the problem. 

Constraints $ \mathrm{C_{1}} $, $ \mathrm{C_{2}} $, $ \mathrm{\bar{C}_{4}} $, $ \mathrm{\bar{C}_{5}} $, $ \mathrm{C_{6}} $, $ \mathrm{C_{7}} $, $ \mathrm{C_{8}} $, $ \mathrm{C_{9}} $, $ \mathrm{C_{10}} $, $ \mathrm{C_{14}} $ are related to the access network, $ \mathrm{C_{3}} $, $ \mathrm{C_{11}} $, $ \mathrm{C_{12}} $, $ \mathrm{\bar{C}_{15}} $ are related to the backhaul network whereas $ \mathrm{C_{13}} $ is related to both networks. Constraints $ \mathrm{C_1} - \mathrm{C_2} $ depict the rate selection for all UEs, constraint $ \mathrm{C_3} $ restricts the transmit power of the MBS, constraint $ \mathrm{\bar{C}_4} $ restricts the transmit power of the SBSs, constraint $ \mathrm{\bar{C}_5} $ guarantees that the unicast SINR is larger than the corresponding target SINR (specified in Table \ref{table_SBS_rates_sinr}), constraints $ \mathrm{C_6} - \mathrm{C_8} $ ensure that each SBS serves at least one UE but cannot serve more UEs than the number of streams it can handle, constraints $ \mathrm{C_9} - \mathrm{C_{10}} $ ensure that each admitted UE is served by at least $ B_\mathrm{min} $ and by at most $ B_\mathrm{max} $ SBSs, constraints $ \mathrm{C_{11}} - \mathrm{C_{12}} $ guarantee a rate selection for every SBS cluster, constraint $ \mathrm{C_{13}} $ guarantees that the total access throughput in a cluster does not exceed the throughput of the corresponding serving backhaul link, $ \mathrm{C_{14}} $ ensures that there are $ U_\mathrm{served} $ served UEs per cluster, constraint $ \mathrm{\bar{C}_{15}} $ guarantees that the SINR per SBS cluster is larger than the selected target SINR (specified in Table \ref{table_SBS_rates_sinr}). 

% Remark
\textit{\textsc{Remark:} In the strict sense, the integrality constraints (i.e., $ \mathrm{C_1} $, $ \mathrm{C_6} $, $ \mathrm{C_{11}} $) make $ \mathcal{P}' $ nonconvex. Nevertheless, in the MINLP literature, a MINLP is referred to as nonconvex if it remains nonconvex even after excluding the integral variables. Otherwise, it is called convex \cite{sahinidis2019:mixed-integer-nonlinearp-rogramming}. In general, both convex and nonconvex MINLPs are NP-hard but the latter ones are more challenging to solve. Specifically, $ \mathcal{P}' $ is a nonconvex MINLP and the nonconvexity nature is conferred by the constraints $ \mathrm{\bar{C}_4} $, $ \mathrm{\bar{C}_5} $, $ \mathrm{\bar{C}_{15}} $.}

% Section 4: Proposed Problem Reformulation
\section{Proposed Problem Reformulation} \label{section_proposed_reformulation}

In this section, we propose a series of transformations to simplify the nonconvex constraints $ \mathrm{\bar{C}_4} $, $ \mathrm{\bar{C}_5} $, $ \mathrm{\bar{C}_{15}} $. The resulting reformulation $ \mathcal{P} $ (shown in Section \ref{subsection_redefining_problem}) is used in Section \ref{section_BNBC_MISOCP}, Section \ref{section_RNP_SOCP_1}, Section \ref{section_RNP_SOCP_2}, where we propose three algorithms: \texttt{BnC-MISOCP}, \texttt{RnP-SOCP-1} and \texttt{RnP-SOCP-2}.

% Section 4.1
\subsection{Eliminating Additive Coupling between Binary Variables} \label{subsection_eliminating_additive_coupling_binary_variables}

To deal with the additive coupling of the binary variables at the right-hand side (RHS) of $ \mathrm{\bar{C}_5} $ (i.e. sum of variables), we separate $ \mathrm{\bar{C}_5} $ into multiple equivalent constraints, as described in \emph{Proposition 1}.

% Section 4.2
\subsection{Eliminating the Multiplicative Coupling between Continuous and Binary Variables} \label{subsection_eliminating_multiplicative_coupling_continuous_binary_variables}

To deal with the multiplicative coupling between the unicast beamforming vectors and binary variables (in the form $ \mathbf{w}_{b,u} \kappa_{b,u} $) in $ \mathrm{\bar{C}_4} - \mathrm{C_5} $, we reformulate such interdependencies as equivalent additive couplings, which are simpler to handle, as described in \emph{Proposition 2}. In addition, note that $ \mathrm{C_{17} - C_{20}} $ are convex, whereas $ \mathrm{\bar{C}_{21}} $ is a nonconvex mixed-integer nonlinear constraint. To circumvent the involved structure $ \mathrm{\bar{C}_{21}} $, we remodel it (without loss of optimality) harnessing the \textit{big-M} method \cite{dantzig1948:programming-linear-structure}, which allows to remove the multiplicative tie between the beamformers and binary variables, as described in \emph{Proposition 3}. Finally, because constraint $ \mathrm{\bar{C}_{15}} $ has a similar structure as $ \mathrm{\bar{C}_{5}} $, we can reformulate it in an equivalent manner, as described in \emph{Proposition 4}.

\begin{figure*}[!t]
	% ensure that we have normalsize text
	\normalsize
	
	% Proposition 4
	\begin{proposition}
		Due to existence of $ \mathrm{C_{11}} $, constraint $ \mathrm{\bar{C}_{15}} $ can be equivalently recast as $ \mathrm{C_{15}} $, where
		% Constraints
		\begin{align} \nonumber
			& \mathrm{C_{15}}: \sum_{l' \in \mathcal{L}} \left| \mathbf{g}_b^H \mathbf{m}_{l'} \right|^2 + \sigma^2_\mathrm{SBS} \leq \left( 1 + {\Gamma_j^\mathrm{SBS} }^{-1} \right) \left| \mathbf{g}_b^H \mathbf{m}_l \right|^2 + \left( 1 - \beta_{l,j} \right)^2 Q_b^2, \forall l \in \mathcal{L}, b \in \mathcal{B}_l, j \in \mathcal{J}^{\mathrm{SBS}},	 				 
		\end{align}
		and $ Q_b^2 = P_\mathrm{tx}^\mathrm{MBS} \left\| \mathbf{g}_b \right\|_2^2 + \sigma^2_\mathrm{SBS} $ is an upper bound for the LHS term of $ \mathrm{C_{15}}$. 
		
		% Proof 4
		\noindent \textit{Proof: The proof is along the same lines as the procedures adopted in Proposition 1, Proposition 2 and Proposition 3. Therefore, it is omitted.}
	\end{proposition}
	
	% Proposition 5
	\begin{proposition}
		The nonconvex constraints $ \mathrm{C_{21}} - \mathrm{C_{22}} $ can be equivalently expressed as SOC constraints $ \mathrm{C_{23} - C_{25}} $, i.e., 
		\begin{align}
		  & \mathrm{C_{21} - C_{22}} =
		  \begin{cases}
		  	\mathrm{C_{23}}: \left\| \left[ \bar{\mathbf{h}}_u^H \mathbf{W}, \sigma_\mathrm{UE} \right] \right\|_2 \leq \sqrt{1 + {\Gamma_j^\mathrm{UE} }^{-1} } \mathsf{Re} \left\lbrace \mathbf{h}_u^H \mathbf{w}_u \right\rbrace  + \left(1 - \alpha_{u,j} \right) Q_u, \forall l \in \mathcal{L}, u \in \mathcal{U}_l, j \in \mathcal{J}^{\mathrm{UE}}, 
		  	\\
		   	\mathrm{C_{24}}: \mathsf{Re} \left\lbrace \mathbf{h}_u^H \mathbf{w}_u \right\rbrace \geq \alpha_{u,j} \sqrt{\Gamma_j^\mathrm{UE}} \sigma_\mathrm{UE}, \forall l \in \mathcal{L}, u \in \mathcal{U}_l, j \in \mathcal{J}^{\mathrm{UE}}, 
		   	\\
		   	\mathrm{C_{25}}: \mathsf{Im} \left\lbrace \mathbf{h}_u^H \mathbf{w}_u \right\rbrace = 0, \forall l \in \mathcal{L}, u \in \mathcal{U}_l, j \in \mathcal{J}^{\mathrm{UE}}. \nonumber \\				 	 				 
		  \end{cases}
		\end{align}
		
		% Proof 5
		\noindent \textit{Proof: See Appendix \ref{appendix_proposition_5}.}
	\end{proposition}
	% Proposition 6
	\begin{proposition}
		The nonconvex constraints $ \mathrm{C_{15}} - \mathrm{C_{16}} $ can be recast as the more conservative SOC constraints $ \mathrm{C_{26} - C_{27}} $, where
		\begin{align} \nonumber
		  & \mathrm{C_{15} - C_{16}} =
		  \begin{cases}
		  	\mathrm{C_{26}}: \left\| \left[ \mathbf{g}_b^H \mathbf{M}, \sigma_\mathrm{SBS} \right] \right\|_2 \leq \sqrt{ 1 + {\Gamma_j^\mathrm{SBS} }^{-1} } \mathsf{Re} \left\lbrace \mathbf{g}_b^H \mathbf{m}_l \right\rbrace + \left( 1 - \beta_{l,j} \right) Q_b, \forall l \in \mathcal{L}, b \in \mathcal{B}_l, j \in \mathcal{J}^{\mathrm{SBS}}, 
		  	\\
		   	\mathrm{C_{27}}: \mathsf{Re} \left\lbrace \mathbf{g}_b^H \mathbf{m}_l \right\rbrace \geq \beta_{l,j} \sqrt{\Gamma_j^\mathrm{SBS}} \sigma_\mathrm{SBS}, \forall l \in \mathcal{L}, b \in \mathcal{B}_l, j \in \mathcal{J}^{\mathrm{SBS}}.
		  \end{cases}
		\end{align}
		
		% Proof 6
		\noindent \textit{Proof: See Appendix \ref{appendix_proposition_6}.}
	\end{proposition}

% The spacer can be tweaked to stop underfull vboxes.
\vspace*{4pt}
\end{figure*}

% Section 4.3
\subsection{Adding Cuts to Tighten the Feasible Set} \label{subsection_adding_cuts_tighten_feasible_set}

To reduce the number of branches to be evaluated by MINLP solvers, we include valid inequalities (cuts) for certain constraints involving integer variables. Thus, we add the constraints $ \mathrm{C_{16}} $ and $ \mathrm{C_{22}} $, defined as

\noindent{
\resizebox{1.01\columnwidth}{!}{
\begin{minipage}{1.01\columnwidth}
% Constraints
\begin{align}
    & \mathrm{C_{16}}: \left| \mathbf{g}_b^H \mathbf{m}_l \right|^2 \geq \beta_{l,j} \Gamma_j^\mathrm{SBS} \sigma^2_\mathrm{SBS}, \forall l \in \mathcal{L}, b \in \mathcal{B}_l, j \in \mathcal{J}^{\mathrm{SBS}}, \nonumber \\ 
    & \mathrm{C_{22}}: \Big|  \sum_{b \in \mathcal{B}_l} \mathbf{h}_{b,u}^H \mathbf{w}_{b,u} \Big|^2 \geq \alpha_{u,j} \Gamma_j^\mathrm{UE} \sigma^2_\mathrm{UE}, \forall l \in \mathcal{L}, u \in \mathcal{U}_l, j \in \mathcal{J}^{\mathrm{UE}}. \nonumber  	
\end{align}
\end{minipage}
}}

Note that $ \mathrm{C_{16}} $ is a lower bound for the multicast SINR numerator, which becomes tight when the interference term is zero. This constraint is always satisfied when $ \beta_{l,j} $ are binary thus reducing the feasible set and tightening the problem relaxation when the binary variables are recast as real values (as in the proposed algorithms in Section \ref{section_RNP_SOCP_1} and Section \ref{section_RNP_SOCP_2}). Adding $ \mathrm{C_{16}} $ does not change the nature of the problem nor affects its optimality. Similarly, $ \mathrm{C_{22}} $ is a lower bound for the unicast SINR numerator.

% Section 4.4
\subsection{Redefining the Problem} \label{subsection_redefining_problem}

After applying the transformations in Section \ref{subsection_eliminating_additive_coupling_binary_variables}, Section  \ref{subsection_eliminating_multiplicative_coupling_continuous_binary_variables} and Section \ref{subsection_adding_cuts_tighten_feasible_set}, the nonconvex constraints $ \mathrm{\bar{C}_4} $, $ \mathrm{\bar{C}_5} $, $ \mathrm{\bar{C}_{15}} $ have been replaced by the convex constraints $ \mathrm{C_{17}} $, $ \mathrm{C_{18}} $, $ \mathrm{C_{19}} $, $ \mathrm{C_{20}} $ and the nonconvex constraints $ \mathrm{C_{15}} $, $ \mathrm{C_{21}} $. In addition, the nonconvex constraints $ \mathrm{C_{16}} $, $ \mathrm{C_{22}} $ have been added to contract the feasible set. Collecting these outcomes, we define $ \mathcal{P} $ as
% Problem P
\begin{align}
	% Objective
	\mathcal{P} : & \max_{
			\substack{
						\mathbf{m}_l, \mathbf{w}_{b,u}, p_{b,u}, \\
						\alpha_{u,j}, \beta_{l,j}, \kappa_{b,u}
				 }
  	} 
  	& \text{\footnotesize{convex:}} ~ & R_\mathrm{w-sum}^\mathrm{access} \left( \boldsymbol{\alpha} \right) \nonumber
  	\\
	% Constraint 1
	& ~~~~~~ \mathrm{s.t.} & \text{\footnotesize{convex:}} ~ & \mathrm{C_2} - \mathrm{C_3}, \mathrm{C_7} - \mathrm{C_{10}}, \mathrm{C_{12}} - \mathrm{C_{14}}, \nonumber
	\\
	& & & \mathrm{C_{17}} - \mathrm{C_{20}}, \nonumber
	\\
	% Constraint 2
	& & \text{\footnotesize{nonconvex:}} ~ & \mathrm{C_{15}} - \mathrm{C_{16}}, \mathrm{C_{21}} - \mathrm{C_{22}}, \nonumber
	\\
	% Constraint 3
	& & \text{\footnotesize{binary:}} ~ & \mathrm{C_1}, \mathrm{C_6}, \mathrm{C_{11}}. \nonumber
\end{align}

% Remark
\textit{\textsc{Remark:} Notice that $ \mathcal{P} $ is also a nonconvex MINLP and has the same optimal solution as $ \mathcal{P}' $ since the introduced transformations do not affect the original feasible set. However, the structure of $ \mathcal{P} $ is simpler, thus allowing us to tailor algorithms for solving the problem more efficiently.}

% Section 5
\section{\texttt{BnC-MISOCP}: Proposed MISOCP Formulation} \label{section_BNBC_MISOCP}
In this section, we recast $ \mathcal{P} $ as a MISOCP by transforming the nonconvex constraints into convex ones. We remodel $ \mathrm{C_{21}} - \mathrm{C_{22}} $ as convex constraints and replace $ \mathrm{C_{15}} - \mathrm{C_{16}} $ with convex inner surrogates.

% Section 5.1
\subsection{Transforming Nonconvex Constraints into Convex Constraints} \label{subsection_transforming_nonconvex_constraints_convex_constraints}

To deal with the nonconvex constraints $ \mathrm{C_{21}} - \mathrm{C_{22}} $, we recast them as convex conic constraints as they have hidden convexity. To simplify notation, we first rewrite $ \mathrm{C_{21}} - \mathrm{C_{22}} $ as
% Constraints
\begin{align}
	% Constraint C21
	& \mathrm{C_{21}}: \left| \bar{\mathbf{h}}_u^H \mathbf{W} \right|^2 + \sigma^2_\mathrm{UE} \leq \left( 1 + {\Gamma_j^\mathrm{UE}}^{-1} \right) \left| \mathbf{h}_u^H \mathbf{w}_u \right|^2 \nonumber
	\\
	& ~~~~~~~~~~~~~~~~~ + \left( 1 - \alpha_{u,j} \right)^2 Q_u^2, \forall l \in \mathcal{L}, u \in \mathcal{U}_l, j \in \mathcal{J}^{\mathrm{UE}}, \nonumber 
	\\
	% Constraint C22
	& \mathrm{C_{22}}: \alpha_{u,j} \Gamma_j^\mathrm{UE} \sigma^2_\mathrm{UE} \leq \left| \mathbf{h}_u^H \mathbf{w}_u \right|^2, \forall l \in \mathcal{L}, u \in \mathcal{U}_l, j \in \mathcal{J}^{\mathrm{UE}}, \nonumber 
\end{align}
where $ \big| \sum_{b \in \mathcal{B}_l} \mathbf{h}_{b,u}^H \mathbf{w}_{b,u} \big|^2 = \left| \mathbf{h}_u^H \mathbf{w}_u \right|^2, u \in \mathcal{U}_l $ and $ \sum_{\substack{l' \in \mathcal{L}}} \sum_{u' \in \mathcal{U}_{l'}} \big| \sum_{b' \in \mathcal{B}_{l'}} \mathbf{h}_{b',u}^H \mathbf{w}_{b',u'} \big|^2 = \left| \bar{\mathbf{h}}_u^H \mathbf{W} \right|^2 $. In particular, $ \mathbf{h}_u = \left[ \mathbf{h}_{b_1,u}^H, \cdots, \mathbf{h}_{b_B,u}^H \right]^H $ and $ \mathbf{w}_u = \left[ \mathbf{w}_{b_1,u}^H, \cdots, \mathbf{w}_{b_B,u}^H \right]^H$, denote respectively the channels and beamforming vectors from all SBS in the same cluster that UE $ u $ is located. Further, $ \bar{\mathbf{h}}_u $ denotes the channel between UE $ u $ and all SBSs in the system whereas $ \mathbf{W} $ is a block diagonal matrix collecting all beamforming vectors between SBSs and UEs. After applying these changes, we are in the position of expressing the nonconvex constraints $ \mathrm{C_{21}} - \mathrm{C_{22}} $ as exactly equivalent SOC constraints, as described in \emph{Proposition 5}.

% Section 5.2
\subsection{Recasting Nonconvex Constraints as Convex Inner Approximations} \label{subsection_recasting_nonconvex_constraints_convex_inner_approximations}

To circumvent the nonconvex constraints $ \mathrm{C_{15}} $, $ \mathrm{C_{16}} $, we replace them by convex surrogates. Assuming that $ \mathbf{M} = \left[ \mathbf{m}_1, \cdots, \mathbf{m}_L \right]  $, we express $ \mathrm{C_{15}} $ as
% Constraints
\begin{align} \nonumber
	% Constraint C14
	& \mathrm{C_{15}}: \left\| \mathbf{g}_b^H \mathbf{M} \right\|^2_2 + \sigma^2_\mathrm{SBS} \leq \left( 1 + {\Gamma_j^\mathrm{SBS}}^{-1} \right) \left| \mathbf{g}_b^H \mathbf{m}_l \right|^2 \nonumber
	\\
	& ~~~~~~~~~~~~ + \left( 1 - \beta_{l,j} \right)^2 Q_b^2, \forall l \in \mathcal{L}, b \in \mathcal{B}_l, j \in \mathcal{J}^{\mathrm{SBS}}. \nonumber
\end{align}

Using this expression, we reformulate $ \mathrm{C_{15}} - \mathrm{C_{16}} $ as convex inner SOC approximations, as stated in \emph{Proposition 6}. If constraints $ \mathrm{C_{26}} - \mathrm{C_{27}} $ are satisfied, then $ \mathrm{C_{15}} - \mathrm{C_{16}} $ are automatically guaranteed because the feasible set of $ \mathrm{C_{26}} - \mathrm{C_{27}} $ is contained in that of $ \mathrm{C_{15}} - \mathrm{C_{16}} $. Therefore, they are called inner approximations.

% Section 5.3
\subsection{Summarizing the Changes} \label{subsection_summarizing_chnages_P0}

After applying the transformations above, we define the following problem,
% Problem P0
\begin{align}
	% Objective
	\mathcal{P}_0 : & \max_{
			\substack{
						\mathbf{m}_l, \mathbf{w}_{b,u}, p_{b,u}, \\
						\alpha_{u,j}, \beta_{l,j}, \kappa_{b,u} 
				 }
  	} 
  	& \text{\footnotesize{convex:}} ~ & R_\mathrm{w-sum}^\mathrm{access} \left( \boldsymbol{\alpha} \right) \nonumber
  	\\
	% Constraint 1
	& ~~~~~~ \mathrm{s.t.} & \text{\footnotesize{convex:}} ~ & \mathrm{C_2} - \mathrm{C_3}, \mathrm{C_7} - \mathrm{C_{10}}, \mathrm{C_{12}} - \mathrm{C_{14}}, \nonumber
	\\
	& & & \mathrm{C_{17}} - \mathrm{C_{20}}, \mathrm{C_{23}} - \mathrm{C_{27}}, \nonumber
	\\ 
	% Constraint 2
	& & \text{\footnotesize{binary:}} ~ & \mathrm{C_1}, \mathrm{C_6}, \mathrm{C_{11}}, \nonumber
\end{align}
which is an inner approximation of problem $ \mathcal{P} $ due to convexification of its original feasible set upon replacing $ \mathrm{C_{15}} - \mathrm{C_{16}} $ by $ \mathrm{C_{26}} - \mathrm{C_{27}} $. Thus, any feasible solution to $ \mathcal{P}_0 $ will also be feasible to $ \mathcal{P}' $ and $ \mathcal{P} $. Here, $ \mathcal{P}_0 $ has $ N_v = 2 L N_\mathrm{tx}^\mathrm{MBS} + 2 L B U N_\mathrm{tx}^\mathrm{SBS} + 2 L B U + L J^\mathrm{SBS} + L U J^\mathrm{UE} $ variables, $ N_l = 3L + 2LU + 3LB + 2LBU + 3LUJ^\mathrm{UE} + LBJ^\mathrm{SBS} $ linear constraints and $ N_c = 1 + LBU + LUJ^\mathrm{UE} + LBJ^\mathrm{SBS} $ convex constraints. The complexity is $ \mathcal{O} \left( N_s (N_v)^3 (N_l + N_c) \right) $, where $ N_s $ is the total number of evaluations needed by the mixed-integer (MIP) solver.

% Remark
\textit{\textsc{Remark:} Note that $ \mathcal{P}_0 $ is a convex MINLP, and as such it can be solved optimally by MIP solvers which exploit BnC techniques to prune infeasible solutions thus reducing the search space of the problem. Although BnC techniques can explore the binary space more efficiently and are faster than exhaustive search, they may still require a considerable amount of time to find the optimum, specially when the number of integral variables is large as in $ \mathcal{P}_0 $. Thus, in order to expedite this process, we propose suboptimal algorithms in Section \ref{section_RNP_SOCP_1} and Section \ref{section_RNP_SOCP_2} based on integrality relaxation and penalization.}

% Section 6
\section{Proposed Bounds} \label{section_proposed_bounds}

We derive an \emph{upper bound} and a \emph{lower bound} for $ \mathcal{P}_0 $. The upper bound is defined as a MISOCP whereas the lower bound is a system- and problem-specific rate value. When not possible to obtain a solution for $ \mathcal{P}_0 $ (due to high time complexity), the upper and lower bounds will be used as benchmarks for the developed algorithms in Section \ref{section_RNP_SOCP_1} and Section \ref{section_RNP_SOCP_2}. 

\noindent {\textbf{{Upper Bound (\texttt{UB})}:}} While the weighted sum-rate is a mechanism to balance rates, i.e., to give higher priorities to the least favored UEs, the actual aggregate rate in the network is given by the sum-rate $ R_\mathrm{sum}^\mathrm{access} \left( \boldsymbol{\alpha} \right) = W_\mathrm{BW}^\mathrm{access} \sum_{l \in \mathcal{L}} \sum_{u \in \mathcal{U}_l} \sum_{j \in \mathcal{J}^{\mathrm{UE}}} \alpha_{u,j} R_j^\mathrm{UE} $ (without the weights). Thus, note that $ R_\mathrm{sum}^\mathrm{access} \left( \boldsymbol{\alpha} \right) $ is related to constraint $ \mathrm{C_{13}} $, which ensures that the access sum-rate per cluster does not exceed the rate of the serving backhaul link. Therefore, the access sum-rate $ R_\mathrm{sum}^\mathrm{access} \left( \boldsymbol{\alpha} \right) $ is bounded from above by the backhaul sum-rate, defined as $ R_\mathrm{sum}^\mathrm{backhaul} \left( \boldsymbol{\beta} \right) \triangleq W_\mathrm{BW}^\mathrm{backhaul} \sum_{l \in \mathcal{L} } \sum_{j \in \mathcal{J}^\mathrm{SBS} } \beta_{l,j} R_j^\mathrm{SBS} $, i.e., $ R_\mathrm{sum}^\mathrm{access} \left( \boldsymbol{\alpha} \right) \leq  R_\mathrm{sum}^\mathrm{backhaul} \left( \boldsymbol{\beta} \right) $. Since the backhaul sum-rate depends only on $ \mathbf{m}_l $ and $ \beta_{l,j} $, the upper bound is given by 
% Problem UB
\begin{subequations}
	\begin{align}
		% Objective
		\mathcal{P}_\mathrm{UB}: & \max_{
				\substack{
							\mathbf{m}_l, \beta_{l,j}
						 }
			   } ~ R_\mathrm{sum}^\mathrm{backhaul} \left( \boldsymbol{\beta} \right) ~~\mathrm{s.t.} ~~ \mathrm{C_3}, \mathrm{C_{11}}, \mathrm{C_{12}}, \mathrm{{C}_{26}}, \mathrm{{C}_{27}},\nonumber
	\end{align}
\end{subequations}
which is a MISOCP that can be solved optimally. \emph{The upper bound essentially maximizes the backhaul network throughput without considering the access network requirements.} Note that $ \mathcal{P}_\mathrm{UB} $ has $ N_v = L J^\mathrm{SBS} + 2 L N_\mathrm{tx}^\mathrm{MBS} $ variables, $ N_l = L + LBJ^\mathrm{SBS} $ linear constraints and $ N_c = 1 + L B J^\mathrm{SBS} $ convex constraints. Thus, its complexity is $ \mathcal{O} \left(N_s (N_v)^3 (N_l + N_c) \right) $, where $ N_s $ represents the total number of evaluations needed by the MIP solver.

% Remark
\textit{\textsc{Remark:} $ \mathcal{P}_\mathrm{UB} $ can be interpreted as joint multigroup multicast beamforming and rate selection, which has not been investigated before. A similar problem was studied in \cite{christopoulos2015:multicast-multigroup-precoding-scheduling-satellite-communications} but with continuous rates. Although we do not investigate this new problem alone but in conjunction with the additional access network constraints, we believe it is important to highlight its novelty as it represents the discrete counterpart of the aforementioned problem thus filling a gap in the existing literature and opening new avenues of research. }

\noindent {\textbf{{Lower Bound (\texttt{LB})}:}} The lower bound is based on the analysis of $ \mathcal{P}_0 $. From constraint $ \mathrm{{C}_{13}} $, a number of $ U_\mathrm{served} $ UEs per cluster needs to be served. In the worst case, these UEs are allocated the lowest rate possible, which based on Table \ref{table_SBS_rates_sinr}, corresponds to $ R^\mathrm{UE}_1 = 0.2344$ bps/Hz. With $ L $ clusters, the minimum sum-rate at the access network is defined as $ R_\mathrm{sum-min}^\mathrm{access} = R^\mathrm{UE}_1 \cdot W_\mathrm{BW}^\mathrm{access} \cdot U_\mathrm{served} \cdot L $ bps. We underline that this bound corresponds to the worst possible case in which the UEs are minimally served while still satisfying the system constraints.

% Section 7
\section{\texttt{RnP-SOCP-1}: Proposed SOCP Formulation} \label{section_RNP_SOCP_1}

\emph{This formulation is derived from problem $ \mathcal{P}_0 $}. We propose a relax-and-penalize SOCP algorithm denoted by \texttt{RnP-SOCP-1}, which iteratively optimizes a SOCP. To cope with the integrality constraints $ \mathrm{C_1} $, $ \mathrm{C_6} $, $ \mathrm{C_{11}} $, we replace them with the intersection of two continuous sets \cite{che2014:joint-optimization-cooperative-beamforming-relay-assignment-multi-user-networks}, as described in \emph{Proposition 7}.
% Proposition 7
\begin{proposition}
	The constraints $ \mathrm{C_1} $, $ \mathrm{C_6} $, $ \mathrm{C_{11}} $ can be equivalently expressed as,
	\begin{align}
	  & \mathrm{C_1} =
	  \begin{cases}
	  	\mathrm{X_1}: 0 \leq \alpha_{u,j} \leq 1, \\
	   	\mathrm{Z_1}: \sum_{l,u,j}  \alpha_{u,j} - \alpha_{u,j}^2 \leq 0,	 				 
	  \end{cases} \nonumber
	  \\
	  & \mathrm{C_6} = 
	  \begin{cases}
	  	\mathrm{X_2}: 0 \leq \kappa_{b,u} \leq 1, \\
	   	\mathrm{Z_2}:\sum_{l,b,u} \kappa_{b,u} - \kappa_{b,u}^2 \leq 0,	 				 
	  \end{cases} \nonumber
	  \\
	  & \mathrm{C_{11}} = 
	  \begin{cases}
	  	\mathrm{X_3}: 0 \leq \beta_{l,j} \leq 1, \\
	   	\mathrm{Z_3}: \sum_{l,j} \beta_{l,j} - \beta_{l,j}^2 \leq 0. 	 				 
	  \end{cases} \nonumber
	\end{align}
	
	% Proof 7
	\noindent \textit{Proof: It is straightforward to see that $ \mathrm{X_1} $ and $ \mathrm{Z_1} $ intersect only at $  \left\lbrace 0, 1\right\rbrace $. Thus, we omit further details.}
\end{proposition}

Notice that constraints $ \mathrm{X_1} - \mathrm{X_3} $ are convex whereas $ \mathrm{Z_1} - \mathrm{Z_3} $ are nonconvex. Considering \emph{Proposition 7}, we define

\begin{align} \label{P_1}
	% Objective
	{\mathcal{P}}_1: \max_{
			\substack{
						\boldsymbol{\Theta}
				 }
  } 
	~~ R_\mathrm{w-sum}^\mathrm{access} \left( \boldsymbol{\alpha} \right) ~~ \mathrm{s.t.} ~~ \underbrace{\boldsymbol{\Theta} \in \mathscr{D}}_{\text{{convex}}}, \underbrace{\mathrm{Z_1} - \mathrm{Z_3}}_{\text{{nonconvex}}} \nonumber
\end{align}

which is equivalent to $ \mathcal{P}_0 $. Here, $ \boldsymbol{\Theta} = \left( \mathbf{M}, \mathbf{W}, \mathbf{p}, \boldsymbol{\alpha}, \boldsymbol{\beta}, \boldsymbol{\kappa} \right) $ groups all the optimization variables and $ \mathscr{D} $ denotes the feasible set spanned by the convex constraints $ \mathrm{X_1} - \mathrm{X_3}, \mathrm{C_2} - \mathrm{C_3}, \mathrm{C_7} - \mathrm{C_{10}}, \mathrm{C_{12}} - \mathrm{C_{14}}, \mathrm{C_{17}} - \mathrm{C_{20}}, \mathrm{C_{23}} - \mathrm{C_{27}} $. Although $ {\mathcal{P}}_1 $ is a nonconvex MINLP, its nonconvexity is only due to simple polynomial constraints $ \mathrm{Z_1} - \mathrm{Z_3} $, which belong to the class of difference of convex (DC) functions. 

Since $ {\mathcal{P}}_1 $ is challenging to solve optimally, we aim to obtain a locally optimal solution. To find a solution for $ {\mathcal{P}}_1 $, we devise an algorithm based on the minorization-maximization (MM) principle. To cope with $ \mathrm{Z_1} - \mathrm{Z_3} $, we include them as penalty terms in the objective function \cite{nocedal2008:numerical-optimization}. Thus, we define $ \widetilde{\mathcal{P}}_1 $ in (\ref{P_1_penalty})
\begin{figure*}[!t]
\normalsize
% Problem P_1_penalty
\begin{equation} \label{P_1_penalty}
	% Objective
	\widetilde{\mathcal{P}}_1: \max_{
			\substack{
						\boldsymbol{\Theta} \in \mathscr{D}
				 }
  } 
	~~ R \left( \boldsymbol{\alpha}, \boldsymbol{\beta}, \boldsymbol{\kappa} \right) \triangleq R_\mathrm{w-sum}^\mathrm{access} \left( \boldsymbol{\alpha} \right) \underbrace{ - \lambda_\alpha f_\alpha \left( \boldsymbol{\alpha} \right) - \lambda_\beta f_\beta \left( \boldsymbol{\beta} \right) - \lambda_\kappa f_\kappa \left( \boldsymbol{\kappa} \right) }_\text{nonconvex DC functions}
\end{equation}
% Functions
\begin{align} \nonumber
	& f_\alpha \left( \boldsymbol{\alpha} \right) \triangleq p_\alpha \left( \boldsymbol{\alpha} \right) + q_\alpha \left( \boldsymbol{\alpha} \right), & & p_\alpha \left( \boldsymbol{\alpha} \right) \triangleq \sum_{l \in \mathcal{L}} \sum_{u \in \mathcal{U}_l} \sum_{j \in \mathcal{J}^{\mathrm{UE}}} \alpha_{u,j}, & & q_\alpha \left( \boldsymbol{\alpha} \right) \triangleq - \sum_{l \in \mathcal{L}} \sum_{u \in \mathcal{U}_l} \sum_{j \in \mathcal{J}^{\mathrm{UE}}} \alpha_{u,j}^2, \nonumber \\
	& f_\beta \left( \boldsymbol{\beta} \right) \triangleq p_\beta \left( \boldsymbol{\beta} \right) + q_\beta \left( \boldsymbol{\beta} \right), & & p_\beta \left( \boldsymbol{\beta} \right) \triangleq \sum_{l \in \mathcal{L}} \sum_{j \in \mathcal{J}^\mathrm{SBS}} \beta_{l,j}, & & q_\beta \left( \boldsymbol{\beta} \right) \triangleq - \sum_{l \in \mathcal{L}} \sum_{j \in \mathcal{J}^\mathrm{SBS}} \beta_{l,j}^2, \nonumber \\
	& f_\kappa \left( \boldsymbol{\kappa} \right) \triangleq p_\kappa \left( \boldsymbol{\kappa} \right) + q_\kappa \left( \boldsymbol{\kappa} \right), & & p_\kappa \left( \boldsymbol{\kappa} \right) \triangleq \sum_{l \in \mathcal{L}} \sum_{b \in \mathcal{B}_l} \sum_{u \in \mathcal{U}_l} \kappa_{b,u}, & & q_\kappa \left( \boldsymbol{\kappa} \right) \triangleq - \sum_{l \in \mathcal{L}} \sum_{b \in \mathcal{B}_l} \sum_{u \in \mathcal{U}_l} \kappa_{b,u}^2. \nonumber
\end{align}
\hrulefill
% Problem P_1_penalty_iterative
\begin{equation} \label{P_1_penalty_iterative}
	% Objective
	\widetilde{\mathcal{P}}_1^{(t)}: \max_{
			\substack{
						\boldsymbol{\Theta} \in \mathscr{D}
				 }
  } 
	~~ \tilde{R}^{(t)} \left( \boldsymbol{\alpha}, \boldsymbol{\beta}, \boldsymbol{\kappa} \right) \triangleq R_\mathrm{w-sum}^\mathrm{access} \left( \boldsymbol{\alpha} \right) - \lambda_\alpha \tilde{f}_\alpha^{(t)} \left( \boldsymbol{\alpha} \right) - \lambda_\beta \tilde{f}_\beta^{(t)} \left( \boldsymbol{\beta} \right) - \lambda_\kappa \tilde{f}_\kappa^{(t)} \left( \boldsymbol{\kappa} \right)
\end{equation}
% Functions
\begin{align} \nonumber
	\tilde{f}_\alpha^{(t)} \left( \boldsymbol{\alpha} \right) \triangleq p_\alpha \left( \boldsymbol{\alpha} \right) + \tilde{q}_\alpha^{(t)} \left( \boldsymbol{\alpha} \right), 
	~~~~~~~~~~ \tilde{f}_\beta^{(t)} \left( \boldsymbol{\beta} \right) \triangleq p_\beta \left( \boldsymbol{\beta} \right) + \tilde{q}_\beta^{(t)} \left( \boldsymbol{\beta} \right), 
	~~~~~~~~~~ \tilde{f}_\kappa^{(t)} \left( \boldsymbol{\kappa} \right) \triangleq p_\kappa \left( \boldsymbol{\kappa} \right) + \tilde{q}_\kappa^{(t)} \left( \boldsymbol{\kappa} \right).
\end{align}

\hrulefill
\vspace*{4pt}
\end{figure*}
where $ \lambda_\alpha \geq 0 $, $ \lambda_\beta \geq 0 $, $ \lambda_\kappa \geq 0 $. Whenever $ \boldsymbol{\alpha} $, $ \boldsymbol{\beta} $, $ \boldsymbol{\kappa} $ are not binary, the functions $ f_\alpha \left( \boldsymbol{\alpha} \right) $, $f_\beta \left( \boldsymbol{\beta} \right) $, $ f_\kappa \left( \boldsymbol{\kappa} \right) $ are positive. By including them in the objective, they can be used as a measure of the degree of satisfaction of the binary constraints, with $ \lambda_\alpha $, $ \lambda_\beta $, $ \lambda_\kappa $ representing penalty factors. Problems $ \mathcal{P}_1 $ and $ \widetilde{\mathcal{P}}_1 $ are related in the following sense. If \emph{Proposition 8} is satisfied, $ \mathcal{P}_1 $ and $ \widetilde{\mathcal{P}}_1 $ become equivalent \cite{phan2012:nonsmooth-optimization-efficient-beamforming-cognitive-radio-multicasting, nocedal2008:numerical-optimization}.

%, and $ f_\alpha \left( \boldsymbol{\alpha} \right) $, $f_\beta \left( \boldsymbol{\beta} \right) $, $ f_\kappa \left( \boldsymbol{\kappa} \right) $ are defined as,

% Proposition 8
\begin{proposition} 
	The optimization problems $ \mathcal{P}_1 $ and $ \widetilde{\mathcal{P}}_1 $ are equivalent for sufficiently large values of $ \lambda_\alpha $, $ \lambda_\beta $, $ \lambda_\kappa $, in which case both problems attain the same optimal value and solution.
	
	% Proof 8
	\noindent \textit{Proof: See Appendix \ref{appendix_proposition_8}.}
\end{proposition}

To solve $ \widetilde{\mathcal{P}}_1 $, the complication is in the objective since $ f_\alpha \left( \boldsymbol{\alpha} \right) $, $ f_\beta \left( \boldsymbol{\beta} \right) $, $ f_\kappa \left( \boldsymbol{\kappa} \right) $ are nonconvex DC functions. Thus, we apply first-order approximations to $ q_\alpha \left( \boldsymbol{\alpha} \right) $, $ q_\beta \left( \boldsymbol{\beta} \right) $, $ q_\kappa \left( \boldsymbol{\kappa} \right) $, and define
% Functions
\begin{align} 
	& \tilde{q}_\alpha^{(t)} \left( \boldsymbol{\alpha} \right) \triangleq q_\alpha \left( \boldsymbol{\alpha}^{(t-1)} \right) + \nabla_{\boldsymbol{\alpha}} q_\alpha^T \left( \boldsymbol{\alpha}^{(t-1)} \right) \left( \boldsymbol{\alpha} - \boldsymbol{\alpha}^{(t-1)} \right), \nonumber 
	\\
	& \tilde{q}_\beta^{(t)} \left( \boldsymbol{\beta} \right) \triangleq q_\beta \left( \boldsymbol{\beta}^{(t-1)} \right) + \nabla_{\boldsymbol{\beta}} q_\beta^T \left( \boldsymbol{\beta}^{(t-1)} \right) \left( \boldsymbol{\beta} - \boldsymbol{\beta}^{(t-1)} \right), \nonumber 
	\\
	& \tilde{q}_\kappa^{(t)} \left( \boldsymbol{\kappa} \right) \triangleq q_\kappa \left( \boldsymbol{\kappa}^{(t-1)} \right) + \nabla_{\boldsymbol{\kappa}} q_\kappa^T \left( \boldsymbol{\kappa}^{(t-1)} \right) \left( \boldsymbol{\kappa} - \boldsymbol{\kappa}^{(t-1)} \right), \nonumber
\end{align}
where $ \tilde{q}_\alpha^{(t)} \left( \boldsymbol{\alpha} \right) \geq q_\alpha \left( \boldsymbol{\alpha} \right) $, $ \tilde{q}_\beta^{(t)} \left( \boldsymbol{\beta} \right) \geq q_\beta\left( \boldsymbol{\beta} \right) $, $ \tilde{q}_\kappa^{(t)} \left( \boldsymbol{\kappa} \right) \geq q_\kappa \left( \boldsymbol{\kappa} \right) $ are outer linear approximations for $ q_\alpha \left( \boldsymbol{\alpha} \right) $, $ q_\beta \left( \boldsymbol{\beta} \right) $, $ q_\kappa \left( \boldsymbol{\kappa} \right) $, respectively. 

Here, $ \boldsymbol{\alpha}^{\left( t-1 \right) } $, $ \boldsymbol{\beta}^{\left( t-1 \right) } $, $ \boldsymbol{\kappa}^{\left( t-1 \right) } $ denote a feasible solution (i.e. reference point for linearization) whereas $ \nabla_{\mathbf{x}} $ represents the derivative with respect to variable $ \mathbf{x} $. Using the MM principle and constructing a sequence of surrogate functions $ \tilde{q}_\alpha^{(t)} \left( \boldsymbol{\alpha} \right) $, $ \tilde{q}_\beta^{(t)} \left( \boldsymbol{\beta} \right) $, $ \tilde{q}_\kappa^{(t)} \left( \boldsymbol{\kappa} \right) $ at every iteration $ t $, we solve problem $ \widetilde{\mathcal{P}}_1^{(t)} $ defined in (\ref{P_1_penalty_iterative}), which is a SOCP where $ \tilde{f}_\alpha^{(t)} \left( \boldsymbol{\alpha} \right) \geq f_\alpha \left( \boldsymbol{\alpha} \right) $, $ \tilde{f}_\beta^{(t)} \left( \boldsymbol{\beta} \right) \geq f_\beta \left( \boldsymbol{\beta} \right) $, $ \tilde{f}_\kappa^{(t)} \left( \boldsymbol{\kappa} \right) \geq f_\kappa \left( \boldsymbol{\kappa} \right) $. In particular, problem $ \widetilde{\mathcal{P}}_1^{(t)} $ is convex and can be solved using interior-point methods. By solving $ \widetilde{\mathcal{P}}_1^{(t)} $ iteratively, we show in \emph{Proposition 9} and \emph{Proposition 10}, that $ \widetilde{\mathcal{P}}_1^{(t)} $ is a global lower bound of $ \widetilde{\mathcal{P}}_1 $ and the obtained solution is a KKT point.

% Proposition 9
\begin{proposition} 
	Problem $ \widetilde{\mathcal{P}}_1^{(t)} $ is a global lower bound for $ \widetilde{\mathcal{P}}_1 $ since $ \tilde{R}^{(t)} \left( \boldsymbol{\alpha}, \boldsymbol{\beta}, \boldsymbol{\kappa} \right) \leq R \left( \boldsymbol{\alpha}, \boldsymbol{\beta}, \boldsymbol{\kappa} \right) $.
	
	% Proof 9
	\noindent \textit{Proof: See Appendix \ref{appendix_proposition_9}.}
\end{proposition}

% Proposition 10
\begin{proposition} 
	Starting from a feasible point $ \boldsymbol{\Theta}^{(0)} = \left( \cdot, \cdot, \cdot, \boldsymbol{\alpha}^{(0)}, \boldsymbol{\beta}^{(0)}, \boldsymbol{\kappa}^{(0)} \right) $, the sequence of solutions $ \boldsymbol{\Theta}^{(t)} = \left( \mathbf{M}^{(t)}, \mathbf{W}^{(t)}, \mathbf{p}^{(t)}, \boldsymbol{\alpha}^{(t)}, \boldsymbol{\beta}^{(t)}, \boldsymbol{\kappa}^{(t)} \right) $, for $ t \geq 1 $, obtained by iteratively solving $ \widetilde{\mathcal{P}}_1^{(t)} $ constitutes a sequence of enhanced points for $ \widetilde{\mathcal{P}}_1 $, which converges to a KKT point.
	
	% Proof 10
	\noindent \textit{Proof: See Appendix \ref{appendix_proposition_10}.}
\end{proposition}

To solve $ \widetilde{\mathcal{P}}^{(t)}_1 $, a feasible point $ \boldsymbol{\Theta}^{(0)} $ is needed to guarantee convergence as explained in \emph{Proposition 10}. We generate random initial feasible points and test them for feasibility, as described in \cite{bandi2020:joint-solution-scheduling-precoding-multiuser-miso}. We use the best of these points as the initial $ \boldsymbol{\Theta}^{(0)} $, and iteratively solve $ \widetilde{\mathcal{P}}^{(t)}_1 $ as shown in Algorithm \ref{algorithm_1}. 

% Algorithm
% Algorithm 1
\setlength{\textfloatsep}{10pt plus 1.0pt minus 2.0pt}
\captionsetup[algorithm]{labelformat=empty}
\begin{algorithm}[!h]
		\footnotesize
		\centering
		\begin{tabular}{m{1.2cm} m{11cm}}
			\emph{Step 1:} & Define $ N_\mathrm{iter} $, $ \delta $, $ \lambda_{\alpha}$, $ \lambda_{\beta} $, $ \lambda_{\kappa} $. \\
			\emph{Step 2:} & Find an initial point $ \boldsymbol{\Theta}^{(0)} = \left( \cdot, \cdot, \cdot, \boldsymbol{\alpha}^{(0)}, \boldsymbol{\beta}^{(0)}, \boldsymbol{\kappa}^{(0)} \right) $ \\
			  			   & using $ \left\lbrace 0, 1 \right\rbrace $ values. \\
			\emph{Step 3:} & Initialize $ t = 1 $. \\
			\emph{Step 4:} & Solve $ \mathcal{P}_1^{(t)} $ using $ \boldsymbol{\Theta}^{(t-1)} $. \\
			\emph{Step 5:} & Assign $ \boldsymbol{\Theta}^{(t)} \leftarrow \boldsymbol{\Theta}^{(t-1)} $. \\
			\emph{Step 6:} & Update the iteration index $ t $ by one, i.e. $ t = t + 1 $.  \\
			\emph{Step 7:} & Verify if the stop criterion is attained. Otherwise, \\
						   & return to \emph{Step 4}.  \\
		\end{tabular}
	\caption{Optimization of $ \mathcal{P}_1 $}
	\label{algorithm_1}
\end{algorithm}

We stop the iterative process when a criterion has been met, i.e., $ t = N_\mathrm{iter} $ or $ \tilde{R}^{(t)} \left( \boldsymbol{\alpha}, \boldsymbol{\beta}, \boldsymbol{\kappa} \right) - \tilde{R}^{(t-1)} \left( \boldsymbol{\alpha}, \boldsymbol{\beta}, \boldsymbol{\kappa} \right) \leq \delta $. The computational complexity of $ \widetilde{\mathcal{P}}^{(t)}_1 $ is similar to that of one evaluation of $ \mathcal{P}_0 $. In particular, $ N_v = 2 L N_\mathrm{tx}^\mathrm{MBS} + 2 L B U N_\mathrm{tx}^\mathrm{SBS} + 2 L B U + L J^\mathrm{SBS} + L U J^\mathrm{UE} $ variables, $ N_l = 3L + 2LU + 3LB + 2LJ^\mathrm{SBS} + 4LBU + 5LUJ^\mathrm{UE} + LBJ^\mathrm{SBS} $ linear constraints and $ N_c = 1 + LBU + LUJ^\mathrm{UE} + LBJ^\mathrm{SBS} $ convex constraints. Therefore, the complexity is $ \mathcal{O} \left( N_\mathrm{iter} (N_v)^3 (N_l + N_c) \right) $, where $ N_\mathrm{iter} $ is the number of iterations.

% Section 8
\section{\texttt{RnP-SOCP-2}: Proposed SOCP Formulation} \label{section_RNP_SOCP_2}

\emph{This formulation is derived from problem $ \mathcal{P}_1 $}. We propose an alternative relax-and-penalize SOCP algorithm, denoted by \texttt{RnP-SOCP-2}, whose main characteristic is the reduced number of optimization variables compared to \texttt{RnP-SOCP-1}, thus allowing to obtain solutions faster. To decrease the large number of optimization variables in $ \mathcal{P}_1 $, (essentially dominated by the number of antennas at the MBS and SBSs) we adopt a simpler approach in which instead of optimizing high-dimensional beamforming vectors, we only optimize their gains. 

In particular, we define the variables $ v_{b,u} $ and $ t_l $ as the gains (i.e., amplitude and phase) of predefined unicast (i.e., access) and multicast (i.e., backhaul) beamforming vectors $ \widehat{\mathbf{w}}_{b,u} $ and $ \widehat{\mathbf{m}}_{l} $, respectively, such that $ \mathbf{m}_{l} = {t_l} \widehat{\mathbf{m}}_{l} $, $ \mathbf{w}_{b,u} = {v_{b,u}} \widehat{\mathbf{w}}_{b,u} $, $ \left\| \widehat{\mathbf{m}}_{l} \right\|^2_2 = 1 $, $ \left\| \widehat{\mathbf{w}}_{b,u} \right\|^2_2 = 1 $. We design the unit-norm unicast beamforming vectors $ \widehat{\mathbf{w}}_{b,u} $ using the zero-forcing (ZF) criterion. On the other hand, the unit-norm multicast beamforming vectors $ \widehat{\mathbf{m}}_{l} $ are obtained experimentally upon evaluating the upper bound $ \mathcal{P}_\mathrm{UB} $ for multiple realizations with varying degrees of shadowing and small-scale fading, and then taking the average of all these beamforming vectors. This procedure allows us to obtain a fair estimation of the multicast beamforming vectors because the SBSs are stationary and therefore the MBS-SBS channels geometry do not change substantially. Thus, the constraints that are affected by $ \mathbf{m}_{l} = {t_l} \widehat{\mathbf{m}}_{l} $, $ \mathbf{w}_{b,u} = {v_{b,u}} \widehat{\mathbf{w}}_{b,u} $ are $ \mathrm{C_{3}} $, $ \mathrm{C_{20}} $,$ \mathrm{C_{23}} - \mathrm{C_{27}} $ which are redefined at the top of this page, where $ \mathbf{S}_b $ is a block diagonal matrix containing the combinations of beamformers $ \widehat{\mathbf{w}}_{b,u} $ and channels for UE $ u $, $ c_{b,u} = \mathbf{h}_{b,u}^H \widehat{\mathbf{w}}_{b,u}  $, $ \mathbf{R}_b = \mathrm{diag} \left( \mathbf{g}_b^H \widehat{\mathbf{M}} \right) $, $ r_{b,l} = \mathsf{Re} \left\lbrace \mathbf{g}_b^H \widehat{\mathbf{m}}_l \right\rbrace $. 
\begin{figure*}[!t]
% ensure that we have normalsize text
\normalsize

% Constraints L1-L7
\begin{align} \nonumber
  	% Constraint L1
  	(\mathrm{C_{3}}) &~\mathrm{L_{1}}: \sum_{l \in \mathcal{L}} t_l^2 \leq P_\mathrm{tx}^\mathrm{MBS}, \nonumber 
  	\\
  	% Constraint L2
  	(\mathrm{C_{20}}) &~\mathrm{L_{2}}: \left\| \left[ 2 \widehat{\mathbf{w}}_{b,u}^H v_{b,u},  \kappa_{b,u} - p_{b,u} \right] \right\|_2 \leq \kappa_{b,u} + p_{b,u}, \forall l \in \mathcal{L}, b \in \mathcal{B}_l, u \in \mathcal{U}_l, \nonumber
  	\\
  	% Constraint L3
   	(\mathrm{C_{23}}) &~\mathrm{L_{3}}: \left\| \left[ \mathbf{S}_b \mathbf{v}, \sigma_\mathrm{UE} \right] \right\|_2 \leq \sqrt{1 + {\Gamma_j^\mathrm{UE} }^{-1} } \mathsf{Re}~\Big\{ \sum_{b \in \mathcal{B}_l} c_{b,u} v_{b,u} \Big\} + \left(1 - \alpha_{u,j} \right) Q_u, \forall l \in \mathcal{L}, u \in \mathcal{U}_l, j \in \mathcal{J}^{\mathrm{UE}}, \nonumber 
   	\\	
   	% Constraint L4
   	(\mathrm{C_{24}}) &~\mathrm{L_{4}}: \mathsf{Re}~\Big\{ \sum_{b \in \mathcal{B}_l} c_{b,u} v_{b,u} \Big\} \geq \alpha_{u,j} \sqrt{\Gamma_j^\mathrm{UE}} \sigma_\mathrm{UE}, \forall l \in \mathcal{L}, u \in \mathcal{U}_l, j \in \mathcal{J}^{\mathrm{UE}}, \nonumber 
   	\\
   	% Constraint L5
   	(\mathrm{C_{25}}) &~\mathrm{L_{5}}: \mathsf{Im}~\Big\{ \sum_{b \in \mathcal{B}_l} c_{b,u} v_{b,u} \Big\} = 0, \forall l \in \mathcal{L}, u \in \mathcal{U}_l, j \in \mathcal{J}^{\mathrm{UE}}. \nonumber 
	\\
	% Constraint L6
   	(\mathrm{C_{26}}) &~\mathrm{L_{6}}: \left\| \left[ \mathbf{R}_b \mathbf{t}, \sigma_\mathrm{SBS} \right] \right\|_2 \leq \sqrt{ 1 + {\Gamma_j^\mathrm{SBS} }^{-1} } r_{b,l} t_l + \left( 1 - \beta_{l,j} \right) Q_b, \forall l \in \mathcal{L}, b \in \mathcal{B}_l, j \in \mathcal{J}^{\mathrm{SBS}}, \nonumber 
   	\\
   	% Constraint L7
   	(\mathrm{C_{27}}) &~\mathrm{L_{7}}: r_{b,l} t_l \geq \beta_{l,j} \sqrt{\Gamma_j^\mathrm{SBS}} \sigma_\mathrm{SBS}, \forall l \in \mathcal{L}, b \in \mathcal{B}_l, j \in \mathcal{J}^{\mathrm{SBS}}, \nonumber 
\end{align}

\hrulefill
\vspace*{4pt}
\end{figure*}

% Table IV
% Table: Simulation settings
\begin{table*}[t!]
	\setlength\tabcolsep{3.0pt} % default value: 6pt
	\renewcommand{\arraystretch}{1.1}% Wider
	\scriptsize
	\centering
	\caption{Simulation settings}
	\begin{tabular}{|c||>{\columncolor[gray]{0.9}}c c >{\columncolor[gray]{0.9}}c c >{\columncolor[gray]{0.9}}c c|>{\columncolor[gray]{0.9}}c c >{\columncolor[gray]{0.9}}c c >{\columncolor[gray]{0.9}}c|}
		% Header
		\hline
		% Rows
		\multirow{2}{*}{\bf \scriptsize Scenario} & \multicolumn{6}{c|}{\bf \scriptsize Backhaul network} & \multicolumn{5}{c|}{\bf \scriptsize Access network} \\ 
		%\cline{2-12}
		& $ N^{\mathrm{MBS}}_\mathrm{tx} $ & $ P^{\mathrm{MBS}}_\mathrm{tx} $ [dBm] & $ L $ & $ B $ & $ B_\mathrm{total} $ & $ \chi_\mathrm{backhaul} $ & $ P^{\mathrm{SBS}}_\mathrm{tx} $ [dBm] & $ U $ & $ U_\mathrm{total} $ & $ U_\mathrm{served} $ & $ \chi_\mathrm{access} $  \\
		\hline
		S\textsubscript{1} & $ 64 $ & $ 9, 12, \dots, 27 $ & $ 2 $ & $ 3 $ & $ 6 $ & $ 0 $ & $ 6, 10, 14 $ & $ 6 $ & $ 12 $ & $ 3 $ & $ 0 $ \\
		\hline
		\multirow{2}{*}{S\textsubscript{2}} & $ 16, 32, 48, 64 $ & $ 15, 18, \dots, 36 $ & $ 1, 2, \dots, 6 $ & $ 3 $ & $ 3, 6, \dots, 18 $ & $ 0 $ & $ - $ & $ - $ & $ - $ & $ - $ & $ - $ \\
											& $ 64 $ & $ 15, 18, \dots, 36 $ & $ 5 $ & $ 1, 2, \dots, 6 $ & $ 5, 10, \dots, 30 $ & $ 0 $ & $ - $ & $ - $ & $ - $ & $ - $ & $ - $ \\
		\hline
		S\textsubscript{3} & $ 64 $ & $ 15, 18, \dots, 36 $ & $ 5 $ & $ 3 $ & $ 15 $ & $ 0 $ & $ 0, 2, \dots, 14 $ & $ 20 $ & $ 100 $ & $ 4 $ & $ 0 $ \\
		                   %& $ 64 $ & $ 36 $ & $ 5 $ & $ 3 $ & $ 0 $ & $ 0, 2, \dots, 14 $ & $ 20 $ & $ 4 $ & $ 0 $ \\
		\hline
		S\textsubscript{4} & $ 64 $ & $ 18, 27, 36 $ & $  2, 3, \dots, 6 $ & $ 3 $ & $ 6, 9, \dots, 18 $ & $ 0 $ & $ 14 $ & $ 20 $ & $ 100 $ & $ 4 $ & $ 0 $ \\
		\hline
		%S\textsubscript{4} & $ 64 $ & $ 36 $ & $ 5 $ & $ 3 $ & $ 0 $ & $ 8 $ & $ 20 $ & $ 2, 3, \dots, 8 $ & $ 0 $ \\
		%\hline
		S\textsubscript{5} & $ 64 $ & $ 36 $ & $ 5 $ & $ 3 $ & $ 15 $ & $ [0, 1] $ & $ 14 $ & $ 20 $ & $ 100 $ & $ 4 $ & $ [0, 1] $ \\
		\hline
		S\textsubscript{6} & $ 64 $ & $ 36 $ & $ 5 $ & $ 3 $ & $ 15 $ & $ 0 $ & $ 14 $ & $ 20 $ & $ 100 $ & $ 4 $ (slotted) & $ 0 $ \\
		\hline
	\end{tabular}
	\label{table_simulation_settings}
\end{table*}

After applying these changes, we define,
% Problem P_1
\begin{align}
	% Objective
	\mathcal{P}_2: \max_{
				\boldsymbol{\Theta}
  	} 
	~~ R_\mathrm{w-sum}^\mathrm{access} \left( \boldsymbol{\alpha} \right) ~~ \mathrm{s.t.} ~~ \underbrace{\boldsymbol{\Theta} \in \mathscr{D}}_{\text{{convex}}}, \underbrace{\mathrm{Z_1} - \mathrm{Z_3}}_{\text{{nonconvex}}}, \nonumber
\end{align}
where $ \boldsymbol{\Theta} = \left( \mathbf{t}, \mathbf{v}, \mathbf{p}, \boldsymbol{\alpha}, \boldsymbol{\beta}, \boldsymbol{\kappa} \right) $ with $ \mathscr{D} $ denoting the feasible set spanned by the constraints $ \mathrm{L_1} - \mathrm{L_7},  \mathrm{X_1} - \mathrm{X_3}, \mathrm{C_2}, \mathrm{C_7} - \mathrm{C_{10}}, \mathrm{C_{12}} - \mathrm{C_{14}}, \mathrm{C_{17}} - \mathrm{C_{19}} $. In a similar manner as with $ \widetilde{\mathcal{P}}_1 $, we define $ \widetilde{\mathcal{P}}_2 $, and thereupon its linearized version as $ \widetilde{\mathcal{P}}_2^{(t)} $, which can be solved via Algorithm \ref{algorithm_1}. $ \widetilde{\mathcal{P}}_2^{(t)} $ is a SOCP program with $ N_v = 2L + 4LBU + LJ^\mathrm{SBS} + L U J^\mathrm{UE} $ decision variables, which roughly represent half of that used in \texttt{BnC-MISOCP} and \texttt{RnP-SOCP-1} (for the evaluated settings). In addition, $ \widetilde{\mathcal{P}}_2^{(t)} $ has $ N_l = 3L + 2LU + 3LB + 2LBU + 3LUJ^\mathrm{UE} + LBJ^\mathrm{SBS} $ linear constraints and $ N_c = 1 + LBU + LUJ^\mathrm{UE} + LBJ^\mathrm{SBS} $ convex constraints. Thus, the complexity of \texttt{RnP-SOCP-2} is $ \mathcal{O} \left( N_\mathrm{iter} (N_v)^3 (N_l + N_c) \right) $, with $ N_\mathrm{iter} $ denoting the number of iterations. Further, we note that \texttt{RnP-SOCP-2} exhibits reduced complexity compared to \texttt{RnP-SOCP-1}.

% Section 9: Simulation Results
\section{Simulation Results} \label{section_simulation_results}

We evaluate the performance of \textsf{RadiOrchestra} in different scenarios with varying conditions. Throughout all simulations, we consider the following default parameters, unless specified otherwise. The carrier frequency is $ f_c = 41 $ GHz (V-band in FR2) with $ W_\mathrm{BW}^\mathrm{access} = W_\mathrm{BW}^\mathrm{backhaul} = 100 $ MHz bandwidth \cite{3gpp2017:38.901}. The channel models are UMa LOS for the backhaul and UMi LOS/NLOS for the access \cite{3gpp2017:38.901}, which include path-loss, shadowing and small-scale fading. In the system, there are $ L = 5 $ clusters each having $ B = 3 $ SBSs and $ U = 20 $ UEs, thus making a total of $ B_\mathrm{total} =  15 $ SBSs and $ U_\mathrm{total} = 100 $ UEs. The MBS has a maximum transmit power of $ P^{\mathrm{MBS}}_\mathrm{tx} = 36 $ dBm and is equipped with a $ 16 \times 4 $ antenna array ($ N_\mathrm{tx}^\mathrm{MBS} = 64 $) whereas the SBSs can transmit at a maximum power of $ P^{\mathrm{SBS}}_\mathrm{tx} = 14 $ dBm and have smaller $ 4 \times 4 $ arrays ($ N_\mathrm{tx}^\mathrm{SBS} = 16 $). We assume that SBSs can support up to four UEs ($ N^\mathrm{SBS}_\mathrm{streams} = 4 $) simultaneously, and there are $ U_\mathrm{served} = 4 $ UEs served concurrently (i.e., in one slot) in each cluster. Further, all UEs have the same priority, i.e., $ \omega_u = \frac{1}{L*U} $ and $ \sum_{l \in \mathcal{L} } \sum_{u \in \mathcal{U}_l} \omega_{u} = 1 $. In Table \ref{table_simulation_settings}, we show the parameters for each scenario. The algorithms have been implemented using \textsf{CVX} and \textsf{MOSEK} on a computer with 16GB RAM and a Intel Core i7-6700 processor.

\begin{figure*}[!t]
	\centering
	% Legend
 	\begin{subfigure}[b]{\textwidth}
 		\centering
		\begin{center}
			\begin{tikzpicture}
			    \begin{axis}[%
			    hide axis,
			    xmin=10,
			    xmax=50,
			    ymin=0,
			    ymax=0.4,
			    legend style={draw=white!15!black,legend cell align=left, font=\fontsize{9}{8}\selectfont,},
			    legend columns = 5,
			    ]
			    \addlegendimage{color = blue2, mark = *, line width = 1pt, densely dotted, mark options = {fill = blue2, solid}}
			    \addlegendentry{$ \mathtt{UB} $};
			    
			    \addlegendimage{color = gray, mark = *, line width = 1pt, densely dotted, mark options = {fill = gray, solid}}
			    \addlegendentry{$ \mathtt{LB} $};
			    
			    \addlegendimage{color = dcolor6, mark = pentagon*, line width = 1pt, densely dotted, mark options = {fill = dcolor6, solid}}
			    \addlegendentry{$ \mathtt{BnC-MISOCP} $};
			    
			    \addlegendimage{color = dcolor5, mark = diamond*, line width = 1pt, densely dotted, mark options = {fill = dcolor5, solid}}
			    \addlegendentry{$ \mathtt{RnP-SOCP-1} $};
			    
			    \addlegendimage{color = dcolor4, mark = triangle*, line width = 1pt, densely dotted, mark options = {fill = dcolor4, solid}}
			    \addlegendentry{$ \mathtt{RnP-SOCP-2} $};
			    \end{axis}
			\end{tikzpicture}
		\end{center}
 	\end{subfigure}
	% Scenario S1a
 	\begin{subfigure}[b]{.32\textwidth}
 		\centering
		\begin{center}
			\resizebox{\textwidth}{!}
			{%
			\includegraphics[]{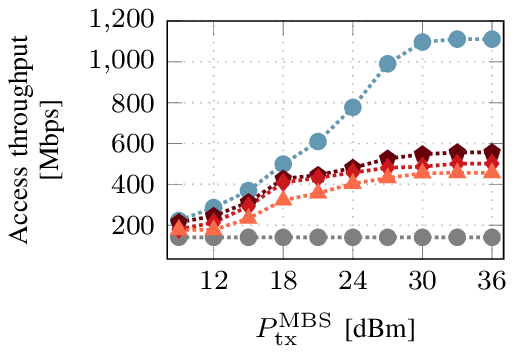}
			}
			\caption{Varying $ P^\mathrm{MBS}_\mathrm{tx} $ when $ P^\mathrm{SBS}_\mathrm{tx} = 6 $ dBm}
			\label{figure_scenario_s1a}
		\end{center}
 	\end{subfigure}
    \hfill 
	% Scenario S1b
 	\begin{subfigure}[b]{.32\textwidth}
 		\centering
		\begin{center}
			\resizebox{\textwidth}{!}
			{%
			\includegraphics[]{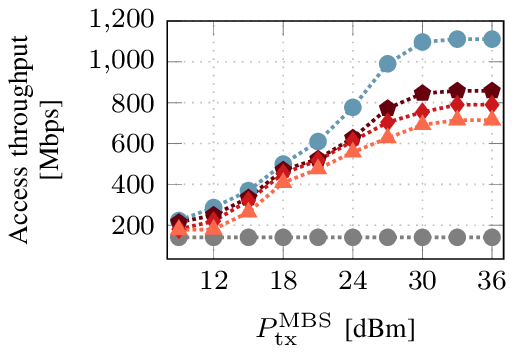}
			}
			\caption{Varying $ P^\mathrm{MBS}_\mathrm{tx} $ when $ P^\mathrm{SBS}_\mathrm{tx} = 10 $ dBm}
			\label{figure_scenario_s1b}
		\end{center}
 	\end{subfigure}
    \hfill 
	% Scenario S1c
 	\begin{subfigure}[b]{.32\textwidth}
 		\centering
		\begin{center}
			\resizebox{\textwidth}{!}
			{%
			\includegraphics[]{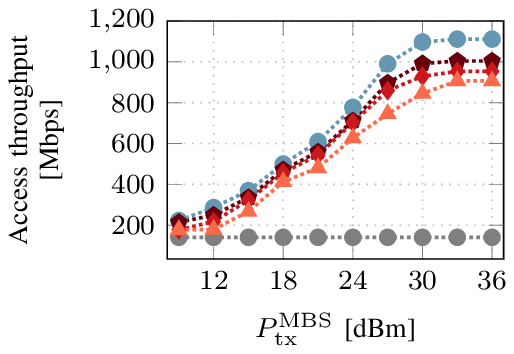}
			}
			\caption{Varying $ P^\mathrm{MBS}_\mathrm{tx} $ when $ P^\mathrm{SBS}_\mathrm{tx} = 14 $ dBm}
			\label{figure_scenario_s1c}
		\end{center}
 	\end{subfigure}
    \hfill 
	% Scenario S1d
 	\begin{subfigure}[b]{.32\textwidth}
 		\centering
		\begin{center}
			\resizebox{\textwidth}{!}
			{%
			\includegraphics[]{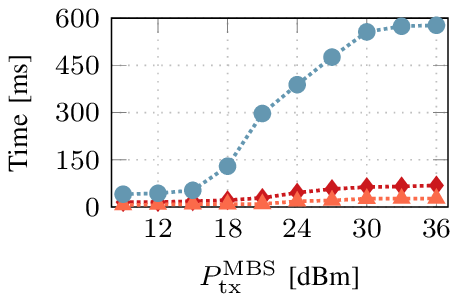}
			}
			\caption{Time complexity when $ P^\mathrm{SBS}_\mathrm{tx} = 14 $ dBm}
			\label{figure_scenario_s1d}
		\end{center}
 	\end{subfigure}
    \hfill 
	% Scenario S1e
 	\begin{subfigure}[b]{.32\textwidth}
 		\centering
		\begin{center}
			\resizebox{\textwidth}{!}
			{%
			\includegraphics[]{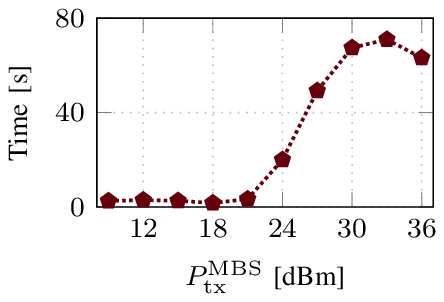}
			}
			\caption{Time complexity when $ P^\mathrm{SBS}_\mathrm{tx} = 14 $ dBm}
			\label{figure_scenario_s1e}
		\end{center}
 	\end{subfigure}
    \hfill 
	% Scenario S1f
 	\begin{subfigure}[b]{.33\textwidth}
 		\centering
		\begin{center}
			\resizebox{\textwidth}{!}
			{%
			\includegraphics[]{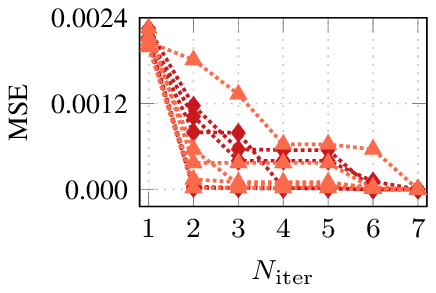}
			}
			\caption{Convergence}
			\label{figure_scenario_s1f}
		\end{center}
 	\end{subfigure}
 	\label{figure_scenario_s1}
 	\caption{Evaluation of Scenario S\textsubscript{1}. \emph{We notice the small performance gap of \texttt{RnP-SOCP-1} and \texttt{RnP-SOCP-2} with respect to \texttt{BnC-MISOCP}, which is reasonable considering that their time complexities are smaller by 3 orders of magnitude. Because \textsf{CVX} needs to parse the mathematical model into a suitable structure for \textsf{MOSEK}, the results showing time complexity consider the raw solving time while neglecting the parsing time. Besides, we note that \texttt{UB} can be used for quick benchmarking when the access throughput bottleneck is originated by the backhaul network. In addition, we note that \texttt{LB} is loose as it is agnostic to the network conditions but provides an idea of the worst-case scenario without solving any problem. It becomes valuable when evaluating cases wherein the transmit power at the MBS or SBSs are limited as in Fig. \ref{figure_scenario_s1a} because under such conditions the lowest rates will very likely be allocated.}}
\end{figure*}

% Scenario S1
\noindent {\textbf{{Scenario S\textsubscript{1}: Optimality gap and computational complexity.}} We benchmark the algorithms considering a small setting, with the purpose of obtaining an optimal solution for \texttt{BnC-MISOCP} within a reasonable amount of time and compare its performance against that of \texttt{RnP-SOCP-1} and \texttt{RnP-SOCP-2}. Fig. \ref{figure_scenario_s1a}, Fig. \ref{figure_scenario_s1b}, Fig. \ref{figure_scenario_s1c} show the access throughput with various MBS and SBSs transmit powers. In particular, \texttt{RnP-SOCP-1} and \texttt{RnP-SOCP-2} are $ 5.1\% $ and $ 9.7\% $ below \texttt{BnC-MISOCP} when $ P^\mathrm{SBS}_\mathrm{tx} = 14 $ dBm (see Fig. \ref{figure_scenario_s1c}). Also, \texttt{UB} becomes tighter with increasing $ P^\mathrm{SBS}_\mathrm{tx} $, e.g., within only $ 9.6\% $ with respect to \texttt{BnC-MISOCP} in Fig. \ref{figure_scenario_s1c}. This occurs because \texttt{UB} only considers the backhaul throughput optimization, which depends on $ P^\mathrm{MBS}_\mathrm{tx} $. Thus, as long as the bottleneck is originated in the access network (due to low transmit power at the SBSs), \texttt{UB} will not capture such limitations. With higher $ P^\mathrm{SBS}_\mathrm{tx} $, as shown in Fig. \ref{figure_scenario_s1c}, the access throughput limitation is removed and is shifted to the backhaul network, where $ P^\mathrm{MBS}_\mathrm{tx} $ is varied from a low to a high transmit power. As a result, in Fig. \ref{figure_scenario_s1c} the access throughput limitation is dominated by $ P^\mathrm{MBS}_\mathrm{tx} $, where we recognize a high degree of similarity between \texttt{UB} and \texttt{BnC-MISOCP}. Therefore, \texttt{UB} can be used as a tight bound to evaluate the performance of the system whenever the SBSs can transmit at sufficiently high power. 

On the other hand, Fig. \ref{figure_scenario_s1d} and  Fig. \ref{figure_scenario_s1e} provide the time complexities when $ P^\mathrm{SBS}_\mathrm{tx} = 14 $ (as in Fig. \ref{figure_scenario_s1c}) showing that \texttt{RnP-SOCP-1} and \texttt{RnP-SOCP-2} are roughly $ 1000 $ and $ 2000 $ times computationally faster than \texttt{BnC-MISOCP}, respectively. Similarly, the time complexity of \texttt{UB} is approximately $ 100 $ times lighter than that of \texttt{BnC-MISOCP}. This huge difference is because the complexity of \texttt{BnC-MISOCP} is combinatorial, i.e., collapsing to exhaustive search in the worst case. Although this case may not be reached in practice, \texttt{BnC-MISOCP} requires to solve multiple convex problems to prune the infeasible branches and thus abridge the search process. However, \texttt{RnP-SOCP-1} and \texttt{RnP-SOCP-2} circumvent this issue by relaxing the binary variables, penalizing them and solving the problem in the continuous domain, which explains their reduced complexity. Besides, \texttt{UB} has a small number of optimization variables compared to \texttt{BnC-MISOCP}, explaining its faster solving time. Note that the time complexities grow with increasing $ P^\mathrm{MBS}_\mathrm{tx} $ because a higher $ P^\mathrm{MBS}_\mathrm{tx} $ enables the allocation of a wider range of rates thus needing more evaluations, specially by \texttt{BnC-MISOCP} and \texttt{UB}. Further, Fig. \ref{figure_scenario_s1f} shows the convergence of \texttt{RnP-SOCP-1} and \texttt{RnP-SOCP-2} for $ 5 $ different realizations. Here, we measured the error of the binary variables with respect to their rounded versions and computed the mean squared error (MSE), showing that after $ 6 $ or $ 7 $ iterations the error converges to zero, i.e., the relaxed binary variables values become integer.

% Figure: Scenario S2
\setcounter{figure}{5}
\begin{figure*}[!t]
  \centering
  	% Scenario S2a
	\begin{subfigure}[b]{.48\textwidth}
		\begin{center}
			\resizebox{\textwidth}{!}
			{%
			\includegraphics[]{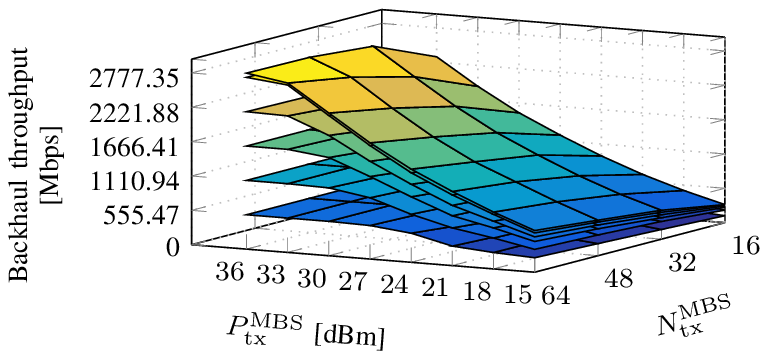}
			}
			\caption{Backhaul throughput for varying $ P_{\mathrm{tx}}^\mathrm{MBS} $,  $ N_{\mathrm{tx}}^\mathrm{MBS} $ and $ L $ when $ B = 3 $.} 
		  	\label{figure_scenario_s2a}
		\end{center}
    \end{subfigure}%
    \hfill 
	% Scenario S2b
    \begin{subfigure}[b]{.48\textwidth}
		\begin{center}
			\resizebox{\textwidth}{!}
			{%
			\includegraphics[]{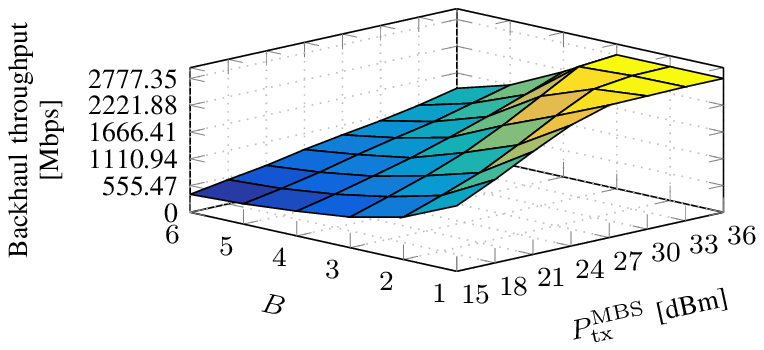}
			}
			\caption{Backhaul throughput for varying $ B $ and $ N_{\mathrm{tx}}^\mathrm{MBS} $ when $ L = 5 $.} 
			\label{figure_scenario_s2b}
		\end{center}
    \end{subfigure}
	\caption{Evaluation of Scenario S\textsubscript{2}. \emph{We note that \texttt{UB} can be used to evaluate multiple network configurations, thus providing insights of potentially optimal operations points that can be adopted in the planning phase of the network.}}
\end{figure*}

% Figure: Scenario S3
\begin{figure*}[!t]
	\centering
	% Legend
 	\begin{subfigure}[b]{\textwidth}
 		\centering
		\begin{center}
			\includegraphics[]{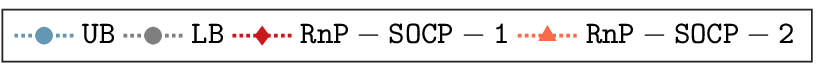}		
		\end{center}
 	\end{subfigure}
	% Scenario S3a
 	\begin{subfigure}[b]{.22\textwidth}
 		\centering
		\begin{center}
			\resizebox{\textwidth}{!}
			{%
			\includegraphics[]{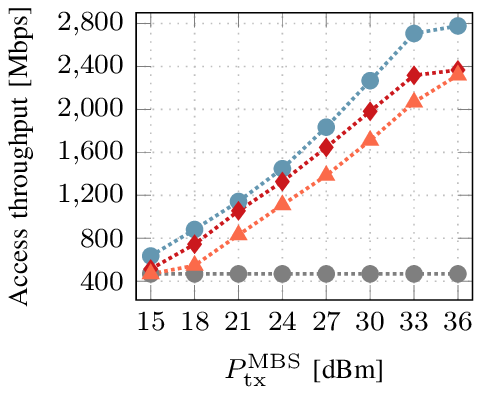}
			}
			\caption{{\tiny Varying $ P^\mathrm{MBS}_\mathrm{tx} | P^\mathrm{SBS}_\mathrm{tx} = 14 $ dBm}}
			\label{figure_scenario_s3a}
		\end{center}
 	\end{subfigure}
    \hfill 
	% Scenario S3b
 	\begin{subfigure}[b]{.22\textwidth}
 		\centering
		\begin{center}
			\resizebox{\textwidth}{!}
			{%
			\includegraphics[]{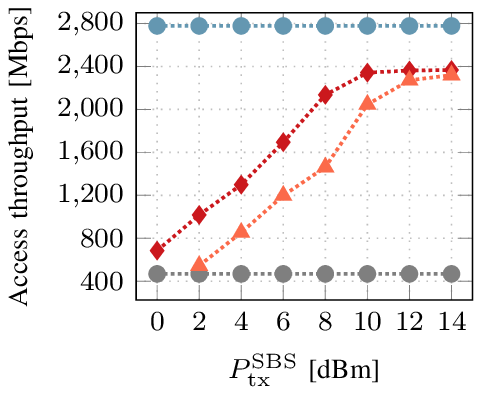}
			}
			\caption{{\tiny Varying $ P^\mathrm{SBS}_\mathrm{tx} | P^\mathrm{MBS}_\mathrm{tx} = 36 $ dBm}}
			\label{figure_scenario_s3b}
		\end{center}
 	\end{subfigure}
    \hfill 
	% Scenario S3c
 	\begin{subfigure}[b]{.26\textwidth}
 		\centering
		\begin{center}
			\resizebox{\textwidth}{!}
			{%
			\includegraphics[]{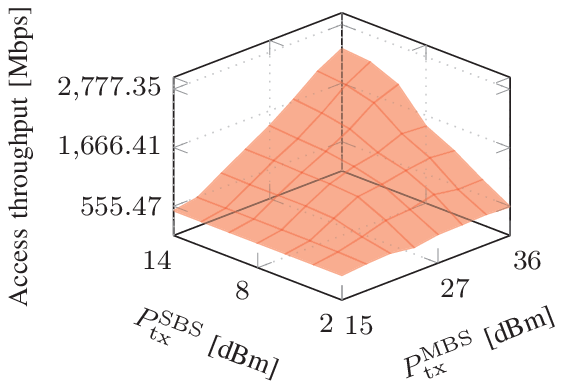}
			}
			\caption{{\tiny Varying $ P^\mathrm{MBS}_\mathrm{tx} $ and $ P^\mathrm{SBS}_\mathrm{tx} $}}
			\label{figure_scenario_s3c}
		\end{center}
 	\end{subfigure}
    \hfill 
	% Scenario S3d
 	\begin{subfigure}[b]{.26\textwidth}
 		\centering
		\begin{center}
			\resizebox{\textwidth}{!}
			{%
			\includegraphics[]{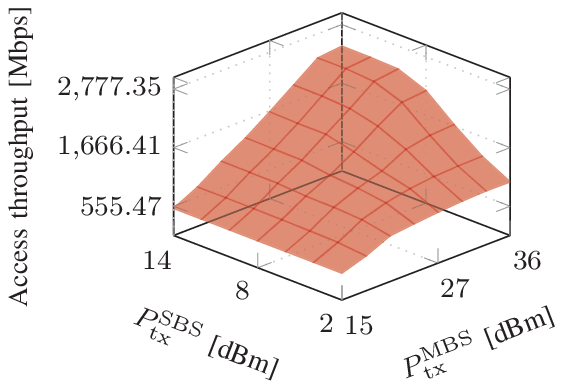}
			}
			\caption{{\tiny Varying $ P^\mathrm{MBS}_\mathrm{tx} $ and $ P^\mathrm{SBS}_\mathrm{tx} $}}
			\label{figure_scenario_s3d}
		\end{center}
 	\end{subfigure}
 	\hfill
	% Scenario S3e
 	\begin{subfigure}[b]{.24\textwidth}
		\begin{center}
			\resizebox{\textwidth}{!}
			{%
			\includegraphics[]{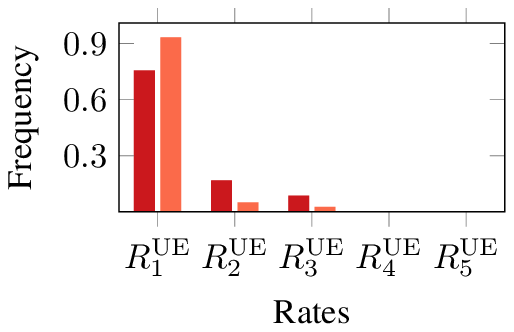}
			}
			\caption{{\tiny  $ {P_{\mathrm{tx}}^\mathrm{MBS} =  36} $ dBm $ ~|~ {P_{\mathrm{tx}}^\mathrm{SBS} =  2} $ dBm}}
			\label{figure_scenario_s3e}
		\end{center}
 	\end{subfigure}
    \hfill 
	% Scenario S3f
 	\begin{subfigure}[b]{.24\textwidth}
		\begin{center}
			\resizebox{\textwidth}{!}
			{%
			\includegraphics[]{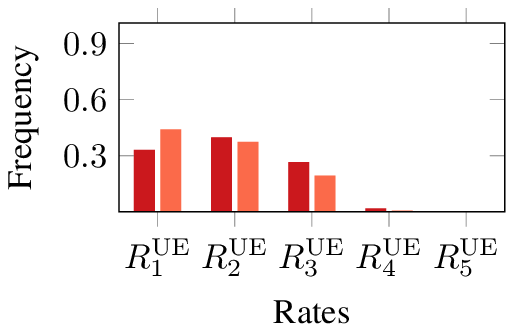}
			}
			\caption{{\tiny  $ {P_{\mathrm{tx}}^\mathrm{MBS} =  36} $ dBm $ ~|~ {P_{\mathrm{tx}}^\mathrm{SBS} =  6} $ dBm}}
			\label{figure_scenario_s3f}
		\end{center}
 	\end{subfigure}
    \hfill 
	% Scenario S3g
 	\begin{subfigure}[b]{.24\textwidth}
		\begin{center}
			\resizebox{\textwidth}{!}
			{%
			\includegraphics[]{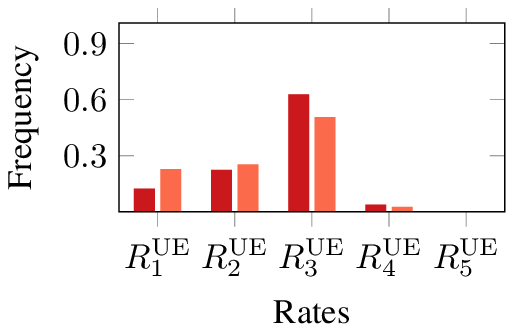}
			}
			\caption{{\tiny  $ {P_{\mathrm{tx}}^\mathrm{MBS} =  36} $ dBm $ ~|~ {P_{\mathrm{tx}}^\mathrm{SBS} =  10} $ dBm}}
			\label{figure_scenario_s3g}
		\end{center}
 	\end{subfigure}
    \hfill 
	% Scenario S3h
 	\begin{subfigure}[b]{.24\textwidth}
		\begin{center}
			\resizebox{\textwidth}{!}
			{%
			\includegraphics[]{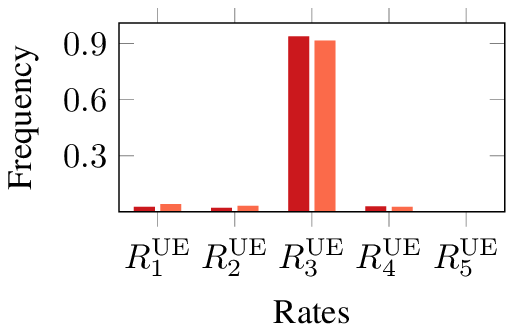}
			}
			\caption{{\tiny  $ {P_{\mathrm{tx}}^\mathrm{MBS} =  36} $ dBm $ ~|~ {P_{\mathrm{tx}}^\mathrm{SBS} =  14} $ dBm}}
			\label{figure_scenario_s3h}
		\end{center}
 	\end{subfigure}
 	\label{figure_scenario_s3}
 	\caption{Evaluation of Scenario S\textsubscript{3}. \emph{We note that maximizing the access throughput is highly dependent on both backhaul and access network parameters, which highlights the importance of jointly optimizing them.}}
\end{figure*}

% Scenario S2
\noindent {\textbf{{Scenario S\textsubscript{2}: Upper bound as a means of network planning.}}} Since \texttt{UB} is much simpler to solve than \texttt{BnC-MISOCP} (as shown in Fig. \ref{figure_scenario_s1d} and Fig. \ref{figure_scenario_s1e}), we can use \texttt{UB} in larger settings to examine multiple configurations of number of antennas, transmit power, number of clusters and cluster size. From the planning perspective, these results are valuable as they allow us to choose suitable operation points for the network. In Fig. \ref{figure_scenario_s2a}, we show the backhaul throughput (i.e., the objective of \texttt{UB}) for various combinations of $ P^\mathrm{MBS}_\mathrm{tx} $, $ N^\mathrm{MBS}_\mathrm{tx} $, $ L $, where the bottommost and uppermost layers represent $ L = 1 $ (one cluster) and $ L = 6 $ (six clusters), respectively. We observe that the backhaul throughput improves with increasing number of antennas and transmit power because more antennas enhance the multiplexing capability while a higher power allows transmitting at higher rates. However, when the number of clusters grows from $ L = 5 $ to $ L = 6 $, the throughput saturates showing marginal improvement because the scenario becomes more interference limited (due to more SBSs deployed). \emph{We realize that with $ N^\mathrm{MBS}_\mathrm{tx} = 64 $ antennas, $ P^\mathrm{MBS}_\mathrm{tx} = 36 $ dBm transmit power and $ L = 5 $ clusters, the backhaul network can be operated at its full capacity}. In Fig. \ref{figure_scenario_s2a}, we considered $ B = 3 $, but we validate such decision in Fig. \ref{figure_scenario_s2b}, where we illustrate the backhaul throughput for various combinations of $ P^\mathrm{MBS}_\mathrm{tx} $ and $ B $ when $ L = 5 $. We note that the throughput decreases when the cluster size increases from $ B = 1 $ to $ B = 6 $ because, to reach more SBSs, higher MBS power is consumed but also more interference is generated due to more SBSs scattered. However, a larger SBS cluster is preferred because \emph{(i)} more UEs can be served (each SBS can serve a limited number of UEs) and  \emph{(ii)} UEs can be allocated higher rates by being connected to more SBSs. With $ B = 3 $, still the maximum backhaul throughput can be achieved.

% Remark: Scenario S2 
%\textit{\textsc{Remark:} We note that \texttt{UB} can be used to evaluate multiple network configurations, thus providing insights of potentially optimal operations points that can be adopted in the planning phase of the network.}

% Figure: Scenario S4
\begin{figure*}[!t]
	\centering
	% Legend
 	\begin{subfigure}[b]{\textwidth}
 		\centering
		\begin{center}
			\includegraphics[]{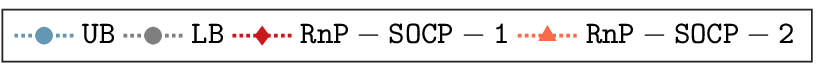}
		\end{center}
 	\end{subfigure}
	% Scenario S5a
 	\begin{subfigure}[b]{.32\textwidth}
 		\centering
		\begin{center}
			\resizebox{\textwidth}{!}
			{%
			\includegraphics[]{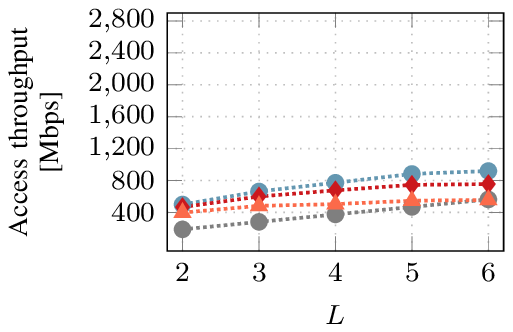}
			}
			\caption{$ P_{\mathrm{tx}}^\mathrm{MBS} = 18 $ dBm and $ P_{\mathrm{tx}}^\mathrm{SBS} = 14 $ dBm}
			\label{figure_scenario_s5a}
		\end{center}
 	\end{subfigure}
    \hfill 
	% Scenario S5b
 	\begin{subfigure}[b]{.32\textwidth}
 		\centering
		\begin{center}
			\resizebox{\textwidth}{!}
			{%
			\includegraphics[]{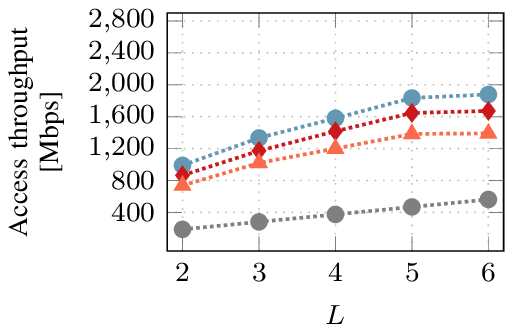}
			}
			\caption{$ P_{\mathrm{tx}}^\mathrm{MBS} = 27 $ dBm and $ P_{\mathrm{tx}}^\mathrm{SBS} = 14 $ dBm}
			\label{figure_scenario_s5b}
		\end{center}
 	\end{subfigure}
    \hfill 
	% Scenario S5c
 	\begin{subfigure}[b]{.32\textwidth}
 		\centering
		\begin{center}
			\resizebox{\textwidth}{!}
			{%
			\includegraphics[]{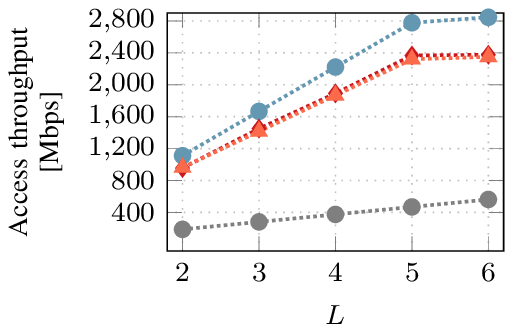}
			}
			\caption{$ P_{\mathrm{tx}}^\mathrm{MBS} = 36 $ dBm and $ P_{\mathrm{tx}}^\mathrm{SBS} = 14 $ dBm}
			\label{figure_scenario_s5c}
		\end{center}
 	\end{subfigure}
 	\hfill
	% Scenario S5d
 	\begin{subfigure}[b]{.24\textwidth}
 		\centering
		\begin{center}
			\resizebox{\textwidth}{!}
			{%
			\includegraphics[]{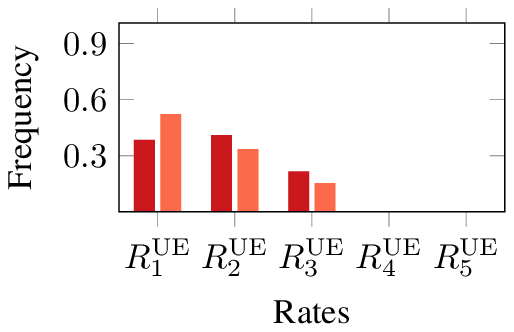}
			}
			\caption{\centering $ L = 2~|~P_{\mathrm{tx}}^\mathrm{MBS} = 18 $ dBm and $ P_{\mathrm{tx}}^\mathrm{SBS} = 14 $ dBm}
			\label{figure_scenario_s5d}
		\end{center}
 	\end{subfigure}
    \hfill 
	% Scenario S5e
 	\begin{subfigure}[b]{.24\textwidth}
 		\centering
		\begin{center}
			\resizebox{\textwidth}{!}
			{%
			\includegraphics[]{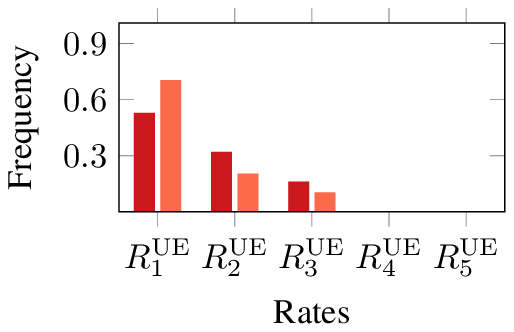}
			}
			\caption{\centering $ L = 3~|~P_{\mathrm{tx}}^\mathrm{MBS} = 18 $ dBm and $ P_{\mathrm{tx}}^\mathrm{SBS} = 14 $ dBm}
			\label{figure_scenario_s5e}
		\end{center}
 	\end{subfigure}
    \hfill 
	% Scenario S5f
 	\begin{subfigure}[b]{.24\textwidth}
 		\centering
		\begin{center}
			\resizebox{\textwidth}{!}
			{%
			\includegraphics[]{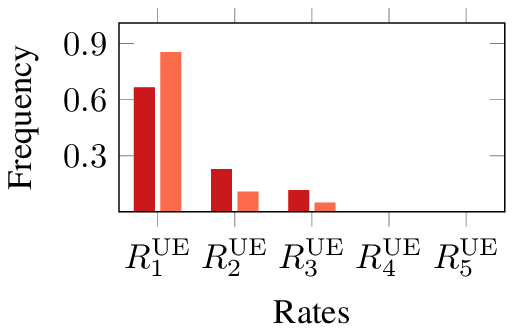}
			}
			\caption{\centering $ L = 4~|~P_{\mathrm{tx}}^\mathrm{MBS} = 18 $ dBm and $ P_{\mathrm{tx}}^\mathrm{SBS} = 14 $ dBm} 
			\label{figure_scenario_s5f}
		\end{center}
 	\end{subfigure}
 	\hfill
	% Scenario S5g
 	\begin{subfigure}[b]{.24\textwidth}
 		\centering
		\begin{center}
			\resizebox{\textwidth}{!}
			{%
			\includegraphics[]{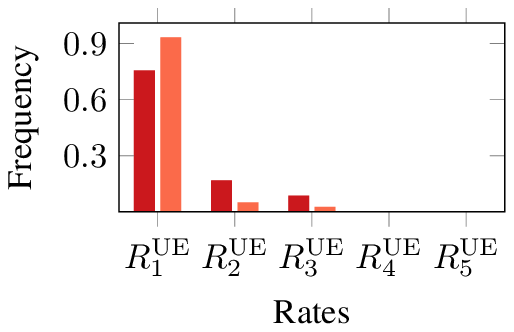}	
			}
			\caption{\centering $ L = 5~|~P_{\mathrm{tx}}^\mathrm{MBS} = 18 $ dBm and $ P_{\mathrm{tx}}^\mathrm{SBS} = 14 $ dBm}
			\label{figure_scenario_s5g}
		\end{center}
 	\end{subfigure}
 	\hfill
	% Scenario S5h
 	\begin{subfigure}[b]{.24\textwidth}
 		\centering
		\begin{center}
			\resizebox{\textwidth}{!}
			{%
			\includegraphics[]{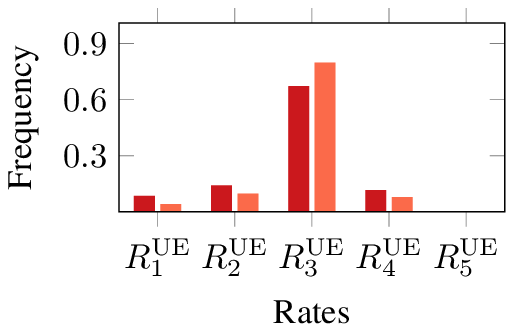}
			}
			\caption{\centering $ L = 2~|~P_{\mathrm{tx}}^\mathrm{MBS} = 36 $ dBm and $ P_{\mathrm{tx}}^\mathrm{SBS} = 14 $ dBm}
			\label{figure_scenario_s5h}
		\end{center}
 	\end{subfigure}
 	\hfill
	% Scenario S5i
 	\begin{subfigure}[b]{.24\textwidth}
 		\centering
		\begin{center}
			\resizebox{\textwidth}{!}
			{%
			\includegraphics[]{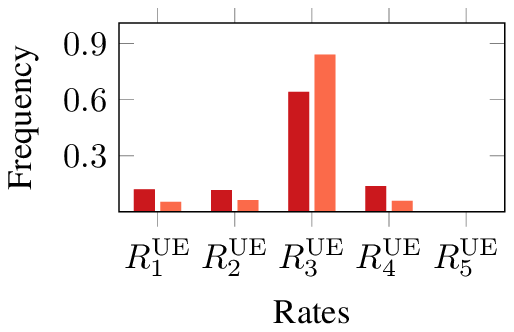}
			}
			\caption{\centering $ L = 3~|~P_{\mathrm{tx}}^\mathrm{MBS} = 36 $ dBm and $ P_{\mathrm{tx}}^\mathrm{SBS} = 14 $ dBm}
			\label{figure_scenario_s5i}
		\end{center}
 	\end{subfigure}
 	\hfill
	% Scenario S5j
 	\begin{subfigure}[b]{.24\textwidth}
 		\centering
		\begin{center}
			\resizebox{\textwidth}{!}
			{%
			\includegraphics[]{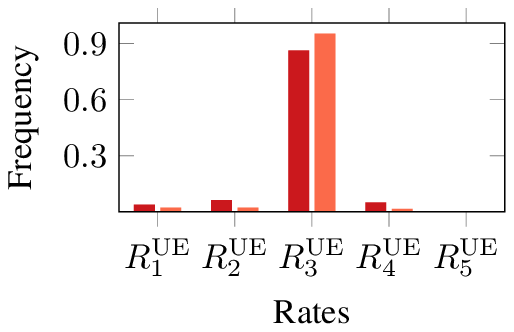}
			}
			\caption{\centering $ L = 4~|~P_{\mathrm{tx}}^\mathrm{MBS} = 36 $ dBm and $ P_{\mathrm{tx}}^\mathrm{SBS} = 14 $ dBm}
			\label{figure_scenario_s5j}
		\end{center}
 	\end{subfigure}
 	\hfill
	% Scenario S5k
 	\begin{subfigure}[b]{.24\textwidth}
 		\centering
		\begin{center}
			\resizebox{\textwidth}{!}
			{%
			\includegraphics[]{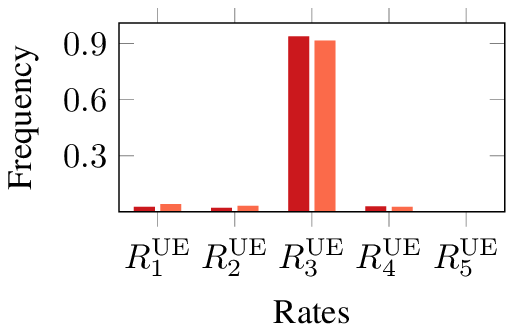}
			}
			\caption{\centering $ L = 5~|~P_{\mathrm{tx}}^\mathrm{MBS} = 36 $ dBm and $ P_{\mathrm{tx}}^\mathrm{SBS} = 14 $ dBm}
			\label{figure_scenario_s5k}
		\end{center}
 	\end{subfigure}
 	\label{figure_scenario_s5}
 	\caption{Evaluation of Scenario S\textsubscript{4}. \emph{We note that the overall access throughput can be expanded with more clusters (i.e., more SBSs and UEs). However, this improvement may saturate beyond a number of clusters due to more interference or insufficient transmit power.}}
\end{figure*}

% Scenario S3
\noindent {\textbf{{Scenario S\textsubscript{3}: Impact of the transmit power.}}} Fig. \ref{figure_scenario_s3a}, Fig. \ref{figure_scenario_s3b}, Fig. \ref{figure_scenario_s3c} and Fig. \ref{figure_scenario_s3d} illustrate how the variation of transmit power at the MBS and SBSs impacts the access network throughput. Fig. \ref{figure_scenario_s3a} shows the case when $ P^\mathrm{SBS}_\mathrm{tx} = 14 $ dBm and $ P^\mathrm{MBS}_\mathrm{tx} $ is varied. As observed, the access throughput improves as the MBS increases its transmit power, which is logical since the backhaul capacity is naturally expanded with higher power. Similarly, Fig. \ref{figure_scenario_s3b} shows the case when $ P^\mathrm{MBS}_\mathrm{tx} = 36 $ dBm and $ P^\mathrm{SBS}_\mathrm{tx} $ is varied. We note that the access throughput improves as the SBSs increase their transmit power. This occurs because higher SBSs power enables UEs to be assigned higher rates. We observe in Fig. \ref{figure_scenario_s3a} and Fig. \ref{figure_scenario_s3b} that when $ P^\mathrm{MBS}_\mathrm{tx} = 36 $ dBm and $ P^\mathrm{SBS}_\mathrm{tx} = 14 $, both \texttt{RnP-SOCP-1} and \texttt{RnP-SOCP-2} achieve nearly the same performance although \texttt{RnP-SOCP-2} grows at a slower rate. This slower improvement stems from the fact that the beamforming vectors for \texttt{RnP-SOCP-2} are predesigned and only their gains can be optimized, thus allowing for less flexibility compared to \texttt{RnP-SOCP-1}. Thus, their performance meet only in the presence of high MBS/SBSs transmit power. At this point, the gap compared to \texttt{UB} is $ 14.8\% $ and $ 16.5\% $ for \texttt{RnP-SOCP-1} and \texttt{RnP-SOCP-2}, respectively. Fig. \ref{figure_scenario_s3c} and Fig. \ref{figure_scenario_s3d} show the effect of varying both $ P^\mathrm{SBS}_\mathrm{tx} $ and $ P^\mathrm{MBS}_\mathrm{tx} $. In Fig. \ref{figure_scenario_s3e}, Fig. \ref{figure_scenario_s3f}, Fig. \ref{figure_scenario_s3g}, Fig. \ref{figure_scenario_s3h}, we show the allocation of UE rates when $ P^\mathrm{MBS}_\mathrm{tx} = 36 $ and $ P^\mathrm{SBS}_\mathrm{tx} $ is varied gradually from a low to a high power. At lower $ P^\mathrm{SBS}_\mathrm{tx} $ as in Fig. \ref{figure_scenario_s3e}, the UEs are mainly assigned the lowest rates. As $ P^\mathrm{SBS}_\mathrm{tx} $ becomes higher, it becomes possible to allocate higher rates to the UEs, as observed in Fig. \ref{figure_scenario_s3h}.

\noindent {\textbf{{Scenario S\textsubscript{4}: Impact of the number of clusters.}}} Fig. \ref{figure_scenario_s5a}, Fig. \ref{figure_scenario_s5b}, Fig. \ref{figure_scenario_s5c} show the access throughput when $ P_{\mathrm{tx}}^\mathrm{SBS} = 14 $ dBm and the number of clusters is varied from $ L = 2 $ to $ L = 6 $ for different $ P_{\mathrm{tx}}^\mathrm{MBS} $ values. The access throughput improves with increasing $ L $ because more clusters translates to more served UEs (there are $ U_\mathrm{served} $ UEs per cluster), and hence the higher aggregate rate. Besides, higher $ P_{\mathrm{tx}}^\mathrm{MBS} $ also improves the access throughput because it boosts the backhaul network capacity. In particular, we observe throughput saturation when increasing from $ L = 5 $ to $ L = 6 $, which is consistent with the behavior observed in Fig. \ref{figure_scenario_s2a} where the backhaul network throughput was evaluated. Further, we note that \texttt{RnP-SOCP-1} outperforms \texttt{RnP-SOCP-2} when $ P_{\mathrm{tx}}^\mathrm{MBS} = \left\lbrace 18, 27 \right\rbrace $ dBm. However, for sufficiently high $ P_{\mathrm{tx}}^\mathrm{MBS} = 36 $ dBm, the performance of both are comparable. Besides, we examine the UE rate allocation in Fig. \ref{figure_scenario_s5d}, Fig. \ref{figure_scenario_s5e}, Fig. \ref{figure_scenario_s5f} and Fig. \ref{figure_scenario_s5g} assuming $ P_{\mathrm{tx}}^\mathrm{MBS} = 18 $ dBm, $ P_{\mathrm{tx}}^\mathrm{SBS} = 14 $ dBm. We observe that when the number of clusters is small, e.g. $ L = 2 $ (see Fig. \ref{figure_scenario_s5d}), the rates assigned to the UEs span a wider range compared to the case when $ L = 5 $ (see Fig. \ref{figure_scenario_s5g}). The reason for this behavior is that more interference is generated in the backhaul network with $ L = 5 $ than with $ L = 2 $. In particular, with $ L = 2 $, only two signals are transmitted whereas with $ L = 5 $, five different signals are sent from the MBS, thus generating more interference at the receiving SBSs. In Fig. \ref{figure_scenario_s5h}, Fig. \ref{figure_scenario_s5i}, Fig. \ref{figure_scenario_s5j} and Fig. \ref{figure_scenario_s5k} we also examine the UE rates assuming $ P_{\mathrm{tx}}^\mathrm{MBS} = 36 $ dBm, $ P_{\mathrm{tx}}^\mathrm{SBS} = 14 $ dBm. In this case, the backhaul network has sufficiently high power. As a result, throughout Fig. \ref{figure_scenario_s5h}, Fig. \ref{figure_scenario_s5i}, Fig. \ref{figure_scenario_s5j} and Fig. \ref{figure_scenario_s5k}, the distribution of rates remains more or less similar.

% Figure: Scenario S6
\begin{figure*}[!t]
	\centering
	% Legend
 	\begin{subfigure}[b]{\textwidth}
 		\centering
		\begin{center}
			\includegraphics[]{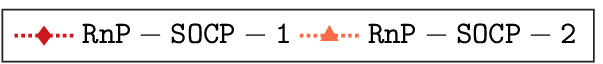}
		\end{center}
 	\end{subfigure}
	% Scenario S6a
 	\begin{subfigure}[b]{.22\textwidth}
		\begin{center}
			\resizebox{\textwidth}{!}
			{%
			\includegraphics[]{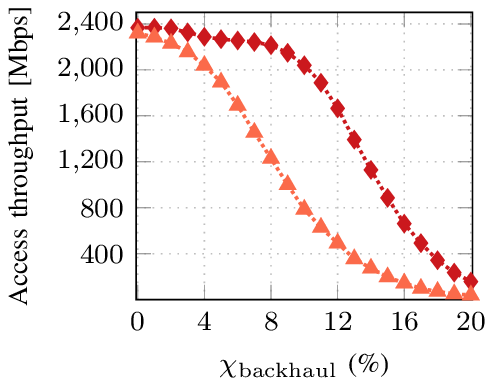}
			}
			\caption{Imprecise backhaul channel}
			\label{figure_scenario_s6a}
		\end{center}
 	\end{subfigure}
    \hfill 
	% Scenario S6b
 	\begin{subfigure}[b]{.22\textwidth}
		\begin{center}
			\resizebox{\textwidth}{!}
			{%
			\includegraphics[]{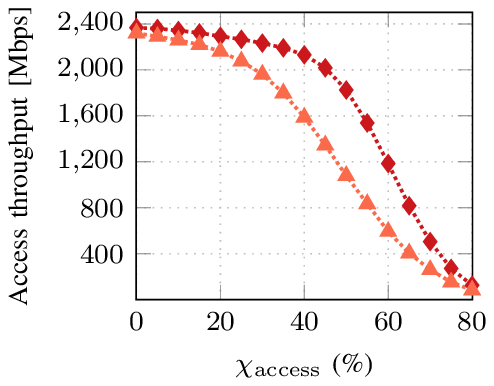}
			}
			\caption{Imprecise access channel}
			\label{figure_scenario_s6b}
		\end{center}
 	\end{subfigure}
    \hfill 
	% Scenario S6c
 	\begin{subfigure}[b]{.26\textwidth}
		\begin{center}
			\resizebox{\textwidth}{!}
			{%
			\includegraphics[]{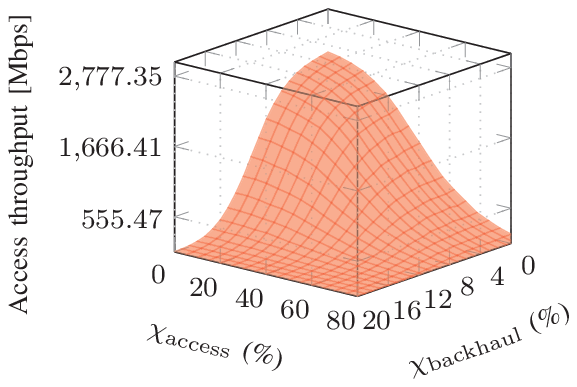}
			}
			\caption{Imprecise channels}
			\label{figure_scenario_s6c}
		\end{center}
 	\end{subfigure}
	% Scenario S6d
 	\begin{subfigure}[b]{.26\textwidth}
		\begin{center}
			\resizebox{\textwidth}{!}
			{%
			\includegraphics[]{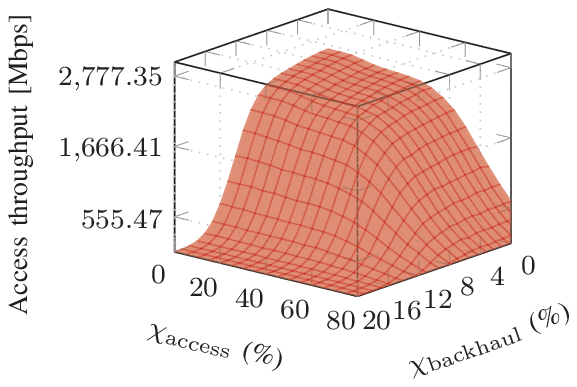}
			}
			\caption{Imprecise channels}
			\label{figure_scenario_s6d}
		\end{center}
 	\end{subfigure}
 	%\label{figure_scenario_s6}
 	\caption{Evaluation of Scenario S\textsubscript{5}. \emph{We have used the model $ \mathbf{c} = \sqrt{1-\chi^2} \hat{\mathbf{c}} + \chi \mathbf{p} $ to emulate imprecise channel conditions, where $ \mathbf{c} $ is the estimated channel, $\hat{\mathbf{c}} $ is the exact access/backhaul channel (but unknown), $ \chi \in \left[ 0, 1\right]  $ is the degree at which the perturbation contaminates the channel, and $ \mathbf{p} \sim \left( \mathbf{0}, \left\| \hat{\mathbf{c}} \right\|^2_2 \mathbf{I}/K \right) $ is a random perturbation, where $ K $ is the length of $ \hat{\mathbf{c}} $. We note the importance of careful provision of the backhaul network because it is the link with highest importance delivering data to the UEs. A potential disruption affecting this link causes a degradation of the whole network whereas impairments in the individual access links do not have a significant impact on the overall network performance. We underline a fundamental difference regarding the impact of imperfect CSI in system models assuming discrete or continuous rates. While CSI variations affect both systems, it has more detrimental consequences in the discrete-rate case. For instance, in continuous-rate models, a CSI variation will produce a SINR different from the expected thus also affecting the rate. However, the resulting rate will still be feasible for the model due to being continuous. On the contrary, in discrete-rate models, if the SINR is below the required target, the data will not be decoded by the SBS/UE thus causing the resulting rate to drop to zero.}}
\end{figure*}

% Figure: Scenario S6
\begin{figure*}[!t]
	\centering
	% Legend
 	\begin{subfigure}[b]{\textwidth}
 		\centering
		\begin{center}
			\includegraphics[]{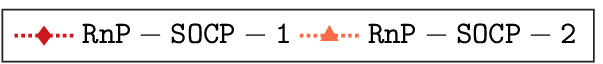}
		\end{center}
 	\end{subfigure}
	% Scenario S7f
 	\begin{subfigure}[b]{0.48\textwidth}
		\begin{center}
			\resizebox{\textwidth}{!}
			{%
			\includegraphics[]{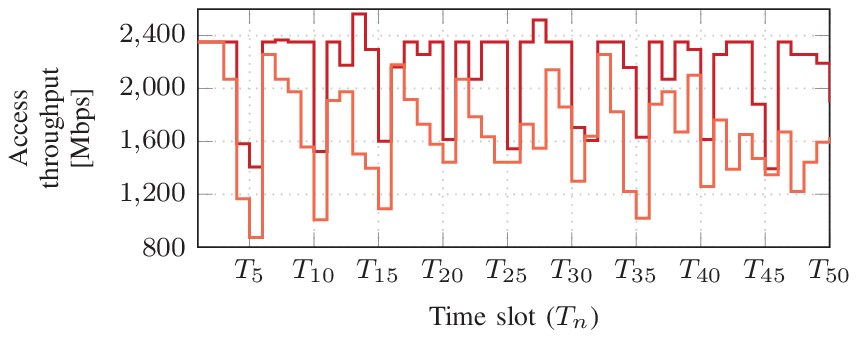}
			}
			\caption{Serving $ U_\mathrm{served} = 4 $ UEs per cluster per slot. }
			\label{figure_scenario_s7f}
		\end{center}
 	\end{subfigure}
    \hfill 
	% Scenario S7g
 	\begin{subfigure}[b]{0.48\textwidth}
		\begin{center}
			\resizebox{\textwidth}{!}
			{%
			\includegraphics[]{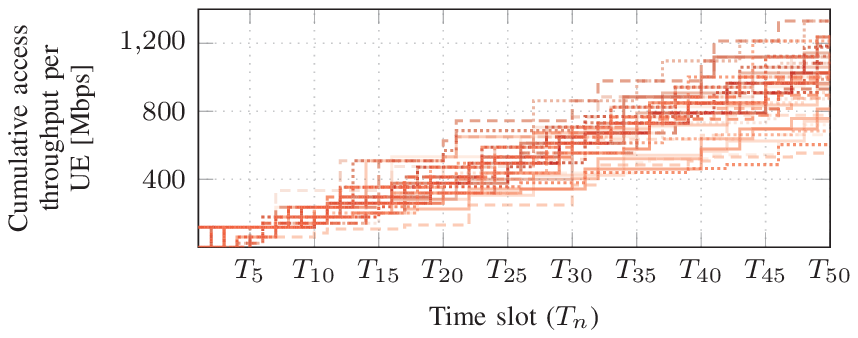}
			}
			\caption{Displaying the individual rates of all UEs in cluster $ \mathcal{U}_1 $. }
			\label{figure_scenario_s7g}
		\end{center}
 	\end{subfigure}
    \caption{Evaluation of Scenario S\textsubscript{6}. \emph{We observe that it is possible to serve all UEs in a system by allocating them in multiple slots, showing that \textsf{RadiOrchestra} is scalable. In addition, the UE rates can be adapted to enforce different priorities based on any network policy of the operator. In this example, we aimed at improving fairness among UEs.}}
 	\label{figure_scenario_s7}
\end{figure*}

% Scenario S6
\noindent {\textbf{{Scenario S\textsubscript{5}: Impact of imprecise channel estimation.}}} Fig. \ref{figure_scenario_s6a} shows the access throughput when the access channels are estimated perfectly but the backhaul channels inaccurately. Here, the channel energy variation is represented by $ \xi_\mathrm{backhaul} $. Although backhaul channels are generally static due to fixed positions of MBS and SBS, it is important to test the network against estimation errors that may arise due to hardware miscalibration or impairments. We observe that as the degree of error in the backhaul channels increases, the access throughput is affected more severely due to information that cannot be decoded by the SBSs and therefore not relayed to the UEs. Further, \texttt{RnP-SOCP-1} is more robust than \texttt{RnP-SOCP-2} to dealing with such imprecisions because \texttt{RnP-SOCP-2} only optimizes the beamformers gains, making it less robust to perturbations. With \texttt{RnP-SOCP-1} and \texttt{RnP-SOCP-2}, the throughput decreases $4.2\%$ and $18.4\%$, respectively when the channel energy varies within $ \xi_\mathrm{backhaul} = 5\%$, and $10.1\%$ and $58.5\%$, respectively when the channel energy varies within $ \xi_\mathrm{backhaul} = 10\%$. Fig. \ref{figure_scenario_s6b} shows the access throughput when the access channels are estimated inaccurately but the backhaul channels perfectly, and the error energy is represented by $ \xi_\mathrm{access} $. The access channel may be inaccurately estimated due to UE mobility, feedback quantization or unmanaged interference from other networks. 

We note that the access throughput with \texttt{RnP-SOCP-1} and \texttt{RnP-SOCP-2} only suffers a decay of $ 9.9\% $ and $31.6\%$, respectively, even when the access channels change within $ \xi_\mathrm{access} = 40\%$,  which is much less compared to the case in Fig. \ref{figure_scenario_s6a}. The reason for this outcome is that a disruption in an access link may cause only a single UE not being able to decode its information (since its SINR may decrease). In contrast, a disruption in a backhaul link may cause many SBSs in a cluster to be automatically unsupplied, thus making them unable to deliver data to the UEs. In addition, the multicast topology of the backhaul network is more susceptible to channel variations, since the link with the weakest condition limits the data rate for the whole SBS cluster. On the other hand, Fig. \ref{figure_scenario_s6c}  and Fig. \ref{figure_scenario_s6d} show the access throughput performance when both the access and backhaul channels contain estimation errors.

% Scenario S6
\noindent {\textbf{{Scenario S\textsubscript{6}: Time-slotted evaluation.}}} We have evaluated the access throughput considering that all UEs have the same priorities. However, the UE priorities (weights) can be adjusted, for instance, to balance the cumulative throughput so that all UEs experience a similar degree of fairness over time. To realize this, we evaluate the algorithms in a slotted manner. Assuming $ L = 5 $, $ U = 20 $, $ U_\mathrm{served} = 4 $, the network needs $ 5 $ slots to allocate the $ 100 $ UEs, i.e., in each slot, $ 20 $ UEs are simultaneously served with $ 4 $ UEs per cluster. In Fig. \ref{figure_scenario_s7f}, we show the access throughput for \texttt{RnP-SOCP-1} and \texttt{RnP-SOCP-2} during $ 50 $ slots of equal duration $ T = T_n - T_{n-1} $ and assuming that the channel is estimated every $ 5 $ slots, i.e., once all the UEs have been served, a new UE scheduling with a different channel is considered. In particular, in every cluster, in time slot $ T_1 $, $ 4 $ UEs out of $ 20 $ are chosen; in slot $ T_2 $, 4 out of $ 16 $ are chosen; in slot $ T_3 $, $ 4 $ out of $ 12 $ are chosen, in slot $ T_4 $, $ 4 $ out of $ 8 $, and in slot $ T_5 $ the remaining $ 4 $ UEs are served. In slot $ T_6 $, the weights are updated based on the cumulative rate the UEs have experienced according to $ w_u^{(n)}  = \frac{1}{T \sum_{i = 1}^n r_u^{(i-1)}} $ (up to normalization), where $ r_u^{(i)} $ is the rate of UE $ u $ in slot $ T_i $. In slot $ T_6 $, another $ 4 $ UEs out of $20 $ are chosen (possibly a different UE batch than in slot $ T_1 $). The process continues in this manner, updating the weights every $ 5 $ slots. In Fig. \ref{figure_scenario_s7g}, we show the individual cumulative throughput for all $ 20 $ UEs in cluster $ \mathcal{U}_1 $. We realize that the throughput experienced by the UEs tend to be similar as the deviation from each other is small, which is achieved due to the adaption of weights.

%\textit{\textsc{Remark:} We observe that it is possible to serve all UEs in a system by allocating them in multiple slots, showing that \textsf{RadiOrchestra} is scalable. In addition, the UE rates can be adapted to enforce different priorities based on any network policy of the operator. In this example, we aimed at improving fairness among UEs.}

% Section 10
\section{Conclusions} \label{section_conclusions}
Self-backhauling millimeter-wave networks are a key enabler for dense deployments by virtue of reducing costs (not needing fiber links) and facilitating higher flexibility through usage of wireless links. However, designing efficient and practical solutions for such systems are extremely complex due to the intertwined nature of backhaul and radio access networks that are not straightforward to model, and intrinsically result in complex problems with coupled optimization variables that are challenging to solve. In this paper, \textsf{RadiOrchestra} demonstrated how to tame this complexity with a series of design choices in the system, and providing mathematical formulation and optimization of radio resources. We proposed three formulations and their respective algorithms, \texttt{BnC-MISOCP}, \texttt{RnP-SOCP-1} and \texttt{RnP-SOCP-2}, to jointly optimize beamforming, user association, rate selection and admission control with the aim of maximizing the access network throughput. Our complexity analysis showed that \texttt{RnP-SOCP-1} and \texttt{RnP-SOCP-2} are less complex than \texttt{BnC-MISOCP} while the simulation results illustrated that their performance remained within $ 16.5\% $ of the upper bound. We believe this attractive complexity-performance trade-off is key to potential adaptation of \textsf{RadiOrchestra} in future systems. \textsf{RadiOrchestra} can be extended in several directions. In \textsf{RadiOrchestra} we considered that both the access and backhaul networks operate over a fixed bandwidth. However, to make the approach more flexible and therefore capable of dealing with unbalanced channel conditions, bandwidth optimization could be incorporated as an additional degree of freedom. Another direction is expanding \textsf{RadiOrchestra} to be robust against channel imprecisions at both the access and backhaul networks to ultimately preserve the integrity of data. While current networks are centralized, enabling distributed optimization algorithms is desirable due to lower latency. Thus, a possible direction of expanding \textsf{RadiOrchestra} is to parallelize the optimization to let each SBS cluster optimize the resources without a central coordinator. In \textsf{RadiOrchestra}, we assumed that the UEs are pre-associated to a given SBS cluster. In dynamic networks, however, this association can change. Therefore, it is interesting to investigate these changes in contexts of transitions between different clusters.

% Acknowledgment
\section*{Acknowledgment} \label{section_acknowledgment}
The research is in part funded by the Deutsche Forschungsgemeinschaft (DFG) within the B5G-Cell project in SFB 1053 MAKI, by the LOEWE initiative (Hesse, Germany) within the emergenCITY center, and by the European Commission through Grant No. 101017109 (DAEMON project).

%The research has been funded by the Deutsche Forschungsgemeinschaft (DFG) within the B5G-Cell project in SFB 1053 MAKI and co-funded by the LOEWE initiative (Hesse, Germany) within the emergenCITY center and by the European Commission through Grant No. 101017109 (DAEMON project).

% References
\bibliographystyle{IEEEtran}
\bibliography{ref}

% Appendices
\appendices

\setcounter{equation}{0}
\renewcommand{\theequation}{A.\arabic{equation}}
\renewcommand{\thesectiondis}[2]{\Alph{section}:}

\section{Proof of Proposition 2} \label{appendix_proposition_2}

In constraints $ \mathrm{\bar{C}_4} $ and $ \mathrm{C_5} $, the beamformer $ \mathbf{w}_{b,u} $ and binary variable $ \kappa_{b,u} $ are tied. This leads to obtain zero-beamformers for unserved UEs. To ensure the same effect after removing the multiplicative coupling between $ \mathbf{w}_{b,u} $ and $ \kappa_{b,u} $, additional constraints are required. First, we define the auxiliary variable $ p_{b,u} $ representing the power of the beamformer from SBS $ b $ to UE $ u $, which leads us to declare the following constraint, $ \mathrm{C_{17}}: p_{b,u} \geq 0, \forall l \in \mathcal{L}, b \in \mathcal{B}_l, u \in \mathcal{U}_l $. Considering, the newly introduced variable, constraint $ \mathrm{\bar{C}_4} $ is redefined as $ \mathrm{C_{18}}: \sum_{u \in \mathcal{U}_l} p_{b,u} \leq P_\mathrm{tx}^\mathrm{SBS}, \forall l \in \mathcal{L}, b \in \mathcal{B}_l $. In addition, the power $ p_{b,u} $ of a beamformer needs to be zero for unserved UEs and positive for served UEs, which is enforced via $ \mathrm{C_{19}}: p_{b,u} \leq \kappa_{b,u} P_\mathrm{tx}^\mathrm{SBS}, \forall l \in \mathcal{L}, b \in \mathcal{B}_l, u \in \mathcal{U}_l $. To connect the beamformer $ \mathbf{w}_{b,u} $ and its power $ p_{b,u} $, we define $ \left\| \mathbf{w}_{b,u} \right\|_2^2 \leq \kappa_{b,u} p_{b,u} $, which ensures that the beamformer is a zero-vector when $ \kappa_{b,u} = 0 $. Note that $ \left\| \mathbf{w}_{b,u} \right\|_2^2 \leq \kappa_{b,u} p_{b,u} $ is nonconvex but it can be recast as a SOC constraint as shown in the following. Using the difference of squares, the product $ \kappa_{b,u} p_{b,u} $ is equivalent to $ \kappa_{b,u} p_{b,u} = \frac{\left( \kappa_{b,u} + p_{b,u} \right)^2 - \left( \kappa_{b,u} - p_{b,u} \right)^2}{4} $, which allows us to rearrange $ \left\| \mathbf{w}_{b,u} \right\|_2^2 \leq \kappa_{b,u} p_{b,u} $ as a new constraint $ \mathrm{C_{20}}: \left\| \left[ 2 \mathbf{w}_{b,u}^H,  \kappa_{b,u} - p_{b,u} \right] \right\|_2 \leq \kappa_{b,u} + p_{b,u}, \forall l \in \mathcal{L}, b \in \mathcal{B}_l, u \in \mathcal{U}_l $. After these changes, $ \mathbf{w}_{b,u} $ and $ \kappa_{b,u} $ have been decoupled while still guaranteeing the same effect as if coupled. Thus,  the product $ \mathbf{w}_{b,u} \kappa_{b,u} $ can be replaced by $ \mathbf{w}_{b,u} $ upon including $ \mathrm{C_{17}} - \mathrm{C_{20}} $. The, constraint $ \mathrm{\bar{C}_{21}} $ is obtained after replacing $ \mathbf{w}_{b,u} \kappa_{b,u} $ by $ \mathbf{w}_{b,u} $ in $ \mathrm{C_5} $.

\setcounter{equation}{0}
\renewcommand{\theequation}{B.\arabic{equation}}
\renewcommand{\thesectiondis}[2]{\Alph{section}:}
\section{Proof of Proposition 3} \label{appendix_proposition_3}

We follow a similar procedure as in \cite{cheng2012:dynamic-rate-adaptation-multiuser-beamforming-mixed-integer-conic-programming}.We exchange positions between the SINR denominator and the right-hand side (RHS) of $ \mathrm{\bar{C}_{21}} $. Then, we add $ \big| \sum_{b \in \mathcal{B}_l} \mathbf{h}_{b,u}^H \mathbf{w}_{b,u} \big|^2 $ to both sides, thus yielding
\noindent{
\resizebox{1.01\columnwidth}{!}{
\begin{minipage}{1.01\columnwidth}
% Constraint C16
\begin{align}
	& \mathrm{\bar{C}_{21}}: \left( 1 + {\alpha_{u,j}}^{-1} {\Gamma_j^\mathrm{UE}}^{-1} \right) \Big| \sum_{b \in \mathcal{B}_l} \mathbf{h}_{b,u}^H \mathbf{w}_{b,u} \Big|^2 \geq \nonumber 
	\\
	&\sum_{\substack{l' \in \mathcal{L}}} \sum_{u' \in \mathcal{U}_{l'}} \Big| \sum_{b' \in \mathcal{B}_{l'}}  \mathbf{h}_{b',u}^H \mathbf{w}_{b',u'} \Big|^2 + \sigma^2_\mathrm{UE}, \forall l \in \mathcal{L}, u \in \mathcal{U}_l, j \in \mathcal{J}^{\mathrm{UE}}. \nonumber 
\end{align}
\end{minipage}
}}

To deal with this nonconvex constraint, we first derive expressions for its two cases.
% Constraint C16
\begin{align}
	\circled{\footnotesize{1}} ~ & \alpha_{u,j} = 0 \Rightarrow & & \sum_{\substack{l' \in \mathcal{L}}} \sum_{u' \in \mathcal{U}_{l'}} \Big| \sum_{b' \in \mathcal{B}_{l'}} \mathbf{h}_{b',u}^H \mathbf{w}_{b',u'} \Big|^2 \nonumber
	\\
	& & & + \sigma^2_\mathrm{UE} \leq \infty, \forall l \in \mathcal{L}, u \in \mathcal{U}_l, j \in \mathcal{J}^{\mathrm{UE}}, \nonumber \\
	\circled{\footnotesize{2}} ~ & \alpha_{u,j} = 1 \Rightarrow & & \sum_{\substack{l' \in \mathcal{L}}} \sum_{u' \in \mathcal{U}_{l'}} \Big| \sum_{b' \in \mathcal{B}_{l'}} \mathbf{h}_{b',u}^H \mathbf{w}_{b',u'} \Big|^2 + \sigma^2_\mathrm{UE} \leq \nonumber
	\\
	& & & \left( 1 + {\Gamma_j^\mathrm{UE} }^{-1} \right) \Big| \sum_{b \in \mathcal{B}_l} \mathbf{h}_{b,u}^H \mathbf{w}_{b,u} \Big|^2, \forall l \in \mathcal{L}, \nonumber 
	\\ 
	& & & u \in \mathcal{U}_l, j \in \mathcal{J}^{\mathrm{UE}}. \nonumber 
\end{align}

In case \circled{\footnotesize{1}}, the inequality is satisfied  by default. Besides, it is possible to find an upper bound $ Q_u^2 $ for $ \sum_{\substack{l' \in \mathcal{L}}} \sum_{u' \in \mathcal{U}_{l'}} \big| \sum_{b' \in \mathcal{B}_{l'}} \mathbf{h}_{b',u}^H \mathbf{w}_{b',u'} \big|^2 + \sigma^2_\mathrm{UE} $ to prevent using $ \infty $. By harnessing the \textit{big-M} method, we can equivalently combine the two cases into $ \mathrm{C_{21}} $, shown at the top of this page. The upper bound $ Q_u^2 = P_\mathrm{tx}^\mathrm{SBS} \sum_{\substack{l' \in \mathcal{L}}} \sum_{b' \in \mathcal{B}_{l'}} \left\| \mathbf{h}_{b',u} \right\|_2^2 + \sigma^2_\mathrm{UE} $ is obtained by maximizing the LHS of $ \mathrm{C_{21}} $.
\begin{figure*}[!t]
	% ensure that we have normalsize text
	\normalsize
	% Constraint C18
	\begin{align}
		& \mathrm{C_{21}}: \sum_{\substack{l' \in \mathcal{L}}} \sum_{u' \in \mathcal{U}_{l'}} \Big| \sum_{b' \in \mathcal{B}_{l'}} \mathbf{h}_{b',u}^H \mathbf{w}_{b',u'} \Big|^2 + \sigma^2_\mathrm{UE} \leq \left( 1 + {\Gamma_j^\mathrm{UE}}^{-1} \right) \Big| \sum_{b \in \mathcal{B}_l} \mathbf{h}_{b,u}^H \mathbf{w}_{b,u} \Big|^2 + \left( 1 - \alpha_{u,j} \right)^2 Q_u^2, \forall l \in \mathcal{L}, u \in \mathcal{U}_l, j \in \mathcal{J}^{\mathrm{UE}}. \nonumber 
	\end{align}
	\hrulefill
	\begin{align}
	  & \mathrm{C_{21} - C_{22}} =
	  \begin{cases}
	  	\mathrm{C_{23}}: \left\| \left[ \bar{\mathbf{h}}_u^H \mathbf{W}, \sigma_\mathrm{UE} \right] \right\|_2 \leq \sqrt{1 + {\Gamma_j^\mathrm{UE}}^{-1}} \mathsf{Re} \left\lbrace \mathbf{h}_u^H \mathbf{w}_u \right\rbrace  + \left(1 - \alpha_{u,j} \right) Q_u, \forall l \in \mathcal{L}, u \in \mathcal{U}_l, j \in \mathcal{J}^{\mathrm{UE}},
	  	\\
	   	\mathrm{C_{24}}: \mathsf{Re} \left\lbrace \mathbf{h}_u^H \mathbf{w}_u \right\rbrace \geq \alpha_{u,j} \sqrt{\Gamma_j^\mathrm{UE}} \sigma_\mathrm{UE}, \forall l \in \mathcal{L}, u \in \mathcal{U}_l, j \in \mathcal{J}^{\mathrm{UE}},
	   	\\
	   	\mathrm{C_{25}}: \mathsf{Im} \left\lbrace \mathbf{h}_u^H \mathbf{w}_u \right\rbrace = 0, \forall l \in \mathcal{L}, u \in \mathcal{U}_l, j \in \mathcal{J}^{\mathrm{UE}}. \nonumber \\				 	 				 
	  \end{cases}
	\end{align}
	\hrulefill
	\vspace*{4pt}
\end{figure*}

\setcounter{equation}{0}
\renewcommand{\theequation}{C.\arabic{equation}}
\renewcommand{\thesectiondis}[2]{\Alph{section}:}
\section{Proof of Proposition 5} \label{appendix_proposition_5}

Assuming that $ \mathbf{x} = \left[ \bar{\mathbf{h}}_u^H \mathbf{W}, \sigma_\mathrm{UE} \right] $, constraint $ \mathrm{C_{21}} $ can be expressed as
% Constraint
\begin{align} \nonumber
  & \left\| \mathbf{x} \right\|^2_2 \leq \left( 1 + {\Gamma_j^\mathrm{UE} }^{-1} \right) \left| \mathbf{h}_u^H \mathbf{w}_u \right|^2 \nonumber
  \\
  & ~~~~~~~~~~~~~~ + \left( 1 - \alpha_{u,j} \right)^2 Q_u^2, \forall l \in \mathcal{L}, u \in \mathcal{U}_l, j \in \mathcal{J}^{\mathrm{UE}}. \nonumber
\end{align}

Taking the square root at both sides and applying the Jensen's inequality to the RHS expression, we obtain
% Constraint
\begin{align} 
  & \sqrt{\left( 1 + {\Gamma_j^\mathrm{UE} }^{-1} \right) \left| \mathbf{h}_u^H \mathbf{w}_u \right|^2 + \left( 1 - \alpha_{u,j} \right)^2 Q_u^2} \leq \nonumber
  \\
  & ~~~~~~~~~~~~~~~~~~~ \sqrt{ 1 + {\Gamma_j^\mathrm{UE} }^{-1} } \left| \mathbf{h}_u^H \mathbf{w}_u \right| + \left( 1 - \alpha_{u,j} \right) Q_u. \nonumber
\end{align} 

When $ \alpha_{u,j} = 1 $, the inequality is tight, because the RHS and LHS of the expression above become equivalent, i.e., $ \left\| \mathbf{x} \right\|_2 \leq \sqrt{ 1 + {\Gamma_j^\mathrm{UE}}^{-1}} \left| \mathbf{h}_u^H \mathbf{w}_u \right| $. When $ \alpha_{u,j} = 0 $, the inequality still remains valid, i.e. $ \left\| \mathbf{x} \right\|_2 \leq \sqrt{ 1 + {\Gamma_j^\mathrm{UE}}^{-1}} \left| \mathbf{h}_u^H \mathbf{w}_u \right| + Q_u $, because $ Q_u $ is an upper bound for $ \left\| \mathbf{x} \right\|_2 $. As a result, the following expression is equivalent to $ \mathrm{C_{21}} $
% Constraint
\begin{align} \nonumber
  & \left\| \mathbf{x} \right\|_2 \leq \sqrt{ 1 + {\Gamma_j^\mathrm{UE} }^{-1} } \left| \mathbf{h}_u^H \mathbf{w}_u \right| \nonumber
  \\
  & ~~~~~~~~~~~~ + \left( 1 - \alpha_{u,j} \right) Q_u, \forall l \in \mathcal{L}, u \in \mathcal{U}_l, j \in \mathcal{J}^{\mathrm{UE}}. \nonumber
\end{align}

Notice that the beamforming vectors are invariant to phase shift. In particular, $ \mathbf{w}_u $ and $ \mathbf{w}_u e^{ j \theta_u } $ yield the same received SINR at the UE $ u $. Thus, it is possible to choose a phase $ e^{ j \theta_u } $ such that $ \mathbf{h}_u^H \mathbf{w}_u  $ becomes purely real and nonnnegative \cite[ch.~18]{bengtsson2001:optimal-suboptimal-transmit-beamforming}. Therefore, the following holds
\begin{align} \nonumber
  & \left\| \mathbf{x} \right\|_2 \leq \sqrt{ 1 + {\Gamma_j^\mathrm{UE}}^{-1} } \left| \mathbf{h}_u^H \mathbf{w}_u \right| + \left( 1 - \alpha_{u,j} \right) Q_u \triangleq\nonumber
  \\ 
  & ~~~~~~ \begin{cases}
  	\left\| \mathbf{x} \right\|_2 \leq \sqrt{ 1 + {\Gamma_j^\mathrm{UE} }^{-1} } \mathsf{Re} \left\lbrace \mathbf{h}_u^H \mathbf{w}_u \right\rbrace + \left( 1 - \alpha_{u,j} \right) Q_u, \\
   	\mathsf{Re} \left\lbrace \mathbf{h}_u^H \mathbf{w}_u \right\rbrace \geq 0, \\
   	\mathsf{Im} \left\lbrace \mathbf{h}_u^H \mathbf{w}_u \right\rbrace = 0.  	 				 
  \end{cases} \nonumber
\end{align}

Similarly, constraint $ \mathrm{C_{22}} $ can be expressed as
\begin{equation} \nonumber
  \alpha_{u,j} \sqrt{\Gamma_j^\mathrm{UE}} \sigma_\mathrm{UE} \leq \left| \mathbf{h}_u^H \mathbf{w}_u \right| \triangleq
  \begin{cases}
  	\alpha_{u,j} \sqrt{\Gamma_j^\mathrm{UE}} \sigma_\mathrm{UE} \leq \mathsf{Re} \left\lbrace \mathbf{h}_u^H \mathbf{w}_u \right\rbrace
  	%, \forall l \in \mathcal{L}, u \in \mathcal{U}_l, j \in \mathcal{J}^{\mathrm{UE}}, 
  	\\
   	\mathsf{Re} \left\lbrace \mathbf{h}_u^H \mathbf{w}_u \right\rbrace \geq 0, 
   	\\
   	\mathsf{Im} \left\lbrace \mathbf{h}_u^H \mathbf{w}_u \right\rbrace = 0. 	 				 
  \end{cases}
\end{equation}

Combining the results above, constraints $ \mathrm{C_{21} - C_{22}} $ are remodeled as $ \mathrm{C_{23}} - \mathrm{C_{25}} $, shown at the top of this page.

\setcounter{equation}{0}
\renewcommand{\theequation}{D.\arabic{equation}}
\renewcommand{\thesectiondis}[2]{\Alph{section}:}
\section{Proof of Proposition 6} \label{appendix_proposition_6}

Note that $ \left| \mathbf{g}_b^H \mathbf{m}_l \right| \geq \mathsf{Re} \left\lbrace \mathbf{g}_b^H \mathbf{m}_l \right\rbrace $ always holds true. The inequality becomes tight when the phase of $ \mathbf{g}_b^H \mathbf{m}_l $ is zero \cite{chen2009:distributed-p2p-beamforming-multiuser-relay-networks, bornhorst2012:distributed-beamforming-multigroup-multicasting-relay-networks}. This is, in general, not true unless there is a single SBS per cluster. Using this conservative relation, we replace $ \mathrm{C_{15}} - \mathrm{C_{16}} $ by $ \mathrm{C_{26}} - \mathrm{C_{27}} $, which are defined as
\noindent{
\resizebox{1.01\columnwidth}{!}{
\begin{minipage}{1.01\columnwidth}
% Constraint C26 - C27
\begin{align}
	& \mathrm{C_{26}}: \left\| \mathbf{g}_b^H \mathbf{M} \right\|^2_2 + \sigma^2_\mathrm{SBS} \leq \left( 1 + {\Gamma_j^\mathrm{SBS} }^{-1} \right) \mathsf{Re} \left\lbrace \mathbf{g}_b^H \mathbf{m}_l \right\rbrace^2 \nonumber
	\\
	& ~~~~~~~~~~~~~~~~~~ + \left( 1 - \beta_{l,j} \right)^2 Q_b^2, \forall l \in \mathcal{L}, b \in \mathcal{B}_l, j \in \mathcal{J}^{\mathrm{SBS}}, \nonumber 
	\\
	& \mathrm{C_{27}}: \beta_{l,j} \Gamma_j^\mathrm{SBS} \sigma^2_\mathrm{SBS} \leq \mathsf{Re} \left\lbrace \mathbf{g}_b^H \mathbf{m}_l \right\rbrace^2, \forall l \in \mathcal{L}, b \in \mathcal{B}_l, j \in \mathcal{J}^{\mathrm{SBS}}, \nonumber
\end{align}
\end{minipage}
}}
where $ Q_b^2 = P_\mathrm{tx}^\mathrm{MBS} \left\| \mathbf{g}_b \right\|_2^2 + \sigma^2_\mathrm{SBS} $. However, these inequalities can be recast as the following convex SOC constraints
\noindent{
\resizebox{1.01\columnwidth}{!}{
\begin{minipage}{1.01\columnwidth}
% Constraint C23 - C24
\begin{align}
	& \mathrm{C_{26}}: \left\| \left[ \mathbf{g}_b^H \mathbf{M}, \sigma_\mathrm{SBS} \right] \right\|_2 \leq \sqrt{ 1 + {\Gamma_j^\mathrm{SBS} }^{-1} } \mathsf{Re} \left\lbrace \mathbf{g}_b^H \mathbf{m}_l \right\rbrace \nonumber
	\\
	& ~~~~~~~~~~~~~~~~~~~~~~ + \left( 1 - \beta_{l,j} \right) Q_b, \forall l \in \mathcal{L}, b \in \mathcal{B}_l, j \in \mathcal{J}^{\mathrm{SBS}}, \nonumber 
	\\
	& \mathrm{C_{27}}: \beta_{l,j} \sqrt{\Gamma_j^\mathrm{SBS}} \sigma_\mathrm{SBS} \leq \mathsf{Re} \left\lbrace \mathbf{g}_b^H \mathbf{m}_l \right\rbrace, \forall l \in \mathcal{L}, b \in \mathcal{B}_l, j \in \mathcal{J}^{\mathrm{SBS}}, \nonumber
\end{align}
\end{minipage}
}}
where the Jensen's inequality has been applied to $ \mathrm{C_{26}} $.

%\section{Proof of Proposition 7} \label{appendix_proposition_7}
%The function $ \hat{f} \left( \alpha_{u,j} \right) = \alpha_{u,j} - \alpha_{u,j}^2 $ is concave and non-negative in the region defined by $ \mathrm{X_1} $. Specifically, it is positive for $ \alpha_{u,j} \in \left\langle 0, 1 \right\rangle  $ and only zero for $ \alpha_{u,j} \in \left\lbrace 0, 1\right\rbrace $. In order for $ \hat{f} \left( \alpha_{u,j} \right) $ to be non-positive (as enforced by $ \mathrm{Z_1} $), then $ \alpha_{u,j} $ can only be $ 0 $ or $ 1 $, in which case $ \hat{f} \left( \alpha_{u,j} \right) = 0 $. Thus, it becomes evident that $ \mathrm{C_1} $ is equivalent to the intersection of constraints $ \mathrm{X_1} $ and $ \mathrm{Z_1} $. This also holds true for the sum of functions $ \hat{f} \left( \alpha_{u,j} \right) $. Similar relations can be obtained for $ \mathrm{C_6} $ and $ \mathrm{C_{10}} $.

\setcounter{equation}{0}
\renewcommand{\theequation}{E.\arabic{equation}}
\renewcommand{\thesectiondis}[2]{\Alph{section}:}
\section{Proof of Proposition 8} \label{appendix_proposition_8}
We define the Lagrange dual function of $ \mathcal{P}_1 $ as $ \phi \left( \lambda_\alpha, \lambda_\beta, \lambda_\kappa \right) = \max_{\boldsymbol{\Theta} \in \mathscr{D}} L\left( \boldsymbol{\alpha}, \boldsymbol{\beta}, \boldsymbol{\kappa}, \lambda_\alpha, \lambda_\beta, \lambda_\kappa \right) $, where $ L\left( \boldsymbol{\alpha}, \boldsymbol{\beta}, \boldsymbol{\kappa}, \lambda_\alpha, \lambda_\beta, \lambda_\kappa \right) = R_\mathrm{w-sum}^\mathrm{access} \left( \boldsymbol{\alpha} \right) - \lambda_\alpha f_\alpha \left( \boldsymbol{\alpha} \right) - \lambda_\beta f_\beta \left( \boldsymbol{\beta} \right) - \lambda_\kappa f_\kappa \left( \boldsymbol{\kappa} \right) $. In addition, we define

\begin{align} \nonumber
	\mathrm{primal:} ~~ p^* & = \max_{\boldsymbol{\Theta} \in \mathscr{D}} \min_{\lambda_\alpha, \lambda_\beta, \lambda_\kappa \geq 0} L\left( \boldsymbol{\alpha}, \boldsymbol{\beta}, \boldsymbol{\kappa}, \lambda_\alpha, \lambda_\beta, \lambda_\kappa \right) \nonumber 
	\\
							& = \max ~ (\mathcal{P}_1 ). \nonumber 
	\\ 
	\mathrm{dual:} ~~ d^*   & = \min_{\lambda_\alpha, \lambda_\beta, \lambda_\kappa \geq 0} \max_{\boldsymbol{\Theta} \in \mathscr{D}} L\left( \boldsymbol{\alpha}, \boldsymbol{\beta}, \boldsymbol{\kappa}, \lambda_\alpha, \lambda_\beta, \lambda_\kappa \right) \nonumber 
	\\ 
							& = \min_{\lambda_\alpha, \lambda_\beta, \lambda_\kappa \geq 0} \phi \left( \lambda_\alpha, \lambda_\beta, \lambda_\kappa \right). \nonumber
\end{align}

According to the weak duality theorem, the following holds
\begin{align} \label{weak_duality_theorem}
	p^* \leq \min_{\lambda_\alpha, \lambda_\beta, \lambda_\kappa \geq 0} \phi \left( \lambda_\alpha, \lambda_\beta, \lambda_\kappa \right).
\end{align}

Note that $ f_\alpha \left( \boldsymbol{\alpha} \right) \geq 0 $, $ f_\beta \left( \boldsymbol{\beta} \right) \geq 0 $, $ f_\kappa \left( \boldsymbol{\kappa} \right) \geq 0 $, for $ \boldsymbol{\Theta} \in \mathscr{D} $. Thus, the Lagrangian $ L\left( \boldsymbol{\alpha}, \boldsymbol{\beta}, \boldsymbol{\kappa}, \lambda_\alpha, \lambda_\beta, \lambda_\kappa \right) $ is monotonically decreasing with respect to $ \lambda_\alpha $, $ \lambda_\beta $, $ \lambda_\kappa $ when $ \boldsymbol{\Theta} \in \mathscr{D} $. Further, this means that $ \phi \left( \lambda_\alpha, \lambda_\beta, \lambda_\kappa \right) $ is monotonically decreasing with respect to $ \lambda_\alpha $, $ \lambda_\beta $, $ \lambda_\kappa $ and is bounded by the optimal value of $ \mathcal{P}_1 $. We distinguish the following two cases.

\textit{Case 1:} Suppose that $ f_\alpha \left( \boldsymbol{\alpha}_0 \right) = 0, f_\beta \left( \boldsymbol{\beta}_0 \right) = 0, f_\kappa \left( \boldsymbol{\kappa}_0 \right) = 0 $
for some $ \lambda_{\alpha_0} < \infty $, $ \lambda_{\beta_0} < \infty $, $ \lambda_{\kappa_0} < \infty $, implying that $ \boldsymbol{\alpha}_0 $, $ \boldsymbol{\beta}_0 $, $ \boldsymbol{\kappa}_0 $ are binary. Therefore, $ \boldsymbol{\alpha}_0 $, $ \boldsymbol{\beta}_0 $, $ \boldsymbol{\kappa}_0 $ are also feasible to $ \mathcal{P}_1 $. Replacing this solution in the primal problem, we obtain $ L\left( \boldsymbol{\alpha}_0, \boldsymbol{\beta}_0, \boldsymbol{\kappa}_0, \lambda_{\alpha_0}, \lambda_{\beta_0}, \lambda_{\kappa_0} \right) = R_\mathrm{w-sum}^\mathrm{access} \left( \boldsymbol{\alpha}_0 \right) \leq p^* $. Now, considering the dual problem and (\ref{weak_duality_theorem}), we have that $ \phi \left( \lambda_{\alpha_0}, \lambda_{\beta_0}, \lambda_{\kappa_0} \right) = L\left( \boldsymbol{\alpha}_0, \boldsymbol{\beta}_0, \boldsymbol{\kappa}_0, \lambda_{\alpha_0}, \lambda_{\beta_0}, \lambda_{\kappa_0} \right) = R_\mathrm{w-sum}^\mathrm{access} \left( \boldsymbol{\alpha}_0 \right) \geq p^* $, which implies that $ p^* = d^* $, i.e. strong duality holds. Based on the previous result, we realize that
\begin{align} 
	\phi \left( \lambda_{\alpha_0}, \lambda_{\beta_0}, \lambda_{\kappa_0} \right) & = \min_{\lambda_\alpha, \lambda_\beta, \lambda_\kappa \geq 0} \phi \left( \lambda_\alpha, \lambda_\beta, \lambda_\kappa \right), \nonumber
	\\
	\phi \left( \lambda_{\alpha}, \lambda_{\beta}, \lambda_{\kappa} \right) & = p^*, \forall \lambda_{\alpha} \geq \lambda_{\alpha_0}, \forall \lambda_{\beta} \geq \lambda_{\beta_0}, \forall \lambda_{\kappa} \geq \lambda_{\kappa_0}, \nonumber
\end{align}
which means that for any $ \lambda_\alpha, \lambda_\beta, \lambda_\kappa $, such that $ \lambda_{\alpha_0} < \lambda_{\alpha} < \infty, \lambda_{\beta_0} < \lambda_{\beta} < \infty, \lambda_{\kappa_0} < \lambda_{\kappa} < \infty $, problems $ \mathcal{P}_1 $ and $ \widetilde{\mathcal{P}}_1 $ share the same optimal value an optimal solution. Thus, $ {\mathcal{P}}_1 $ can be solved by means of $ \widetilde{\mathcal{P}}_1 $ for appropriately chosen large values of $ \lambda_\alpha, \lambda_\beta, \lambda_\kappa $.

\textit{Case 2:} Suppose that $ f_\alpha \left( \boldsymbol{\alpha}_0 \right) > 0, f_\beta \left( \boldsymbol{\beta}_0 \right) > 0, f_\kappa \left( \boldsymbol{\kappa}_0 \right) > 0, $ for $ \lambda_{\alpha_0} > 0 $, $ \lambda_{\beta_0} > 0 $, $ \lambda_{\kappa_0} > 0 $, implying that some elements of $ \boldsymbol{\alpha}_0 $, $ \boldsymbol{\beta}_0 $, $ \boldsymbol{\kappa}_0 $ take values between $ 0 $ and $ 1 $. From the dual problem, we have that $ \phi \left( \lambda_{\alpha_0}, \lambda_{\beta_0}, \lambda_{\kappa_0} \right) \rightarrow -\infty $. However, this contradicts the weak duality theorem, which states that $ \phi \left( \lambda_\alpha, \lambda_\beta, \lambda_\kappa \right) $ is bounded from below by the primal solution, which is at worst zero. Thus, this case is not valid.

\setcounter{equation}{0}
\renewcommand{\theequation}{F.\arabic{equation}}
\renewcommand{\thesectiondis}[2]{\Alph{section}:}
\section{Proof of Proposition 9} \label{appendix_proposition_9}
Note that $ q_\alpha \left( \boldsymbol{\alpha} \right) $, $ q_\beta \left( \boldsymbol{\beta} \right) $, $ q_\kappa \left( \boldsymbol{\kappa} \right) $ are concave. Therefore, their first-order approximations $ \tilde{q}_\alpha^{(t)} \left( \boldsymbol{\alpha} \right) $, $ \tilde{q}_\beta^{(t)} \left( \boldsymbol{\beta} \right) $, $ \tilde{q}_\kappa^{(t)} \left( \boldsymbol{\kappa} \right) $ satisfy $ q_\alpha \left( \boldsymbol{\alpha} \right) \leq \tilde{q}_\alpha^{(t)} \left( \boldsymbol{\alpha} \right) $, $ q_\beta \left( \boldsymbol{\beta} \right) \leq \tilde{q}_\beta^{(t)} \left( \boldsymbol{\beta} \right) $, $ q_\kappa \left( \boldsymbol{\kappa} \right) \leq \tilde{q}_\kappa^{(t)} \left( \boldsymbol{\kappa} \right) $. Now, we define
\begin{align} \nonumber
	g_1 \left( \boldsymbol{\alpha}, \boldsymbol{\beta}, \boldsymbol{\kappa} \right) & \triangleq R_\mathrm{w-sum}^\mathrm{access} \left( \boldsymbol{\alpha} \right) - \lambda_\alpha p_\alpha \left( \boldsymbol{\alpha} \right) - \lambda_\beta p_\beta \left( \boldsymbol{\beta} \right) 
	\\
	& ~~~~~~~~~~~~~~~~~~~~~~~~~~~~~~~~~ - \lambda_\kappa p_\kappa \left( \boldsymbol{\kappa} \right), \nonumber 
	\\
	g_2 \left( \boldsymbol{\alpha}, \boldsymbol{\beta}, \boldsymbol{\kappa} \right) & \triangleq \lambda_\alpha q_\alpha \left( \boldsymbol{\alpha} \right) + \lambda_\beta q_\beta \left( \boldsymbol{\beta} \right) + \lambda_\kappa q_\kappa \left( \boldsymbol{\kappa} \right), \nonumber 
	\\
	\tilde{g}_2^{(t)} \left( \boldsymbol{\alpha}, \boldsymbol{\beta}, \boldsymbol{\kappa} \right) & \triangleq \lambda_\alpha \tilde{q}_\alpha^{(t)} \left( \boldsymbol{\alpha} \right) + \lambda_\beta \tilde{q}_\beta^{(t)} \left( \boldsymbol{\beta} \right) + \lambda_\kappa \tilde{q}_\kappa^{(t)} \left( \boldsymbol{\kappa} \right). \nonumber
\end{align}

Considering the expressions above, the objective function of $ \widetilde{\mathcal{P}}_1 $ can be rewritten as $ R \left( \boldsymbol{\alpha}, \boldsymbol{\beta}, \boldsymbol{\kappa} \right) = g_1 \left( \boldsymbol{\alpha}, \boldsymbol{\beta}, \boldsymbol{\kappa} \right) - g_2 \left( \boldsymbol{\alpha}, \boldsymbol{\beta}, \boldsymbol{\kappa} \right) $ whereas the objective of $ \widetilde{\mathcal{P}}_1^{(t)} $ can be rewritten as $ \tilde{R}^{(t)} \left( \boldsymbol{\alpha}, \boldsymbol{\beta}, \boldsymbol{\kappa} \right) = g_1 \left( \boldsymbol{\alpha}, \boldsymbol{\beta}, \boldsymbol{\kappa} \right) - \tilde{g}_2^{(t)} \left( \boldsymbol{\alpha}, \boldsymbol{\beta}, \boldsymbol{\kappa} \right) $. Since $  g_2 \left( \boldsymbol{\alpha}, \boldsymbol{\beta}, \boldsymbol{\kappa} \right) \leq \tilde{g}_2^{(t)} \left( \boldsymbol{\alpha}, \boldsymbol{\beta}, \boldsymbol{\kappa} \right) $ then $ \tilde{R}^{(t)} \left( \boldsymbol{\alpha}, \boldsymbol{\beta}, \boldsymbol{\kappa} \right) $ is a lower bound for the objective of $ \widetilde{\mathcal{P}}_1 $, i.e. $ \tilde{R}^{(t)} \left( \boldsymbol{\alpha}, \boldsymbol{\beta}, \boldsymbol{\kappa} \right) \leq R \left( \boldsymbol{\alpha}, \boldsymbol{\beta}, \boldsymbol{\kappa} \right) $. Further, the equality holds when $ \boldsymbol{\alpha} = \boldsymbol{\alpha}^{\left( t-1 \right)} $, $ \boldsymbol{\beta} = \boldsymbol{\beta}^{\left( t-1 \right)} $, $ \boldsymbol{\kappa} = \boldsymbol{\kappa}^{\left( t-1 \right)} $ showing the bound tightness.

\setcounter{equation}{0}
\renewcommand{\theequation}{G.\arabic{equation}}
\renewcommand{\thesectiondis}[2]{\Alph{section}:}
\section{Proof of Proposition 10} \label{appendix_proposition_10}
Realize that $ \boldsymbol{\Theta}^{(t-1)} $ is a feasible point for $ \widetilde{\mathcal{P}}_1^{(t)} $ whereas $ \boldsymbol{\Theta}^{(t)} $ is its optimal solution. For iteration $ t $, we have that $ R \left( \boldsymbol{\alpha}, \boldsymbol{\beta}, \boldsymbol{\kappa} \right) \geq \tilde{R}^{(t)} \left( \boldsymbol{\alpha}, \boldsymbol{\beta}, \boldsymbol{\kappa} \right) $ and $ R \left( \boldsymbol{\alpha}^{(t-1)}, \boldsymbol{\beta}^{(t-1)}, \boldsymbol{\kappa}^{(t-1)} \right) = \tilde{R}^{(t)} \left( \boldsymbol{\alpha}^{(t-1)}, \boldsymbol{\beta}^{(t-1)}, \boldsymbol{\kappa}^{(t-1)} \right). $ Using these relations, 
\begin{align} \nonumber
	R \left( \boldsymbol{\alpha}^{(t)}, \boldsymbol{\beta}^{(t)}, \boldsymbol{\kappa}^{(t)} \right) & \geq \tilde{R}^{(t)} \left( \boldsymbol{\alpha}^{(t)}, \boldsymbol{\beta}^{(t)}, \boldsymbol{\kappa}^{(t)} \right) \nonumber
	\\
	& \geq \tilde{R}^{(t)} \left( \boldsymbol{\alpha}^{(t-1)}, \boldsymbol{\beta}^{(t-1)}, \boldsymbol{\kappa}^{(t-1)} \right), \nonumber
	\\
	& = R \left( \boldsymbol{\alpha}^{(t-1)}, \boldsymbol{\beta}^{(t-1)}, \boldsymbol{\kappa}^{(t-1)} \right), \nonumber
\end{align}
which shows that $ \left( \boldsymbol{\alpha}^{(t)}, \boldsymbol{\beta}^{(t)}, \boldsymbol{\kappa}^{(t)} \right) $ is more optimal for $ \mathcal{P}_1 $ than $ \left( \boldsymbol{\alpha}^{(t-1)}, \boldsymbol{\beta}^{(t-1)}, \boldsymbol{\kappa}^{(t-1)} \right) $. Further, $ R \left( \boldsymbol{\alpha}^{(t)}, \boldsymbol{\beta}^{(t)}, \boldsymbol{\kappa}^{(t)} \right) \geq R \left( \boldsymbol{\alpha}^{(t-1)}, \boldsymbol{\beta}^{(t-1)}, \boldsymbol{\kappa}^{(t-1)} \right) $ implies that $ \left( \mathbf{M}^{(t)}, \mathbf{W}^{(t)}, \mathbf{p}^{(t)} \right) $ is equally or more optimal for $ \mathcal{P}_1 $ than $ \left( \mathbf{M}^{(t-1)}, \mathbf{W}^{(t-1)}, \mathbf{p}^{(t-1)} \right) $ due to linkage with $ \mathrm{C_{20}} $, $ \mathrm{C_{23}} - \mathrm{C_{24}} $. Thus, $ \boldsymbol{\Theta}^{(t)} = \left( \mathbf{M}^{(t)}, \mathbf{W}^{(t)}, \mathbf{p}^{(t)}, \boldsymbol{\alpha}^{(t)}, \boldsymbol{\beta}^{(t)}, \boldsymbol{\kappa}^{(t)} \right) $ is more befitting for $ \mathcal{P}_1 $ than $ \boldsymbol{\Theta}^{(t-1)} $. As a result, the sequence of points $ \left\lbrace \boldsymbol{\Theta}^{(t)} \right\rbrace $ constitutes a sequence of enhanced points for $ \mathcal{P}_1 $. In addition, $ \left\lbrace \boldsymbol{\Theta}^{(t)} \right\rbrace $ is bounded because $ \tilde{R}^{(t)} \left( \boldsymbol{\alpha}, \boldsymbol{\beta}, \boldsymbol{\kappa} \right) $ is upper-bounded by $ R \left( \boldsymbol{\alpha}, \boldsymbol{\beta}, \boldsymbol{\kappa} \right) $, and $ R \left( \boldsymbol{\alpha}, \boldsymbol{\beta}, \boldsymbol{\kappa} \right) $ is upper-bounded by the multicast rate, which is ultimately constrained by the maximum transmit power from the MBS. By Cauchy's theorem, there must exist a convergent subsequence $ \left\lbrace \boldsymbol{\Theta}^{(t_n)} \right\rbrace $ such that
\noindent{
\resizebox{1.01\columnwidth}{!}{
\begin{minipage}{1.01\columnwidth}
\begin{align} \label{P10_1}
	\lim_{n \rightarrow \infty} \left[ R \left( \boldsymbol{\alpha}^{(t_n)}, \boldsymbol{\beta}^{(t_n)}, \boldsymbol{\kappa}^{(t_n)} \right) - R \left( \boldsymbol{\alpha}^\star, \boldsymbol{\beta}^\star, \boldsymbol{\kappa}^\star \right) \right] = 0,
\end{align}
\end{minipage}
}}
where $ \boldsymbol{\Theta}^\star = \left( \mathbf{M}^\star, \mathbf{W}^\star, \mathbf{p}^\star, \boldsymbol{\alpha}^\star, \boldsymbol{\beta}^\star, \boldsymbol{\kappa}^\star \right) $ is a limit point for $ \left\lbrace \boldsymbol{\Theta}^{(t_n)} \right\rbrace $. Thus, for each iteration $ t $, there exists some $ n $ such that $ t_n \leq t \leq t_{n+1} $. From (\ref{P10_1}) we obtain
\noindent{
\resizebox{1.01\columnwidth}{!}{
\begin{minipage}{1.01\columnwidth}
\begin{align} 
	& \epsilon^{(t_n)} = \lim_{n \rightarrow \infty} \left[ R \left( \boldsymbol{\alpha}^{(t_n)}, \boldsymbol{\beta}^{(t_n)}, \boldsymbol{\kappa}^{(t_n)} \right) - R \left( \boldsymbol{\alpha}^\star, \boldsymbol{\beta}^\star, \boldsymbol{\kappa}^\star \right) \right] = 0, \nonumber 
	\\
	& \epsilon^{(t_{n+1})} = \lim_{n \rightarrow \infty} \left[ R \left( \boldsymbol{\alpha}^{(t_{n+1})}, \boldsymbol{\beta}^{(t_{n+1})}, \boldsymbol{\kappa}^{(t_{n+1})} \right) - R \left( \boldsymbol{\alpha}^\star, \boldsymbol{\beta}^\star, \boldsymbol{\kappa}^\star \right) \right] = 0, \nonumber 
	\\ 
	& \epsilon^{(t)} = \lim_{n \rightarrow \infty} \left[ R \left( \boldsymbol{\alpha}^{(t)}, \boldsymbol{\beta}^{(t)}, \boldsymbol{\kappa}^{(t)} \right) - R \left( \boldsymbol{\alpha}^\star, \boldsymbol{\beta}^\star, \boldsymbol{\kappa}^\star \right) \right], \nonumber
\end{align}
\end{minipage}
}}
showing that $ \epsilon^{(t_n)} \leq \epsilon^{(t)} \leq \epsilon^{(t_{n+1})} $ and $ \lim_{t \rightarrow \infty} R \left( \boldsymbol{\alpha}^{(t)}, \boldsymbol{\beta}^{(t)}, \boldsymbol{\kappa}^{(t)} \right) = R \left( \boldsymbol{\alpha}^\star, \boldsymbol{\beta}^\star, \boldsymbol{\kappa}^\star \right) $. Therefore, each accumulation point $ \boldsymbol{\Theta}^\star = \left( \mathbf{M}^\star, \mathbf{W}^\star, \mathbf{p}^\star, \boldsymbol{\alpha}^\star, \boldsymbol{\beta}^\star, \boldsymbol{\kappa}^\star \right) $ is a KKT point \cite{marks1978:general-inner-approximation-algorithm-nonconvex-mathematical-programs, tam2017:joint-load-balancing-interference-management-heterogeneous-networks-backhaul-capacity}.

\end{document}